\documentclass[aps,prd,article,superscriptaddress,10pt]{revtex4-2}
\usepackage{color}
\usepackage{xcolor}
\usepackage{graphicx}
\usepackage[utf8]{inputenc} 
\usepackage{amsmath}
\usepackage{amssymb}
\usepackage{mathtools}
\DeclarePairedDelimiter{\abs}{\lvert}{\rvert}
\usepackage{float}
\usepackage{comment}
\usepackage{slashed}
\usepackage[normalem]{ulem}
\usepackage{xfrac}
\usepackage{graphicx}
\usepackage{float}
\definecolor{darkred}{rgb}{0.6,0,0}
\usepackage[colorlinks=true,linkcolor=darkred,citecolor=darkred,urlcolor=darkred, pdfborder={0 0 0}]{hyperref}
\usepackage{cleveref}[capital]
\usepackage{silence}
\usepackage{tikz}
\usepackage[compat=1.1.0]{tikz-feynman}
\usetikzlibrary{positioning}
\tikzfeynmanset{warn luatex=false}

\newcommand{\Ohs}{\Omega h^{2}}

\newcommand{\BR}{\mathrm{BR}}
\newcommand{\GeV}{\mathrm{GeV}}
\newcommand{\TeV}{\mathrm{TeV}}

\newcommand{\vpo}{\varphi_{1}}
\newcommand{\vpt}{\varphi_{2}}
\newcommand{\wvp}{\widetilde\varphi}

\newcommand{\mvpt}{m_{\varphi_{2}}}
\newcommand{\mzp}{M_{Z^{\prime}}}
\newcommand{\mhp}{M_{h_{\phi}}}
\newcommand{\zp}{Z^{\prime}}
\newcommand{\gp}{g^{\prime}}

\usepackage{orcidlink} 

\def\bl#1{\textcolor{blue}{#1}}

\begin{document}
%\preprint{IPM/P-2012/009}
%\vspace*{3mm}
% Allow for up to 15 columns
\setcounter{MaxMatrixCols}{15}

\title{\boldmath \color{darkred} Scotogenic dark matter from gauged \texorpdfstring{$L_\mu+L_\tau$}{Lmu+Ltau}}

\author{J. Echeverria-Puentes~\orcidlink{0000-0002-5832-103X}}
\email{j.echeverriapuentes@uandresbello.edu}
\affiliation{Departamento de Ciencias F\'isicas, Facultad de Ciencias Exactas, Universidad Andr\'es Bello, Sazi\'e 2212, Santiago, Chile}

\author{Sergey Kovalenko~\orcidlink{0000-0002-8518-2282}}
\email{sergey.kovalenko@unab.cl}
\affiliation{Departamento de Ciencias F\'isicas, Facultad de Ciencias Exactas, Universidad Andr\'es Bello, Sazi\'e 2212, Santiago, Chile}

\author{Pablo San Martin Sartori}
   \affiliation{Departamento de Ciencias F\'isicas, Facultad de Ciencias Exactas, Universidad Andr\'es Bello, Sazi\'e 2212, Santiago, Chile}
\author{Carlos A. Vaquera-Araujo~\orcidlink{0000-0001-8578-9263}}\email{vaquera@fisica.ugto.mx}
\affiliation{Secretar\'ia de Ciencia, de Humanidades, Tecnolog\'ia e Innovaci\'on, Insurgentes Sur 1582. Colonia Cr\'edito Constructor, Benito Ju\'arez 03940, Ciudad de M\'exico, Mexico}
\affiliation{Departamento de F\'isica, Divisi\'on de Ciencias e Ingenier\'ias, Campus Le\'on, Universidad de
  Guanajuato, Loma del Bosque 103, Lomas del Campestre, Le\'on 37150, Guanajuato, Mexico}
\affiliation{Dual $CP$ Institute of High Energy Physics, Colima 28045, Colima, Mexico}

\author{Yoxara Villamizar~\orcidlink{0000-0001-6792-1734}}
\email{yoxara@lpthe.jussieu.fr}
\affiliation{International Institute of Physics, Universidade Federal do Rio Grande do Norte, Campus Universitario, Lagoa Nova, Natal-RN 59078-970, Brazil}
\affiliation{ Laboratoire de Physique Theorique et Hautes Energies (LPTHE), UMR 7589,
Sorbonne Universite et CNRS, 4 place Jussieu, 75252 Paris Cedex 05, France}
\author{Jilberto Zamora Saa~\orcidlink{0000-0002-5030-7516}}

\email{jilberto.zamora@unab.cl}
\affiliation{Departamento de Ciencias F\'isicas, Facultad de Ciencias Exactas, Universidad Andr\'es Bello, Sazi\'e 2212, Santiago, Chile}
\affiliation{Millennium Institute for Subatomic Physics at High Energy Frontier (SAPHIR), Santiago, Chile}

%#####################
\begin{abstract}
\vspace{0.5cm}
In this work, we introduce the first scotogenic model based on a gauged $L_{\mu} + L_{\tau}$ symmetry whose spontaneous breaking by two units leaves a remnant matter parity $M_P$. The resulting light active neutrinos are of Majorana nature and the lightest electrically neutral state mediating the scotogenic loop is stabilized by $M_P$ and can be identified as a weakly interacting massive particle dark matter candidate. We analyze the phenomenology of a predominantly singlet-like CP-even scalar dark matter candidate and the representative regions of parameter space compatible with the observed thermal relic abundance. The $Z'$ boson is compatible with a broad range of parameters relevant for dark matter. In particular, electroweak precision constraints do not lead to significant tensions for kinetic mixing values up to $\epsilon\sim 10^{-2}$. We discuss the implications of existing searches for leptophilic gauge bosons and the conditions under which visible and invisible collider signatures become relevant.

%The model displays a rich dark matter phenomenology and can be tested in upcoming collider searches.
%\YV{We identify the predominantly singlet-like CP-even scalar $\vpt$ as dark matter and find regions reproducing the observed thermal abundance, $\Ohs\simeq0.12$. The relic-density phenomenology reflects the interplay of the gauge and scalar sectors, including the opening of $\vpt\vpt\to\zp\zp$, co-annihilation with the nearby CP-odd state $\widetilde{\varphi}_{2}$, and scalar-mediated annihilation near the pole $\mhp\simeq2\mvpt$. These results show that the gauged $L_\mu+L_\tau$ scotogenic framework can simultaneously link radiative neutrino masses with a viable thermal dark sector and provides testable signatures in searches for leptophilic gauge bosons.}
\end{abstract}
%#####################

\maketitle
\noindent

%#####################
\section{Introduction \label{Sect:intro}}
%#####################

Although the Standard Model (SM) has proven remarkably successful in describing the strong and electroweak (EW) interactions, it still leaves several profound puzzles unsolved: it offers no explanation for how neutrinos acquire their masses and fails to supply a viable and enduring candidate for Dark Matter (DM). Such limitations make it clear that an extension of the SM is necessary. A particularly promising avenue is the scotogenic model, which links the origin of neutrino masses directly to a dark-sector particle \cite{Tao:1996vb,Ma:2006km,Escribano:2020iqq}. Within this framework, the masses of active neutrinos are produced radiatively at one-loop order, while the lightest neutral mediator of the loop is rendered stable by a global $Z_2$ symmetry and serves as a Weakly Interacting Massive Particle (WIMP). This elegant structure simultaneously tackles both the dark matter enigma and 
%the neutrino mass hierarchy 
the origin of neutrino masses in a single cohesive picture.

An appealing alternative to the conventional scotogenic scenario is one in which the symmetry that stabilizes dark matter is not artificially appended, but emerges directly from a symmetry intimately related to the neutrinos \cite{AristizabalSierra2015Dynamical,PortilloSanchez2023Ultraviolet,Ma2013NewScotogenic,Brdar2014ScalarTriplet,deBoer2024AnomalyFree,CentellesChulia2020Residual,Nomura2019AlternativeBL,Wang2017BLScotogenic,Ma2019U1chi,Ma2019Cobimaximal,Nomura2022NonAbelian,Leite2020Theory,Dong2024ExtendedElectroweak,VanDong:2023thb,Leite:2020wjl,CarcamoHernandez:2020ehn,CarcamoHernandez2022MatterParity,Garnica:2024wur,CarcamoHernandez:2025yny}. In the original SM, baryon number 
$B$ and lepton number 
$L$ arise as global accidental Abelian symmetries resulting from the particle content of the theory. While these symmetries cannot be directly gauged without introducing new fields to cancel gauge anomalies, certain extensions of the SM gauge group allow gauging without requiring additional particles. This is the case %for one of 
of the three lepton number differences $L_e-L_\mu$, $L_e-L_\tau$ and $L_\mu-L_\tau$ \cite{Baek2015FlavourDependent,Baek2016DarkMatter,Lee2018RadiativelyInduced,Han2019GaugedScotogenic,Asai2019MinimalGauged,Nomura2024GeneralFlavorU1}. Among these, the most widely adopted choice is the latter \cite{Ma:2001md}, since the mediator of $L_{\mu}-L_{\tau}$ naturally evades the most stringent LEP-II constraints on a $\zp$ \cite{Carena:2004xs}, and therefore may be comparatively light. However, this feature is not exclusive of $L_{\mu}-L_{\tau}$, but is shared by every theory equipped with a mediator that couples exclusively to second- and third-generation SM leptons.

In this work, we study a scotogenic extension of the SM with gauged muon \emph{plus} tau number 
$L_\mu+L_\tau$, and its phenomenological implications. This tree-level electrophobic symmetry is not anomaly-free on its own and requires the introduction of additional fermions to cancel gauge anomalies in an analogous way to the individual gauging of $B$ and $L$ studied in \cite{FileviezPerez:2010gw, Duerr:2013dza,FileviezPerez:2024fzc}. Nevertheless, the charge assignment of the new fields are compatible with
masses around the electroweak scale, making it possible for the symmetry to be broken at relatively low energies. Moreover, through the introduction of $L_\mu+L_\tau$ charged mediators in the scotogenic loop, a remnant gauged discrete matter parity $M_P$ is preserved after the spontaneous symmetry breaking (SSB) of the new abelian symmetry, ensuring the stability
of the lightest neutral $M_P$-odd particle and potentially providing a viable WIMP DM candidate.

In the phenomenological analysis, we focus on the scenario in which the predominantly singlet-like CP-even scalar $\vpt$ is the dark matter candidate and study the regions of parameter space compatible with the observed thermal relic abundance, $\Ohs\simeq 0.12$ \cite{Planck:2018vyg}. The resulting relic abundance is controlled by the interplay between the scalar and gauge sectors. In particular, the relevant mechanisms include annihilation into pairs of new gauge bosons, $\vpt\vpt\to\zp\zp$, co-annihilation with the nearby CP-odd state $\widetilde{\varphi}_{2}$, and scalar-mediated annihilation close to the resonance $\mhp\simeq 2\mvpt$. We also examine a reduced set of constraints from electroweak precision observables, 
including the effects induced by kinetic mixing between the two Abelian gauge sectors. We also discuss the implications of existing collider searches \cite{CMS:2018yxg,Belle-II:2022yaw,Belle-II:2023ydz}.

The paper is organized as follows: The model is presented in Section \ref{Sec:model}, together with an analysis of the gauge-boson sector and the scalar potential. The scotogenic mechanism for the generation of light active-neutrino masses is shown in Section \ref{Sec:new-lepton-masses}, as well as the mass generation mechanism for the new leptons. Section \ref{sec:pheno} is devoted to the phenomenology and constraints on the model, including Electroweak Precision Observables (EWPO), charged-lepton flavor violation, dark matter and collider interpretation. Finally, a summary of the obtained results and the conclusions of the work are drawn in Section \ref{sec:conclusions}.

%#####################
\section{The Model \label{Sec:model}}
%#####################

Our model extends the Standard Model (SM) gauge symmetry by an additional gauge group  $U(1)_{L_\mu+L_\tau}$. 
The particle content and charge assignments are shown in \Cref{tab:3211}. The field content displayed in the table provide mechanisms for fermion mass generation and the cancellation of all gauge and mixed gravitational anomalies. The $L_\mu+L_\tau$ charges %in \Cref{tab:3211} 
have been scaled in order to display the minimal integer assignment. 

\begin{table}[ht]
\centering
\begin{tabular}{|c|c|c|c|c|c|}
\hline
\hspace{0.2cm} Field \hspace{0.2cm} & \hspace{0.2cm}SU(3)$_C$ \hspace{0.2cm}
& \hspace{0.2cm} SU(2)$_W$ \hspace{0.2cm} & \hspace{0.2cm}U(1)$_Y$ \hspace{%
0.2cm} & \hspace{0.2cm}U(1)$_{L_\mu+L_\tau}$ \hspace{0.2cm} &\hspace{0.2cm} $M_P$\hspace{0.2cm}   \\ 
\hline\hline
$Q_{iL}$ & \textbf{3} & $\textbf{2}  $ & $\frac{1}{6}$ & $0$ & $+$ \\ 
$u_{iR}$ & \textbf{3} & \textbf{1} & $\frac{2}{3}$ & $0$ & $+$\\ 
$d_{iR}$ & \textbf{3} & \textbf{1} & $-\frac{1}{3}$ & $0$ & $+$\\
$L_{1L}$ & \textbf{1} & \textbf{2} & $-\frac{1}{2}$ & $0$ & $+$\\ 
$e_{1R}$ & \textbf{1} & \textbf{1} & $-1$ & $0$  & $+$\\ 
$L_{aL}$ & \textbf{1} & \textbf{2} & $-\frac{1}{2}$ & $2$ & $+$\\ 
$e_{aR}$ & \textbf{1} & \textbf{1} & $-1$ & $2$  & $+$\\ 
\hline
$L'_{L}$ & \textbf{1} & \textbf{2} & $-\frac{1}{2}$ & $6$ & $+$\\ 
$e'_{R}$ & \textbf{1} & \textbf{1} & $-1$ & $6$  & $+$ \\
$n_{R}$ & \textbf{1} & \textbf{1} & $0$ & $-4$ & $+$\\  
$L''_{R}$ & \textbf{1} & \textbf{2} & $-\frac{1}{2}$ & $10$ & $+$\\ 
$e''_{L}$ & \textbf{1} & \textbf{1} & $-1$ & $10$  & $+$ \\
$n_{L}$ & \textbf{1} & \textbf{1} & $0$ & $12$ & $+$\\ 
$N_{aR}$ & \textbf{1} & \textbf{1} & $0$ & $8$  & $+$ \\ 
\hline
$S_{a\,L,R}$ & \textbf{1} & \textbf{1} & $0$ & $1$ & $-$\\ 
$F_{L,R}$ & \textbf{1} & \textbf{1} & $0$ & $6$ & $+$\\ 
$F'_{L,R}$ & \textbf{1} & \textbf{1} & $0$ & $10$ & $+$\\ 
$Y_{L,R}$ & \textbf{1} & \textbf{2} & $-\frac{1}{2}$ & $8$ & $+$\\ 
$E_{L,R}$ & \textbf{1} & \textbf{1} & $-1$ & $8$ & $+$\\ 
 \hline\hline
$H$ & \textbf{1} & \textbf{2} & $\frac{1}{2}$ & $0$ & $+$\\ 
\hline
$\phi$ & \textbf{1} & \textbf{1} & $0$ & $2$ & $+$\\ 
$\eta$ & \textbf{1} & \textbf{2} & $\       \frac{1}{2}$ & $-1$ & $-$ \\ 
$\sigma$ & \textbf{1} & \textbf{1} & $0$ & $-1$ & $-$ \\
 \hline
\end{tabular}%
\caption{Particle content of the model ($a=2,3 $, $i=1,2,3$).  \label{tab:3211} }
\end{table} 

Anomaly cancellation requires a vector-like family of leptons $L_L'$, $e'_R$, $n_R$, $L_R''$, $e''_L$, $n_L$ with non-trivial $L_\mu+L_\tau$ charges and two additional neutral states $N_{aR}$, $(a=2,3)$ with $Q^{(\mu+\tau)}=8$. The cancellation is shown explicitly in Appendix~\ref{App:anomaly}. The masses of these extra neutral leptons are generated by a tree-level seesaw mechanism mediated by two vector-like singlets $F$, $F'$ and one vector-like doublet $Y$. The scotogenic mechanism for neutrino mass generation requires two vector-like singlets $S_a$
that serve as mediators in the one-loop diagrams. An additional charged vector-like pair $E_{L,R}\sim(\mathbf1,\mathbf1,-1,8)$ is included to allow a suppressed new-fermion contribution to $h\to\gamma\gamma$ in the alignment limit, as discussed in Appendix~\ref{sec:diphoton}.

The scalar sector is extended with
the inclusion of an inert doublet $\eta$ and a singlet $\sigma$, both with $Q^{(\mu+\tau)}=-1$, needed for the closure of the scotogenic loop. The spontaneous symmetry breaking (SSB) of the $U(1)_{L_\mu+L_\tau}$  symmetry is triggered by a singlet scalar field $\phi$. 

The model symmetry $\mathcal{G}=SU(3)_C \otimes SU(2)_L \otimes U(1)_Y \otimes U(1)_{L_\mu+L_\tau}$ is spontaneously broken in two steps
\begin{eqnarray}
\mathcal{G}
&\xrightarrow{\langle \phi \rangle}&
SU(3)_C \otimes SU(2)_L \otimes U(1)_Y \otimes M_P
\nonumber\\
&\xrightarrow{\langle H \rangle}&
SU(3)_C \otimes U(1)_Q \otimes M_P.
\end{eqnarray}
Here $\langle \phi \rangle$ and $\langle H \rangle$ denote the vacuum expectation values of the singlet scalar $\phi$ and the SM Higgs doublet $H$, respectively. As will be shown below, after SSB, $L_\mu+L_\tau$ is broken by two units, and therefore, a remnant unbroken discrete symmetry known as matter parity $M_P=(-1)^{Q^{(\mu+\tau)}}$ is preserved. This symmetry is responsible for the stabilization of our weakly interacting massive particle (WIMP) dark matter (DM) candidate, identified as the lightest electrically neutral $M_P$-odd state.

\subsection{Gauge sector and kinetic mixing}

In the minimal realization considered in this work, the SM Higgs doublet is neutral under
$U(1)_{L_\mu+L_\tau}$, while the new Abelian symmetry is spontaneously broken by the scalar singlet $\phi$.
Therefore, no tree-level mass mixing between the neutral electroweak gauge boson and the new Abelian gauge boson is generated.
We therefore consider only kinetic mixing between the Abelian field strengths.

We denote by $\hat V_\mu$ the gauge field associated with
$U(1)_{L_\mu+L_\tau}$ in a basis in which the Abelian kinetic terms are not yet diagonalized. The neutral Abelian gauge sector is described by
\begin{equation}
\begin{aligned}
\mathcal{L}_{\rm Abelian}
\supset&
-\frac{1}{4}\hat B_{\mu\nu}\hat B^{\mu\nu}
-\frac{1}{4}\hat V_{\mu\nu}\hat V^{\mu\nu}
-\frac{\epsilon}{2}\hat B_{\mu\nu}\hat V^{\mu\nu}
+\frac{g^\prime}{2}\,\hat V_\mu J_{(\mu+\tau)}^\mu ,
\end{aligned}
\label{eq:abelian_kinetic_mixing}
\end{equation}
where $\hat B_\mu$ is the hypercharge gauge boson, $\epsilon$ is the kinetic mixing parameter, and $g^\prime$ is the gauge coupling associated with $U(1)_{L_\mu+L_\tau}$. The factor $1/2$ compensates the integer normalization of the charges adopted in \Cref{tab:3211}. The corresponding current is
\begin{equation}
\begin{aligned}
J_{(\mu+\tau)}^\mu
=
\sum_{\psi}
Q_\psi^{(\mu+\tau)}
\overline{\psi}\gamma^\mu \psi
+i
\sum_{\varphi}
Q_\varphi^{(\mu+\tau)}
\varphi^\dagger
{\overleftrightarrow{D}}^\mu
\varphi .
\end{aligned}
\label{eq:Umuta_current}
\end{equation}
Here $Q_\psi^{(\mu+\tau)}$ and $Q_\varphi^{(\mu+\tau)}$ denote the
$U(1)_{L_\mu+L_\tau}$ charges of the fermion and scalar fields, respectively. The SM quarks and first-generation leptons are neutral under
$U(1)_{L_\mu+L_\tau}$, while the second- and third-generation leptons carry nonzero charge. The exotic fermions and the inert scalar fields also contribute to
$J_{(\mu+\tau)}^\mu$ according to their charge assignments.

After $\phi$ develops a VEV, the new gauge boson acquires a mass
\begin{equation}
\begin{aligned}
m_V^2
=
\frac{{g^\prime}^2}{4}
\left[
Q_\phi^{(\mu+\tau)}
\right]^2
v_\phi^2 .
\end{aligned}
\label{eq:V_mass}
\end{equation}
Following Table~\ref{tab:3211}, one has
$Q_\phi^{(\mu+\tau)}=2$, and therefore
$m_V=g^\prime v_\phi$ at leading order. After diagonalization, and considering small kinetic mixing, this mass will be identified with the physical $\zp$ mass up to corrections of order $\epsilon^2$.

The kinetic mixing can be removed through the field redefinition
\begin{equation}
\begin{aligned}
\hat B_\mu
=
B_\mu-\epsilon V_\mu
+\mathcal{O}(\epsilon^2),
\quad \hat V_\mu
=
V_\mu
+\mathcal{O}(\epsilon^2).
\end{aligned}
\label{eq:kinetic_shift}
\end{equation}

After EWSB, the neutral electroweak combinations in the canonical basis are
\begin{equation}
\begin{aligned}
A_\mu^{\rm SM}
=
s_W W_\mu^3+c_W B_\mu ,
\quad
Z_\mu^{\rm SM}
=
c_W W_\mu^3-s_W B_\mu ,
\end{aligned}
\end{equation}
where $s_W=\sin\theta_W$ and $c_W=\cos\theta_W$. Even in the absence of a tree-level mass-mixing term, the kinetic shift in Eq.~\eqref{eq:kinetic_shift} induces a small mixing between $Z_\mu^{\rm SM}$ and $V_\mu$ after symmetry breaking. The physical states can be written as
\begin{equation}
\begin{aligned}
Z_\mu
=
Z_\mu^{\rm SM}
+\xi V_\mu ,
\quad
Z'_\mu
=
V_\mu
-\xi Z_\mu^{\rm SM} ,
\end{aligned}
\label{eq:physical_Z_Zp}
\end{equation}
with the mixing angle at leading order in $\epsilon$
\begin{equation}
\begin{aligned}
\xi
\simeq
\epsilon s_W
\frac{M_Z^2}{M_Z^2-\mzp^2}.
\end{aligned}
\label{eq:xi_mixing}
\end{equation}

The interactions of the physical $\zp$ with SM fermions are then given by
\begin{equation}
\begin{aligned}
\mathcal{L}_{\zp}
=
\zp_\mu
\sum_f
\overline{f}\gamma^\mu
\left[
g_{\zp f}^{L}P_L
+
g_{\zp f}^{R}P_R
\right]f ,
\end{aligned}
\label{eq:Zp_interactions_chiral}
\end{equation}
where
\begin{equation}
\begin{aligned}
g_{\zp f}^{L,R}
=
\frac{g^\prime}{2}
Q_{f_{L,R}}^{(\mu+\tau)}
-
\epsilon g_Y Y_{f_{L,R}}
-
\xi g_Z
\left(
T_3^{f_{L,R}}
-
s_W^2 q_f
\right).
\end{aligned}
\label{eq:Zp_chiral_couplings}
\end{equation}
Here $g_Z=\sqrt{g^2+g_Y^2}$, $q_f$ is the electric charge, and
$Y_{f_{L,R}}$ is the hypercharge of the corresponding chiral component.
The first term in Eq.~\eqref{eq:Zp_chiral_couplings} is the direct leptophilic interaction fixed by
$U(1)_{L_\mu+L_\tau}$, while the remaining terms arise from kinetic and $Z-Z'$ mixing.

In the light-$\zp$ regime, $M_{Z^\prime}\ll M_Z$, the induced interaction reduces to the dark-photon-like form
\begin{equation}
\begin{aligned}
\mathcal{L}_{\zp}^{\rm ind}
\simeq
\bl{-}\epsilon e c_W \zp_\mu J_{\rm EM}^\mu
+
\mathcal{O}
\left(
\epsilon
\frac{M_{Z^\prime}^2}{M_Z^2}
\right),
\end{aligned}
\label{eq:dark_photon_limit}
\end{equation}
in addition to the direct
$U(1)_{L_\mu+L_\tau}$ current interaction. Thus, the $\zp$ is leptophilic by construction, but acquires suppressed couplings to electrically charged SM fermions through kinetic mixing.

%For the SM fermions, the direct charges under $U(1)_{L_\mu+L_\tau}$ are summarized in Table~\ref{tab:Zp_SM_charges}.

% \begin{table}[t]
% \centering
% \begin{tabular}{c c c}
% \hline
% Field & $Q_f^{(\mu+\tau)}$ & Direct $\zp$ coupling \\
% \hline
% $q_L,u_R,d_R$ & $0$ & absent \\
% $L_{eL},e_R$ & $0$ & absent \\
% $L_{\mu L},\mu_R$ & $1$ & $g_{(\mu+\tau)}$ \\
% $L_{\tau L},\tau_R$ & $1$ & $g_{(\mu+\tau)}$ \\
% \hline
% \end{tabular}
% \caption{Direct charges of the Standard Model fermions under
% $U(1)_{L_\mu+L_\tau}$. Couplings to quarks and first-generation leptons are generated only through kinetic mixing.}
% \label{tab:Zp_SM_charges}
% \end{table}

The new gauge boson also couples directly to the scalar dark sector. For a scalar field $\varphi$ with charge
$Q_\varphi^{(\mu+\tau)}$, the relevant interactions arise from
\begin{equation}
\begin{aligned}
\left|D_\mu\varphi\right|^2
\supset&
i \frac{g^\prime}{2}
Q_\varphi^{(\mu+\tau)}
\zp_\mu
\varphi^\dagger
{\overleftrightarrow{\partial}}^\mu
\varphi+
\frac{{g^\prime}^2}{4}
\left[
Q_\varphi^{(\mu+\tau)}
\right]^2
\zp_\mu Z'^\mu
\varphi^\dagger\varphi .
\end{aligned}
\label{eq:scalar_Zp_interactions}
\end{equation}
Consequently, if the dark matter candidate carries nonzero
$U(1)_{L_\mu+L_\tau}$ charge, annihilation into gauge bosons,
\begin{equation}
\begin{aligned}
{\rm DM}\,{\rm DM}
\to
\zp\zp,
\end{aligned}
\end{equation}
is present. This channel directly connects the relic abundance to the gauge sector.

\subsection{Scalar potential \label{Subsec:potential}}
The most general renormalizable scalar potential consistent with $\mathcal{G}$ 
is
\begin{align}
V(H,\eta,\phi,\sigma) =& \mu_H^2 H^\dagger H + \mu_\eta^2 \eta^\dagger \eta
+ \mu_\phi^2 \phi^\dagger \phi + \mu_\sigma^2 \sigma^\dagger \sigma + \left[\kappa_1\, \eta^\dagger H \sigma + \kappa_2\, \phi \sigma \sigma + \text{h.c.}\right]\nonumber\\
&+ \lambda_1 (H^\dagger H)^2
+ \lambda_2 (\sigma^\dagger \sigma)^2
+ \lambda_3 (\phi^\dagger \phi)^2
+ \lambda_4 (\eta^\dagger \eta)^2 \nonumber\\
&+ \lambda_5 (\sigma^\dagger \sigma)(\phi^\dagger \phi)
+ \lambda_6 (\sigma^\dagger \sigma)(\eta^\dagger \eta)
+ \lambda_7 (\sigma^\dagger \sigma)(H^\dagger H) \nonumber\\
&+ \lambda_8 (\phi^\dagger \phi)(\eta^\dagger \eta)
+ \lambda_9 (\phi^\dagger \phi)(H^\dagger H)
+ \lambda_{10} (\eta^\dagger H)(H^\dagger \eta) + \lambda'_{10} (H^\dagger H)(\eta^\dagger \eta)\nonumber\\
&+ \left[\lambda_{11} H^\dagger \eta \phi \sigma + \mathrm{h.c.}\right]. \label{eq:potential}
\end{align}
%\textcolor{red}{\textbf{Note about scalar CP-violation:} If we define $\kappa_1\to|\kappa_1|e^{i\delta_1}$, $\kappa_2\to|\kappa_2|e^{i\delta_2}$ and $\lambda_{11}\to|\lambda_{11}|e^{i\delta_{11}}$ the combination $\delta_1-\delta_2+\delta_{11}$ is invariant under scalar fields rephase. Very similar to . We have to clarify that we impose a scalar CP-conservation scenario.} 
In the following analysis, We take all scalar-potential parameters to be real and assume a CP-conserving vacuum. However, this potential is compatible with CP-violation along the lines studied in Ref.~\cite{Barreiros:2020gxu}. The CP-violating scenario will be studied in a future paper.
Expanding the scalar fields around their vacuum expectation values (VEVs) as 
\begin{align}
H &= \begin{pmatrix} H^+ \\ \dfrac{1}{\sqrt{2}}(v + \xi + i\zeta) \end{pmatrix},
&
\eta &= \begin{pmatrix} \eta^+ \\ \dfrac{1}{\sqrt{2}}(\xi_\eta + i\zeta_\eta) \end{pmatrix},
&
\phi &= \dfrac{1}{\sqrt{2}}(v_\phi + \xi_\phi + i\zeta_\phi),
&
\sigma &= \dfrac{1}{\sqrt{2}}(\xi_\sigma + i\zeta_\sigma), \label{eq:scalarvevs}
\end{align}
and using the minimization conditions, we obtain
\begin{align}
\mu_\phi^2 &= -\lambda_3 v_\phi^2 - \frac{\lambda_9}{2}v^2,
\\
\mu_H^2 &= -\lambda_1 v^2 - \frac{\lambda_9}{2}v_\phi^2.
\label{eq:muCond}
\end{align}
 From these relations, we can write the vacuum expectation values in terms of the potential parameters as
\begin{equation}
v^2= \frac{2\left(\lambda_9 \mu_{\phi}^2 - 2 \lambda_3 \mu_H^2 \right)}{4 \lambda_1 \lambda_3-\lambda_9^2},
\qquad
v_{\phi}^2=\frac{2\left(\lambda_9 \mu_H^2-2 \lambda_1 \mu_{\phi}^2\right)}{4 \lambda_1 \lambda_3-\lambda_9^2}.
\label{eq:vevs}
\end{equation}

The scalar mass matrices are obtained from the second derivatives of the potential with respect to the neutral field components evaluated at the vacuum. The $CP$-odd basis is $\{\zeta,\zeta_\phi,\zeta_\eta,\zeta_\sigma\}$ while the $CP$-even basis is $\{\xi,\xi_\phi,\xi_\eta,\xi_\sigma\}$. In these bases, the corresponding squared mass matrices are given by
\begin{equation}
\mathcal{M}^2_{\text{CP-odd}} =
\begin{pmatrix}
 0 & 0 & 0 & 0 \\
 0 & 0 & 0 & 0 \\
 0 & 0 & A_{\rm odd}& B_{\rm odd} \\
 0 & 0 & B_{\rm odd} & C_{\rm odd} \\
\end{pmatrix}
,\label{MCPodd2}
\end{equation}
\begin{equation}
\mathcal{M}^2_{\text{CP-even}} =
\begin{pmatrix}
2 \lambda_{1} v^2 &  \lambda_{9} v v_{\phi}  & 0 & 0 \\
 \lambda_{9} v v_{\phi} & 2 \lambda_{3}
   v_{\phi}^2  & 0 & 0 \\
 0 & 0 & A_{\rm even}& B_{\rm even} \\
 0 & 0 & B_{\rm even} & C_{\rm even} \\
\end{pmatrix},\label{MCPeven2}
\end{equation}
with
\begin{align}
    A_{\rm odd} &=\frac{1}{2}\left[ 2\mu_{\eta}^2+  v^{2}(\lambda_{10}+\lambda'_{10})+v_{\phi}^2\lambda_{8}\right],& A_{\rm even} &=\frac{1}{2}\left[2\mu_{\eta}^2+ v^2(\lambda_{10}+\lambda'_{10})+v_{\phi}^2 \lambda_{8}\right], \nonumber\\ B_{\rm odd} &= \frac{1}{2} v \left(\sqrt{2}\kappa_{1}-\lambda_{11} v_{\phi}\right),& B_{\rm even} &=\frac{1}{2} v \left( \sqrt{2}\kappa_{1}+\lambda_{11} v_{\phi}\right),\\C_{\rm odd} &=  \frac{1}{2}
   \left(2 \mu_{\sigma} ^2+\lambda_{7} v^2+\lambda_{5}
   v_{\phi}^2-2\sqrt{2} \kappa_{2} v_{\phi}\right), & C_{\rm even} &= \frac{1}{2}\left(2 \mu_{\sigma} ^2+\lambda_{7} v^2+\lambda_{5} v_{\phi}^2+2 \sqrt{2}\kappa_{2} v_{\phi}\right).
\end{align}
The above matrices can be diagonalized to identify the physical scalar mass eigenstates. In the CP-odd sector, the fields $\zeta$ and $\zeta_\phi$ are identified as the Goldstone bosons associated with the spontaneous breaking of the electroweak and $U(1)_{L_\mu+L_\tau}$ symmetries, and are therefore absorbed by the massive gauge bosons $Z$ and $\zp$, respectively. By contrast, since $\eta$ and $\sigma$ do not develop vacuum expectation values, the fields $\zeta_\eta$ and $\zeta_\sigma$ mix into the physical pseudoscalars 
\begin{equation}
\begin{pmatrix}
    \widetilde{\varphi}_1\\
    \widetilde{\varphi}_2
\end{pmatrix}=O^{\widetilde{\varphi}}\begin{pmatrix}
    \zeta_\eta\\
    \zeta_\sigma
\end{pmatrix}\equiv\begin{pmatrix}
    \cos\theta_{\widetilde{\varphi}}&\sin\theta_{\widetilde{\varphi}}\\
    -\sin\theta_{\widetilde{\varphi}}&\cos\theta_{\widetilde{\varphi}}
\end{pmatrix}\begin{pmatrix}
    \zeta_\eta\\
    \zeta_\sigma
\end{pmatrix},
\end{equation}
with mixing angle given by
\begin{equation}
\tan2\theta_{\widetilde{\varphi}}=\frac{2B_{\rm odd}}{A_{\rm odd}-C_{\rm odd}},
\end{equation}
and squared masses 
\begin{align}
m_{\widetilde{\phi}_{1,2}}^2 &= \frac{1}{2}\left[
A_{\rm odd}+C_{\rm odd}
\pm S_{\rm odd} \sqrt{(A_{\rm odd}-C_{\rm odd})^2+4 B_{\rm odd}^2}
\right], \qquad S_{\rm odd}=\mathrm{sign}(A_{\rm odd}-C_{\rm odd}).
\end{align}
In the CP-even sector, the fields $\xi,\xi_\phi$ combine into
\begin{equation}\label{eq:CP_even_scalars}
\begin{pmatrix}
    h_1\\
    h_2
\end{pmatrix}=O^{h}\begin{pmatrix}
    \xi\\
    \xi_\phi
\end{pmatrix}\equiv\begin{pmatrix}
    \cos\theta_h&\sin\theta_h\\
    -\sin\theta_h&\cos\theta_h
\end{pmatrix}\begin{pmatrix}
    \xi\\
    \xi_\phi
\end{pmatrix},
\end{equation}
with
\begin{equation}
\tan2\theta_h=\frac{\lambda_9 v v_\phi}{\lambda_1 v^2-\lambda_3v_\phi^2},
\end{equation}
and
\begin{align}
m_{h_{1,2}}^2 &= \lambda_{1} v^{2} + \lambda_{3} v_{\phi}^{2} \pm S_h\sqrt{\lambda_{9}^{2} v^{2} v_{\phi}^{2} + \left(\lambda_{1} v^{2} - \lambda_{3} v_{\phi}^{2}\right)^{2}}, \qquad S_h=\mathrm{sign}(\lambda_1 v^2-\lambda_3v_\phi^2), \label{MH12}\end{align}
where the state mostly composed of $\xi$ is identified with the $125$ GeV Higgs boson. The remaining fields $\xi_\eta$ and $\xi_\sigma$ mix into the physical scalars
\begin{equation}
\begin{pmatrix}
    \vpo\\
    \vpt
\end{pmatrix}=O^{\varphi}\begin{pmatrix}
    \xi_\eta\\
    \xi_\sigma
\end{pmatrix}\equiv\begin{pmatrix}
    \cos\theta_\varphi&\sin\theta_\varphi\\
    -\sin\theta_\varphi&\cos\theta_\varphi
\end{pmatrix}\begin{pmatrix}
    \xi_\eta\\
    \xi_\sigma
\end{pmatrix},
\end{equation}
with 
\begin{equation}
\tan2\theta_\varphi=\frac{2B_{\rm even}}{A_{\rm even}-C_{\rm even}},
\end{equation}
and  
\begin{align}
m_{\varphi_{1,2}}^2 &= \frac{1}{2}\left[
A_{\rm even}+C_{\rm even}
\pm S_{\rm even} \sqrt{(A_{\rm even}-C_{\rm even})^2+4 B_{\rm even}^2}
\right], \qquad S_{\rm even}=\mathrm{sign}(A_{\rm even}-C_{\rm even}).
\end{align}

Finally, the mass of the inert charged scalar is

\begin{equation}\label{eq:eta_mass}
m_{\eta^\pm}^2=\mu_{\eta}^2+\frac{\lambda^\prime_{10}}{2}v^2+\frac{\lambda_{8}}{2}v_\phi^2.
\end{equation}

The scalar potential contains both cubic and quartic interactions. We require all quartic couplings to remain perturbative 
$\abs{\lambda}<4\pi$ and require the tree-level potential to be bounded from below along every direction in field space. 
Following the treatment in Refs.~\cite{Kannike:2012pe,Bhattacharyya:2015nca}, the conditions associated with directions in which only one or two fields are nonzero are
\begin{equation}
\begin{aligned}
&\lambda_1>0,
\quad
\lambda_2>0,
\quad
\lambda_3>0,
\quad
\lambda_4>0,
\quad\bar\lambda_{10}
+2\sqrt{\lambda_1\lambda_4}>0,
\quad\lambda_9+2\sqrt{\lambda_1\lambda_3}>0,
\\
&\lambda_7+2\sqrt{\lambda_1\lambda_2}>0,
\quad\lambda_8+2\sqrt{\lambda_4\lambda_3}>0,
\quad
\lambda_6+2\sqrt{\lambda_4\lambda_2}>0,
\quad\lambda_5+2\sqrt{\lambda_3\lambda_2}>0,
\end{aligned}
\label{eq:bfb-one-two-fields}
\end{equation}
where $\bar\lambda_{10}=\lambda_{10}^{\prime}+\min(0,\lambda_{10})$. 

Directions involving three nonvanishing fields additionally require
\begin{equation}
\begin{aligned}
&
2\sqrt{\lambda_1\lambda_4\lambda_3}
+
\bar\lambda_{10}\sqrt{\lambda_3}
+\lambda_9\sqrt{\lambda_4}
+\lambda_8\sqrt{\lambda_1}
+
\sqrt{
\left[
\bar\lambda_{10}
+2\sqrt{\lambda_1\lambda_4}
\right]
\left(
\lambda_9+2\sqrt{\lambda_1\lambda_3}
\right)
\left(
\lambda_8+2\sqrt{\lambda_4\lambda_3}
\right)
}>0,\\
&
2\sqrt{\lambda_1\lambda_4\lambda_2}
+
\bar\lambda_{10}\sqrt{\lambda_2}
+\lambda_7\sqrt{\lambda_4}
+\lambda_6\sqrt{\lambda_1}
+
\sqrt{
\left[
\bar\lambda_{10}
+2\sqrt{\lambda_1\lambda_4}
\right]
\left(
\lambda_7+2\sqrt{\lambda_1\lambda_2}
\right)
\left(
\lambda_6+2\sqrt{\lambda_4\lambda_2}
\right)
}>0,\\
&
2\sqrt{\lambda_4\lambda_3\lambda_2}
+\lambda_8\sqrt{\lambda_2}
+\lambda_6\sqrt{\lambda_3}
+\lambda_5\sqrt{\lambda_4}
+
\sqrt{
\left(
\lambda_8+2\sqrt{\lambda_4\lambda_3}
\right)
\left(
\lambda_6+2\sqrt{\lambda_4\lambda_2}
\right)
\left(
\lambda_5+2\sqrt{\lambda_3\lambda_2}
\right)
}>0.
\end{aligned}
\label{eq:bfb-heta-varphi}
\end{equation}

Finally, the directions in which all four fields are nonzero require
\begin{equation}
\begin{aligned}
\inf_{\substack{r>0,\;t>0\\0\leq\rho\leq1}}
\Bigg\{&
\lambda_9rt
+\lambda_7\frac{r}{t}
+\lambda_8\frac{t}{r}
+\frac{\lambda_6}{rt}
-2|\lambda_{11}|\rho
+
2
\sqrt{
\left(
\lambda_3t^2+\frac{\lambda_2}{t^2}+\lambda_5
\right)
\left(
\lambda_1r^2+\frac{\lambda_4}{r^2}
+\lambda_{10}^{\prime}
+\lambda_{10}\rho^2
\right)
}
\Bigg\}
>0.
\end{aligned}
\label{eq:bfb-four-fields}
\end{equation}

The conditions in Eqs.~\eqref{eq:bfb-one-two-fields}--\eqref{eq:bfb-four-fields} are necessary and sufficient for the quartic part of the potential to be strictly positive along every non-vanishing field direction.  In the simplifying limit $\lambda_9 \to 0$, the $H$--$\phi$ CP-even mass matrix is diagonal at
tree level, with other portal interactions unmodified. This yields $v_\phi^2=-\sfrac{\mu_\phi^2}{\lambda_3}$ and $v^2=-\sfrac{\mu_H^2}{\lambda_1}$, with unmixed mass eigenvalues $m_h^2 = 2\lambda_1 v^2$ and $m_{\phi}^2 = 2\lambda_3 v_\phi^2$.

%#####################
\section{Neutrino and exotic-lepton spectrum}\label{Sec:new-lepton-masses}

%#####################
\subsection{Scotogenic Mechanism \label{scotogenic}}

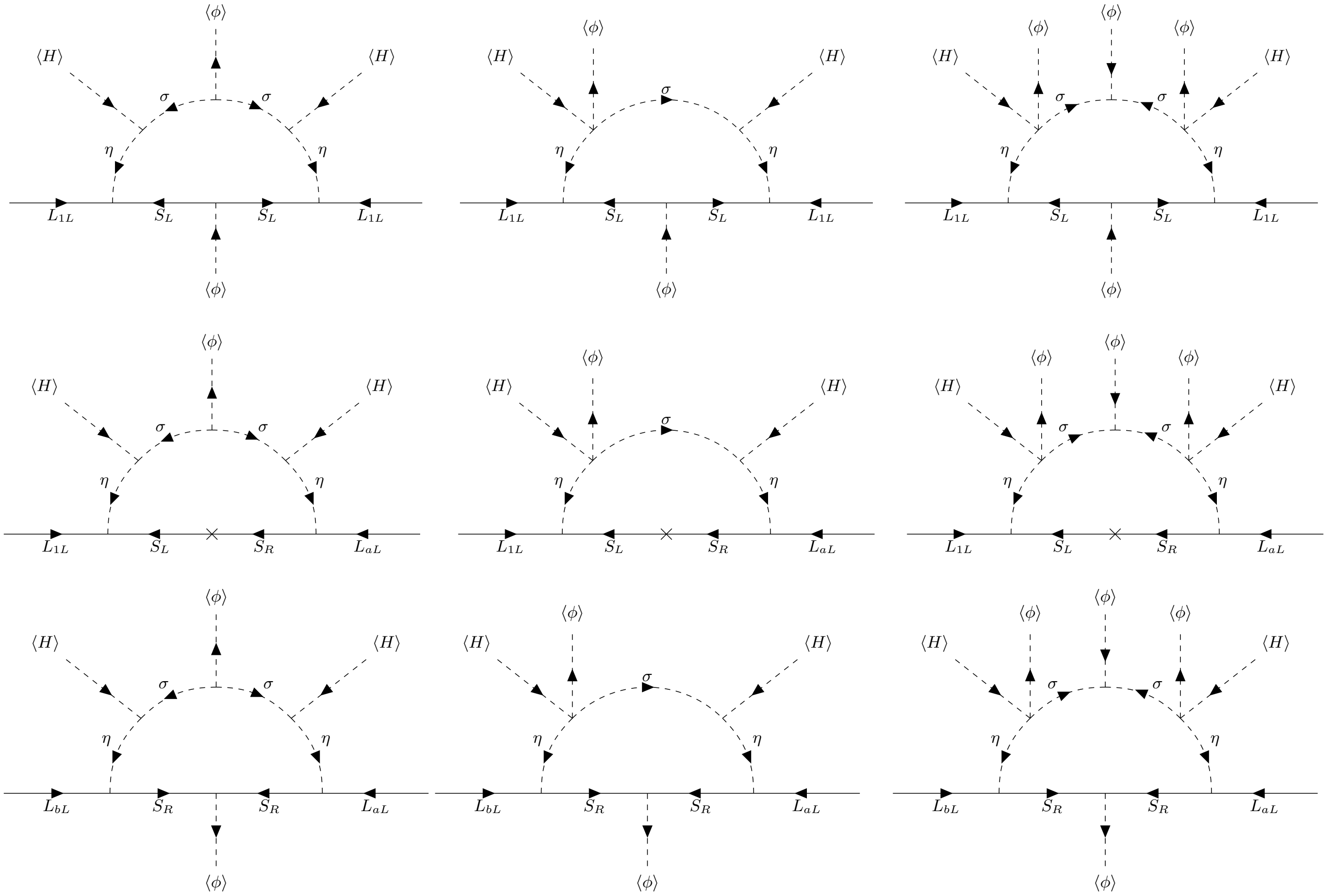
\begin{figure}[h]
    \centering
\resizebox{18 cm}{!}{
 \begin{tikzpicture}
 \begin{feynman}
 \vertex (i1);
 \vertex [right=2cm of i1] (a);
 \vertex [right=2cm of a] (b);
 \vertex [right=2cm of b] (c);
 \vertex [right=2cm of c] (f1);
 \vertex [below=0.001cm of b] (bb);
 \vertex [above=1.414cm of b] (g1);
  \vertex [above=1.414cm of g1] (g2);
  \vertex [above=2cm of b] (tb1);
 \vertex [left=1.414cm of g1] (l1);
  \vertex [left=2.828cm of g2] (l2){$\left\langle H\right\rangle$};
  \vertex [right=1.414cm of g1] (r1);
    \vertex [right=2.828cm of g2] (r2){$\left\langle H\right\rangle$};
  \vertex [above=1.414cm of tb1] (tl) {$\left\langle \phi\right\rangle$};
  \vertex [below=1.414cm of b] (bl) {$\left\langle \phi\right\rangle$}; 
 \diagram* {
 i1 -- [fermion, edge label'=$L_{1L}$] (a) -- [anti fermion, edge label'=$S_L$] (b) -- [fermion, edge label'=$S_L$] (c) -- [anti fermion, edge label'=$L_{1L}$] (f1),
% b -- [insertion=0.0] (bb),
 a -- [anti charged scalar, out=90, in=225, edge label=$\eta$] (l1),
 l1-- [anti charged scalar, out=45, in=180, edge label=$\sigma$] (tb1),
 c -- [anti charged scalar, out=90, in=-45, edge label'=$\eta$] (r1),
tb1-- [charged scalar, out=0, in=45+90, edge label=$\sigma$] (r1),
 tb1 -- [charged scalar] (tl),
 l1 -- [anti charged scalar] (l2),
% l1 -- [charged scalar, insertion=.9999] (l3),
r1 -- [anti charged scalar] (r2),
%r1 -- [charged scalar, insertion=.9999] (r3),
b -- [anti charged scalar] (bl),
 };
 \end{feynman}
 \end{tikzpicture}  \hspace{0.5cm}
 \begin{tikzpicture}
 \begin{feynman}
 \vertex (i1);
 \vertex [right=2cm of i1] (a);
 \vertex [right=2cm of a] (b);
 \vertex [right=2cm of b] (c);
 \vertex [right=2cm of c] (f1);
 \vertex [below=0.001cm of b] (bb);
 \vertex [above=1.414cm of b] (g1);
  \vertex [above=1.414cm of g1] (g2);
  \vertex [above=2cm of b] (tb1);
 \vertex [left=1.414cm of g1] (l1);
  \vertex [left=2.828cm of g2] (l2){$\left\langle H\right\rangle$};
 \vertex [above=1.6 cm of l1] (l3);
  \vertex [above=1.7 cm of l1] (l4){$\left\langle \phi\right\rangle$};
  \vertex [right=1.414cm of g1] (r1);
    \vertex [right=2.828cm of g2] (r2){$\left\langle H\right\rangle$};
  \vertex [below=1.414cm of b] (bl) {$\left\langle \phi\right\rangle$}; 
 \diagram* {
 i1 -- [fermion, edge label'=$L_{1L}$] (a) -- [anti fermion, edge label'=$S_L$] (b) -- [fermion, edge label'=$S_L$] (c) -- [anti fermion, edge label'=$L_{1L}$] (f1),
% b -- [insertion=0.0] (bb),
 a -- [anti charged scalar, out=90, in=225, edge label=$\eta$] (l1),
 l1-- [charged scalar, out=45,in=45+90, edge label=$\sigma$] (r1),
 c -- [anti charged scalar, out=90, in=-45, edge label'=$\eta$] (r1),
 l1 -- [anti charged scalar] (l2),
  l1 -- [charged scalar] (l3),
r1 -- [anti charged scalar] (r2),
b -- [anti charged scalar] (bl),
 };
 \end{feynman}
 \end{tikzpicture}\hspace{0.5 cm}
 \begin{tikzpicture}
 \begin{feynman}
 \vertex (i1);
 \vertex [right=2cm of i1] (a);
 \vertex [right=2cm of a] (b);
 \vertex [right=2cm of b] (c);
 \vertex [right=2cm of c] (f1);
 \vertex [below=0.001cm of b] (bb);
 \vertex [above=1.414cm of b] (g1);
  \vertex [above=1.414cm of g1] (g2);
  \vertex [above=2cm of b] (tb1);
 \vertex [left=1.414cm of g1] (l1);
  \vertex [left=2.828cm of g2] (l2){$\left\langle H\right\rangle$};
  \vertex [right=1.414cm of g1] (r1);
   \vertex [above=1.6 cm of l1] (l3);
  \vertex [above=1.7 cm of l1] (l4){$\left\langle \phi\right\rangle$};
    \vertex [right=2.828cm of g2] (r2){$\left\langle H\right\rangle$};
  \vertex [above=1.6 cm of r1] (r3);
  \vertex [above=1.7 cm of r1] (r4){$\left\langle \phi\right\rangle$};
  \vertex [above=1.414cm of tb1] (tl) {$\left\langle \phi\right\rangle$};
  \vertex [below=1.414cm of b] (bl) {$\left\langle \phi\right\rangle$}; 
 \diagram* {
 i1 -- [fermion, edge label'=$L_{1L}$] (a) -- [anti fermion, edge label'=$S_L$] (b) -- [fermion, edge label'=$S_L$] (c) -- [anti fermion, edge label'=$L_{1L}$] (f1),
 a -- [anti charged scalar, out=90, in=225, edge label=$\eta$] (l1),
 l1-- [charged scalar, out=45, in=180, edge label=$\sigma$] (tb1),
 c -- [anti charged scalar, out=90, in=-45, edge label'=$\eta$] (r1),
tb1-- [anti charged scalar, out=0, in=45+90, edge label=$\sigma$] (r1),
 tb1 -- [anti charged scalar] (tl),
 l1 -- [anti charged scalar] (l2),
 l1 -- [charged scalar] (l3),
 r1 -- [charged scalar] (r3),
r1 -- [anti charged scalar] (r2),
b -- [anti charged scalar] (bl),
 };
 \end{feynman}
 \end{tikzpicture} }   \\
 \vspace{0.3cm}
 \resizebox{18 cm}{!}{\begin{tikzpicture}
 \begin{feynman}
 \vertex (i1);
 \vertex [right=2cm of i1] (a);
 \vertex [right=2cm of a] (b);
 \vertex [right=2cm of b] (c);
 \vertex [right=2cm of c] (f1);
 \vertex [below=0.001cm of b] (bb);
 \vertex [above=1.414cm of b] (g1);
  \vertex [above=1.414cm of g1] (g2);
  \vertex [above=2cm of b] (tb1);
 \vertex [left=1.414cm of g1] (l1);
  \vertex [left=2.828cm of g2] (l2){$\left\langle H\right\rangle$};
  \vertex [right=1.414cm of g1] (r1);
    \vertex [right=2.828cm of g2] (r2){$\left\langle H\right\rangle$};
  \vertex [above=1.414cm of tb1] (tl) {$\left\langle \phi\right\rangle$};
%  \vertex [below=1.414cm of b] (bl) {$\left\langle \phi\right\rangle$}; 
 \diagram* {
 i1 -- [fermion, edge label'=$L_{1L}$] (a) -- [anti fermion, edge label'=$S_L$] (b) -- [anti fermion, edge label'=$S_R$] (c) -- [anti fermion, edge label'=$L_{aL}$] (f1),
 b -- [insertion=0.0] (bb),
 a -- [anti charged scalar, out=90, in=225, edge label=$\eta$] (l1),
 l1-- [anti charged scalar, out=45, in=180, edge label=$\sigma$] (tb1),
 c -- [anti charged scalar, out=90, in=-45, edge label'=$\eta$] (r1),
tb1-- [charged scalar, out=0, in=45+90, edge label=$\sigma$] (r1),
 tb1 -- [charged scalar] (tl),
 l1 -- [anti charged scalar] (l2),
% l1 -- [charged scalar, insertion=.9999] (l3),
r1 -- [anti charged scalar] (r2),
%r1 -- [charged scalar, insertion=.9999] (r3),
%b -- [anti charged scalar, insertion=.9999] (bl),
 };
 \end{feynman}
 \end{tikzpicture} \hspace{0.5 cm}
 \begin{tikzpicture}
 \begin{feynman}
 \vertex (i1);
 \vertex [right=2cm of i1] (a);
 \vertex [right=2cm of a] (b);
 \vertex [right=2cm of b] (c);
 \vertex [right=2cm of c] (f1);
 \vertex [below=0.001cm of b] (bb);
 \vertex [above=1.414cm of b] (g1);
  \vertex [above=1.414cm of g1] (g2);
  \vertex [above=2cm of b] (tb1);
 \vertex [left=1.414cm of g1] (l1);
  \vertex [left=2.828cm of g2] (l2){$\left\langle H\right\rangle$};
 \vertex [above=1.6 cm of l1] (l3);
  \vertex [above=1.7 cm of l1] (l4){$\left\langle \phi\right\rangle$};
  \vertex [right=1.414cm of g1] (r1);
    \vertex [right=2.828cm of g2] (r2){$\left\langle H\right\rangle$};
 % \vertex [below=1.414cm of b] (bl) {$\left\langle \phi\right\rangle$}; 
 \diagram* {
 i1 -- [fermion, edge label'=$L_{1L}$] (a) -- [anti fermion, edge label'=$S_L$] (b) -- [anti fermion, edge label'=$S_R$] (c) -- [anti fermion, edge label'=$L_{aL}$] (f1),
 b -- [insertion=0.0] (bb),
 a -- [anti charged scalar, out=90, in=225, edge label=$\eta$] (l1),
 l1-- [charged scalar, out=45,in=45+90, edge label=$\sigma$] (r1),
 c -- [anti charged scalar, out=90, in=-45, edge label'=$\eta$] (r1),
 l1 -- [anti charged scalar] (l2),
  l1 -- [charged scalar] (l3),
r1 -- [anti charged scalar] (r2),
 };
 \end{feynman}
 \end{tikzpicture}\hspace{0.5 cm}
 \begin{tikzpicture}
 \begin{feynman}
 \vertex (i1);
 \vertex [right=2cm of i1] (a);
 \vertex [right=2cm of a] (b);
 \vertex [right=2cm of b] (c);
 \vertex [right=2cm of c] (f1);
 \vertex [below=0.001cm of b] (bb);
 \vertex [above=1.414cm of b] (g1);
  \vertex [above=1.414cm of g1] (g2);
  \vertex [above=2cm of b] (tb1);
 \vertex [left=1.414cm of g1] (l1);
  \vertex [left=2.828cm of g2] (l2){$\left\langle H\right\rangle$};
  \vertex [right=1.414cm of g1] (r1);
   \vertex [above=1.6 cm of l1] (l3);
  \vertex [above=1.7 cm of l1] (l4){$\left\langle \phi\right\rangle$};
    \vertex [right=2.828cm of g2] (r2){$\left\langle H\right\rangle$};
  \vertex [above=1.6 cm of r1] (r3);
  \vertex [above=1.7 cm of r1] (r4){$\left\langle \phi\right\rangle$};
  \vertex [above=1.414cm of tb1] (tl) {$\left\langle \phi\right\rangle$};
 \diagram* {
 i1 -- [fermion, edge label'=$L_{1L}$] (a) -- [anti fermion, edge label'=$S_L$] (b) -- [anti fermion, edge label'=$S_R$] (c) -- [anti fermion, edge label'=$L_{aL}$] (f1),
  b -- [insertion=0.0] (bb),
 a -- [anti charged scalar, out=90, in=225, edge label=$\eta$] (l1),
 l1-- [charged scalar, out=45, in=180, edge label=$\sigma$] (tb1),
 c -- [anti charged scalar, out=90, in=-45, edge label'=$\eta$] (r1),
tb1-- [anti charged scalar, out=0, in=45+90, edge label=$\sigma$] (r1),
 tb1 -- [anti charged scalar] (tl),
 l1 -- [anti charged scalar] (l2),
 l1 -- [charged scalar] (l3),
 r1 -- [charged scalar] (r3),
r1 -- [anti charged scalar] (r2),
 };
 \end{feynman}
 \end{tikzpicture}} \\ \vspace{0.3cm}
 \resizebox{18 cm}{!}{\begin{tikzpicture}
 \begin{feynman}
 \vertex (i1);
 \vertex [right=2cm of i1] (a);
 \vertex [right=2cm of a] (b);
 \vertex [right=2cm of b] (c);
 \vertex [right=2cm of c] (f1);
 \vertex [below=0.001cm of b] (bb);
 \vertex [above=1.414cm of b] (g1);
  \vertex [above=1.414cm of g1] (g2);
  \vertex [above=2cm of b] (tb1);
 \vertex [left=1.414cm of g1] (l1);
  \vertex [left=2.828cm of g2] (l2){$\left\langle H\right\rangle$};
  \vertex [right=1.414cm of g1] (r1);
    \vertex [right=2.828cm of g2] (r2){$\left\langle H\right\rangle$};
  \vertex [above=1.414cm of tb1] (tl) {$\left\langle \phi\right\rangle$};
  \vertex [below=1.414cm of b] (bl) {$\left\langle \phi\right\rangle$}; 
 \diagram* {
 i1 -- [fermion, edge label'=$L_{bL}$] (a) -- [fermion, edge label'=$S_R$] (b) -- [anti fermion, edge label'=$S_R$] (c) -- [anti fermion, edge label'=$L_{aL}$] (f1),
% b -- [insertion=0.0] (bb),
 a -- [anti charged scalar, out=90, in=225, edge label=$\eta$] (l1),
 l1-- [anti charged scalar, out=45, in=180, edge label=$\sigma$] (tb1),
 c -- [anti charged scalar, out=90, in=-45, edge label'=$\eta$] (r1),
tb1-- [charged scalar, out=0, in=45+90, edge label=$\sigma$] (r1),
 tb1 -- [charged scalar] (tl),
 l1 -- [anti charged scalar] (l2),
% l1 -- [charged scalar, insertion=.9999] (l3),
r1 -- [anti charged scalar] (r2),
%r1 -- [charged scalar, insertion=.9999] (r3),
b -- [charged scalar] (bl),
 };
 \end{feynman}
 \end{tikzpicture} \begin{tikzpicture}
 \begin{feynman}
 \vertex (i1);
 \vertex [right=2cm of i1] (a);
 \vertex [right=2cm of a] (b);
 \vertex [right=2cm of b] (c);
 \vertex [right=2cm of c] (f1);
 \vertex [below=0.001cm of b] (bb);
 \vertex [above=1.414cm of b] (g1);
  \vertex [above=1.414cm of g1] (g2);
  \vertex [above=2cm of b] (tb1);
 \vertex [left=1.414cm of g1] (l1);
  \vertex [left=2.828cm of g2] (l2){$\left\langle H\right\rangle$};
 \vertex [above=1.6 cm of l1] (l3);
  \vertex [above=1.7 cm of l1] (l4){$\left\langle \phi\right\rangle$};
  \vertex [right=1.414cm of g1] (r1);
    \vertex [right=2.828cm of g2] (r2){$\left\langle H\right\rangle$};
  \vertex [below=1.414cm of b] (bl) {$\left\langle \phi\right\rangle$}; 
 \diagram* {
 i1 -- [fermion, edge label'=$L_{bL}$] (a) -- [fermion, edge label'=$S_R$] (b) -- [anti fermion, edge label'=$S_R$] (c) -- [anti fermion, edge label'=$L_{aL}$] (f1),
% b -- [insertion=0.0] (bb),
 a -- [anti charged scalar, out=90, in=225, edge label=$\eta$] (l1),
 l1-- [charged scalar, out=45,in=45+90, edge label=$\sigma$] (r1),
 c -- [anti charged scalar, out=90, in=-45, edge label'=$\eta$] (r1),
 l1 -- [anti charged scalar] (l2),
  l1 -- [charged scalar] (l3),
r1 -- [anti charged scalar] (r2),
b -- [charged scalar] (bl),
 };
 \end{feynman}
 \end{tikzpicture}\hspace{0.5 cm}
 \begin{tikzpicture}
 \begin{feynman}
 \vertex (i1);
 \vertex [right=2cm of i1] (a);
 \vertex [right=2cm of a] (b);
 \vertex [right=2cm of b] (c);
 \vertex [right=2cm of c] (f1);
 \vertex [below=0.001cm of b] (bb);
 \vertex [above=1.414cm of b] (g1);
  \vertex [above=1.414cm of g1] (g2);
  \vertex [above=2cm of b] (tb1);
 \vertex [left=1.414cm of g1] (l1);
  \vertex [left=2.828cm of g2] (l2){$\left\langle H\right\rangle$};
  \vertex [right=1.414cm of g1] (r1);
   \vertex [above=1.6 cm of l1] (l3);
  \vertex [above=1.7 cm of l1] (l4){$\left\langle \phi\right\rangle$};
    \vertex [right=2.828cm of g2] (r2){$\left\langle H\right\rangle$};
  \vertex [above=1.6 cm of r1] (r3);
  \vertex [above=1.7 cm of r1] (r4){$\left\langle \phi\right\rangle$};
  \vertex [above=1.414cm of tb1] (tl) {$\left\langle \phi\right\rangle$};
  \vertex [below=1.414cm of b] (bl) {$\left\langle \phi\right\rangle$}; 
 \diagram* {
 i1 -- [fermion, edge label'=$L_{bL}$] (a) -- [fermion, edge label'=$S_R$] (b) -- [anti fermion, edge label'=$S_R$] (c) -- [anti fermion, edge label'=$L_{aL}$] (f1),
 a -- [anti charged scalar, out=90, in=225, edge label=$\eta$] (l1),
 l1-- [charged scalar, out=45, in=180, edge label=$\sigma$] (tb1),
 c -- [anti charged scalar, out=90, in=-45, edge label'=$\eta$] (r1),
tb1-- [anti charged scalar, out=0, in=45+90, edge label=$\sigma$] (r1),
 tb1 -- [anti charged scalar] (tl),
 l1 -- [anti charged scalar] (l2),
 l1 -- [charged scalar] (l3),
 r1 -- [charged scalar] (r3),
r1 -- [anti charged scalar] (r2),
b -- [charged scalar] (bl),
 };
 \end{feynman}
 \end{tikzpicture} }
    %\caption{One-loop diagram for neutrino masses \label{fig:diagrams}}
    \caption{Representative one-loop topologies generating Majorana neutrino masses. Diagrams related by permutations of the external flavor indices and scalar insertions are not shown.  \label{fig:diagrams}}
\end{figure}

The Yukawa interactions relevant for neutrino mass generation, together with the vector-like mass term for the singlet fermions, are given by
\begin{equation}
-\mathcal{L}_{\text{Yuk}} \supset
y_{a1}\,\, \overline{S^c}_{aL}\widetilde{\eta}^\dagger L_{1L}
+
y_{ab}\,\overline{ S}_{aR}\widetilde{\eta}^\dagger L_{bL}
+y_{Lab}\phi^\dagger \overline{S^c}_{aL} S_{bL}+y_{Rab}\phi\overline{S}_{aR} S^c_{bR}+
m_{D\,ab}\,\overline{S}_{aR}\,S_{bL}
+\text{h.c.},
\label{eq:scotoYuk}
\end{equation}
where $\widetilde{\eta}=i\sigma_2\eta^\ast$ and $a,b=2,3$. These terms are allowed by the gauge symmetry and by the residual matter parity. Representative one-loop topologies generating Majorana neutrino masses are shown in \Cref{fig:diagrams}. In the scotogenic mechanism diagrams, the  scalar insertions of $\langle H\rangle$ and $\langle\phi\rangle$ arise from the interactions proportional to $\kappa_1$, $\kappa_2$, and $\lambda_{11}$ in Eq.(\ref{eq:potential}). 
After spontaneous symmetry breaking, the neutral singlet fermions $S_{aL}$ and $S_{aR}$ mix through the symmetric mass matrix
\begin{equation}
-\mathcal{L}_{\text{Yuk}} \supset\frac{1}{2}\left(\overline{S^c}_{L},\overline{S}_{R}\right)\underbrace{\begin{pmatrix}
   \sqrt{2}y_{L}v_{\phi} & m^T_{D} \\ m_{D}&\sqrt{2}y_{R}v_{\phi}
\end{pmatrix}}_{\mathcal M_S}\begin{pmatrix}
   S_{L}\\
   S^c_{R}
\end{pmatrix} +\mathrm{h.c.}
\end{equation}
Here $y_L=y_L^T$ and $y_R=y_R^T$ denote the symmetric
parts of the corresponding Yukawa matrices. In the ordered
weak basis $n_L=(S_{2L},S_{3L},S_{2R}^c,S_{3R}^c)^T$,
we retain the following convention for the unitary Takagi
transformation:
\begin{equation}
\begin{gathered}
 X_L=U^X n_L,\qquad n_L=(U^X)^\dagger X_L,\\
 (U^X)^*\mathcal M_S(U^X)^\dagger
 =\operatorname{diag}(M_1,M_2,M_3,M_4),\qquad M_\alpha>0.
\end{gathered}
\label{eq:scoto_takagi}
\end{equation}
The fields $X_{\alpha L}$, $\alpha=1,\dots,4$, are the
left-handed components of the physical Majorana mass eigenstates
$X_\alpha=X_{\alpha L}+X_{\alpha L}^c$.
With this ordering, a family
index $a=2,3$ corresponds to column $a-1$ in the left-singlet block
and $a+1$ in the right-singlet block.
 In terms of the physical mediators, the relevant terms for the scotogenic mechanism read
\begin{equation}
\begin{split}
-\mathcal{L}_{\text{Yuk}} \supset&
\frac{y_{a1}}{\sqrt{2}}\,\, \overline{S^c}_{aL}(\xi_\eta+i\zeta_\eta) \nu_{1L}
+
\frac{y_{ab}}{\sqrt{2}}\,\overline{ S}_{aR}(\xi_\eta+i\zeta_\eta) \nu_{bL}
+\text{h.c.}\\=&
\frac{y_{a1}}{\sqrt{2}}\,\, \overline{X^c}_{\alpha L} U^{X*}_{\alpha, a-1}[c_\varphi \vpo-s_\varphi \vpt+i(c_{\widetilde{\varphi}} \widetilde{\varphi}_1-s_{\widetilde{\varphi}} \widetilde{\varphi}_2)] \nu_{1L}
\\&+
\frac{y_{ab}}{\sqrt{2}}\, \overline{X^c}_{\alpha L} U^{X*}_{\alpha, a+1}[c_\varphi \vpo-s_\varphi \vpt+i(c_{\widetilde{\varphi}} \widetilde{\varphi}_1-s_{\widetilde{\varphi}} \widetilde{\varphi}_2)] \nu_{bL}
+\text{h.c.},
\label{eq:scotoYuk2}
\end{split}
\end{equation}
Here $c_\varphi=\cos\theta_\varphi$,
$s_\varphi=\sin\theta_\varphi$, and analogously for
$\theta_{\widetilde\varphi}$. Defining \begin{equation}
\begin{aligned}
\mathcal{Y}_{e\alpha}
&=
\sum_{a=2,3}
y_{a1}U^{X*}_{\alpha,a-1},
\quad
\mathcal{Y}_{b\alpha}=
\sum_{a=2,3}
y_{ab}U^{X*}_{\alpha,a+1},
\qquad
b=\mu,\tau.
\end{aligned}
\label{eq:scoto_effective_Y}
\end{equation}
The corresponding active neutrino mass matrix elements become
\begin{equation}
\begin{split}
m^\nu_{ij}=&\sum_{\alpha=1}^{4}\mathcal{Y}_{i\alpha}
\mathcal{Y}_{j\alpha}\mathcal F_\alpha.
\end{split}
\end{equation}
with loop functions
\begin{equation}
\begin{split}
\mathcal{F}_\alpha=&\frac{M_{\alpha}}{32\pi^2}\bigg\{c^2_\varphi\frac{m^2_{\varphi_1}}{m^2_{\vpo}-M^2_\alpha}\log\left(\frac{M^2_\alpha}{m^2_{\vpo}}\right)+s^2_\varphi\frac{m^2_{\varphi_2}}{m^2_{\vpt}-M^2_\alpha}\log\left(\frac{M^2_\alpha}{m^2_{\vpt}}\right)\\
&\qquad-c^2_{\widetilde{\varphi}}\frac{m^2_{\widetilde{\varphi}_1}}{m^2_{\widetilde{\varphi}_1}-M^2_\alpha}\log\left(\frac{M^2_\alpha}{m^2_{\widetilde{\varphi}_1}}\right)-s^2_{\widetilde{\varphi}}\frac{m^2_{\widetilde{\varphi}_2}}{m^2_{\widetilde{\varphi}_2}-M^2_\alpha}\log\left(\frac{M^2_\alpha}{m^2_{\widetilde{\varphi}_2}}\right)\bigg\}.
\end{split}
\end{equation}

\subsection{Exotic leptons\label{new_leptons}}

Writing $Y=(\nu^Y,E^Y)^T$, the gauge-invariant renormalizable interactions determining the
charged mass matrix are
\begin{align}
 -\mathcal L_C\supset{}&
 y_{e'}\overline e'_R H^\dagger L'_L
 +y_{e''}\overline L''_R H e''_L
 +M_Y\overline Y_LY_R \nonumber\\
 &+y^Y_1\overline Y_R\phi L'_L
 +y^Y_2\overline Y_L\phi^\dagger L''_R
 +\rho_R\overline E_L\phi e'_R
 +\rho_L\overline e''_L\phi E_R
 \nonumber\\
 &+z_L\overline Y_L H E_R
 +z_R\overline E_L H^\dagger Y_R
 +M_E\overline E_LE_R+\mathrm{h.c.}
 \label{eq:rev_charged_yukawa}
\end{align}
Here the vector-like charged singlets are $E_L,E_R$; $E^Y_L,E^Y_R$ denote the charged
components of vector-like doublets, whereas $e'_R$ and $e''_L$ are the singlet members of the original chiral exotic families. For clarity, we display the mass textures in the CP-conserving case, with all
parameters real and $v,v_\phi>0$. After SSB, we have
\begin{align}
 C_L&=(e'_L,e''_L,E^Y_L,E_L)^T,&
 C_R&=(e'_R,e''_R,E^Y_R,E_R)^T,\nonumber\\[6pt]
 -\mathcal L_m&=\overline C_L\mathcal M_C C_R+\mathrm{h.c.},&
 \mathcal M_C&=\begin{pmatrix}
 m_{e'} & 0 & m^Y_1 & 0\\[4pt]
 0 & m_{e''} & 0 & m^\rho_L\\[4pt]
 0 & m^Y_2 & M_Y & m^z_L\\[4pt]
 m^\rho_R & 0 & m^z_R & M_E
 \end{pmatrix},
 \label{eq:disc-charged-matrix}
\end{align}

where
\begin{align}
 m_{e'}&=\frac{y_{e'}v}{\sqrt{2}},&
 m_{e''}&=\frac{y_{e''}v}{\sqrt{2}},&
 m^Y_1&=\frac{y^Y_1v_\phi}{\sqrt{2}},&
 m^Y_2&=\frac{y^Y_2v_\phi}{\sqrt{2}},
 \nonumber\\[6pt]
 m^\rho_L&=\frac{\rho_Lv_\phi}{\sqrt{2}},&
 m^\rho_R&=\frac{\rho_Rv_\phi}{\sqrt{2}},&
 m^z_L&=\frac{z_Lv}{\sqrt{2}},&
 m^z_R&=\frac{z_Rv}{\sqrt{2}}.
 \label{eq:rev_MC_entries}
\end{align}
%Quizas añadir "These quantities denote mass-matrix entries, rather than physical mass eigenvalues."

The positive physical masses are the singular values of $\mathcal{M}_C$, defined by
\begin{equation}\label{eq:new_charged}
U_{CL}^\dagger\mathcal M_CU_{CR}=
\operatorname{diag}(m_{C_1},\ldots,m_{C_4}),
\end{equation} 
ordered increasingly. Bounds from
charged-particle searches must be applied to these masses with the appropriate
production rates, decay modes and lifetimes. The strongest bounds on a sequential charged heavy lepton \(L^\pm\) originate from direct pair-production searches at the L3 detector at LEP \cite{L3:2001xsz}, where the process $e^+e^- \to L^+L^-$ was studied, assuming that the heavy lepton decays through $L^\pm \to W^\pm \nu$,  either to a light neutrino via mixing or to an associated heavy neutral lepton. The absence of any excess of events with the expected $W$-boson and missing-energy signatures led to lower mass limits of $m_{L^\pm} > 100.8~\mathrm{GeV}$, with bounds reaching  $m_{L^\pm} > 101.9~\mathrm{GeV}$  for sufficiently large mass splittings between the charged and neutral heavy leptons. In addition, searches for detector-stable charged heavy leptons yield the slightly stronger constraint $m_{L^\pm} > 102.6~\mathrm{GeV}$, at $95\%$ confidence level. These LEP limits remain the canonical direct constraints quoted by the Particle Data Group (PDG)\cite{ParticleDataGroup:2026mpi} for a heavy charged lepton. 

All exotic neutral states can acquire masses through a Dirac type-I see-saw mechanism mediated by the vector-like fermions  $F_{L,R}$, $F'_{L,R}$, and $Y_{L,R}$. 
The remaining interactions needed for the neutral masses are
\begin{align}
 -\mathcal L_N\supset{}&
 y^F_1\overline F_R\widetilde H^\dagger L'_L
 +y^F_{2a}\overline N_{aR}\phi F_L
 +y^F_3\overline n_R^{\,c}\phi^\dagger F_R \nonumber\\
 &+y^{F'}_1\overline F'_L\widetilde H^\dagger L''_R
 +y^{F'}_2\overline n_L\phi F'_R
 +y^{F'}_{3a}\overline F'_L\phi N_{aR} \nonumber\\
 &+y^Y_{3a}\overline N_{aR}\widetilde H^\dagger Y_L
 +M_F\overline F_LF_R+M_{F'}\overline F'_LF'_R
 +\mathrm{h.c.},
 \label{eq:rev_neutral_yukawa}
\end{align}
in addition to the $y^Y_{1,2}$ and $M_Y$ terms already included in
Eq.~\eqref{eq:rev_charged_yukawa}. Here
$\widetilde H=i\sigma_2H^*$, and the sum over $a$ is understood. 
In the bases
\begin{equation}
 \begin{aligned}
 N_L&=(\nu'_L,n_R^c,n_L,F_L,F'_L,\nu^Y_L)^T,\\
 N_R&=(\nu''_R,N_{2R},N_{3R},F_R,F'_R,\nu^Y_R)^T,
 \end{aligned}
 \label{eq:rev_neutral_bases}
\end{equation}
after SSB, the mass term is $-\mathcal L_m\supset
\overline N_L\mathcal M_NN_R+\mathrm{h.c.}$, with
\begin{equation}
 \mathcal M_N=
 \begin{pmatrix}
 0 & 0 & 0 & m^F_1 & 0 & m^Y_1\\[2pt]
 0 & 0 & 0 & m^F_3 & 0 & 0\\[2pt]
 0 & 0 & 0 & 0 & m^{F'}_2 & 0\\[2pt]
 0 & m^F_{22} & m^F_{23}& M_F & 0 & 0\\[2pt]
 m^{F'}_1 & m^{F'}_{32} & m^{F'}_{33} & 0 & M_{F'} & 0\\[2pt]
 m^Y_2 & m^Y_{32} & m^Y_{33} & 0 & 0 & M_Y
 \end{pmatrix},
 \label{eq:disc-neutral-matrix}
\end{equation}
where, for the real parameters used in the displayed texture,
\begin{align}
 m^F_1&=\frac{y^F_1v}{\sqrt2},&
 m^{F'}_1&=\frac{y^{F'}_1v}{\sqrt2},&
 m^Y_{3a}&=\frac{y^Y_{3a}v}{\sqrt2},\nonumber\\
 m^F_3&=\frac{y^F_3v_\phi}{\sqrt2},&
 m^{F'}_2&=\frac{y^{F'}_2v_\phi}{\sqrt2},&
 m^F_{2a}&=\frac{y^F_{2a}v_\phi}{\sqrt2},&
 m^{F'}_{3a}&=\frac{y^{F'}_{3a}v_\phi}{\sqrt2}.
 \label{eq:rev_MN_entries}
\end{align}
The entries $m^Y_1$, $m^Y_2$ and $M_Y$ are precisely those in Eq.~\eqref{eq:disc-charged-matrix}.
We define $U_{NL}^\dagger\mathcal M_NU_{NR}=
\operatorname{diag}(m_{N_1},\ldots,m_{N_6})$, with positive, ordered masses.

The renormalizable interactions preserve an accidental exotic-fermion number
$D$. The charge assignment is $D=+1$ for
$L'_L,e'_R,L''_R,e''_L,n_L,N_{aR},F_{L,R},F'_{L,R},Y_{L,R},E_{L,R}$,
$D(n_R)=-1$, and $D=0$ for all scalars, SM fields and scotogenic $S$ fields.
The opposite sign for $n_R$ is required by the charge-conjugated field in
Eq.~\eqref{eq:rev_neutral_yukawa}. All components of both bases in
Eq.~\eqref{eq:rev_neutral_bases} then have $D=+1$, and consequently there are six neutral Dirac fermions, described by a symmetric
$12\times12$ Weyl mass matrix, with its Takagi
masses occurring in degenerate pairs when the matrix has full rank.
The electroweak-doublet components of the neutral exotic fermions induce
couplings of the physical mass eigenstates to the $Z$ boson. Constraints
therefore apply to the physical couplings and to all kinematically open
channels $Z\to N_i\bar N_j$.
Requiring $m_{N_i}>M_Z/2$ for all states is a sufficient condition to close
these two-body channels, but is not necessary when the relevant couplings
are sufficiently suppressed. Bounds on lighter states depend on their
mixings and, for the invisible width, on their decay signatures.

\begin{figure}[h]
    \centering   
 \begin{tikzpicture}
 \begin{feynman}
 \vertex (i1);
 \vertex [right=2cm of i1] (a);
 \vertex [right=2cm of a] (b);
 \vertex [right=2cm of b] (c);
 \vertex [right=2cm of c] (f1);
 \vertex [below=0.001cm of b] (bb);
 \vertex [above=2cm of a] (g1){$\left\langle \phi\right\rangle$};
 \vertex [above=2cm of c] (g2){$\left\langle \phi\right\rangle$};
 \diagram* {
 i1 -- [fermion, edge label'=$L''_{R}$] (a) -- [fermion, edge label'=$Y_L$] (b) -- [fermion, edge label'=$Y_R$] (c) -- [fermion, edge label'=$L'_{L}$] (f1),
 b -- [insertion=0.0] (bb),
 a -- [ charged scalar] (g1),
 c-- [charged scalar] (g2),
 };
 \end{feynman}
 \end{tikzpicture} 
 \hspace{0.5cm}
  \begin{tikzpicture}
 \begin{feynman}
 \vertex (i1);
 \vertex [right=2cm of i1] (a);
 \vertex [right=2cm of a] (b);
 \vertex [right=2cm of b] (c);
 \vertex [right=2cm of c] (f1);
 \vertex [below=0.001cm of b] (bb);
 \vertex [above=2cm of a] (g1){$\left\langle \phi\right\rangle$};
 \vertex [above=2cm of c] (g2){$\left\langle H\right\rangle$};
 \diagram* {
 i1 -- [fermion, edge label'=$L'_{L}$] (a) -- [fermion, edge label'=$Y_R$] (b) -- [fermion, edge label'=$Y_L$] (c) -- [fermion, edge label'=$N_{aR}$] (f1),
 b -- [insertion=0.0] (bb),
 a -- [anti charged scalar] (g1),
 c-- [anti charged scalar] (g2),
 };
 \end{feynman}
 \end{tikzpicture}
\\
 \vspace{0.5cm}
 \begin{tikzpicture}
 \begin{feynman}
 \vertex (i1);
 \vertex [right=2cm of i1] (a);
 \vertex [right=2cm of a] (b);
 \vertex [right=2cm of b] (c);
 \vertex [right=2cm of c] (f1);
 \vertex [below=0.001cm of b] (bb);
 \vertex [above=2cm of a] (g1){$\left\langle H\right\rangle$};
 \vertex [above=2cm of c] (g2){$\left\langle \phi\right\rangle$};
 \diagram* {
 i1 -- [fermion, edge label'=$L'_{L}$] (a) -- [fermion, edge label'=$F_R$] (b) -- [fermion, edge label'=$F_L$] (c) -- [fermion, edge label'=$N_{aR}$] (f1),
 b -- [insertion=0.0] (bb),
 a -- [anti charged scalar] (g1),
 c-- [anti charged scalar] (g2),
 };
 \end{feynman}
 \end{tikzpicture} 
 \hspace{0.5cm}
 \begin{tikzpicture}
 \begin{feynman}
 \vertex (i1);
 \vertex [right=2cm of i1] (a);
 \vertex [right=2cm of a] (b);
 \vertex [right=2cm of b] (c);
 \vertex [right=2cm of c] (f1);
 \vertex [below=0.001cm of b] (bb);
 \vertex [above=2cm of a] (g1){$\left\langle H\right\rangle$};
 \vertex [above=2cm of c] (g2){$\left\langle \phi\right\rangle$};
 \diagram* {
 i1 -- [fermion, edge label'=$L''_{R}$] (a) -- [fermion, edge label'=$F'_L$] (b) -- [fermion, edge label'=$F'_R$] (c) -- [fermion, edge label'=$n_{L}$] (f1),
 b -- [insertion=0.0] (bb),
 a -- [anti charged scalar] (g1),
 c-- [anti charged scalar] (g2),
 };
 \end{feynman}
 \end{tikzpicture} 
\\
 \vspace{0.5cm}
\begin{tikzpicture}
 \begin{feynman}
 \vertex (i1);
 \vertex [right=2cm of i1] (a);
 \vertex [right=2cm of a] (b);
 \vertex [right=2cm of b] (c);
 \vertex [right=2cm of c] (f1);
 \vertex [below=0.001cm of b] (bb);
 \vertex [above=2cm of a] (g1){$\left\langle \phi\right\rangle$};
 \vertex [above=2cm of c] (g2){$\left\langle \phi\right\rangle$};
 \diagram* {
 i1 -- [fermion, edge label'=$N_{aR}$] (a) -- [fermion, edge label'=$F_L$] (b) -- [fermion, edge label'=$F_R$] (c) -- [anti fermion, edge label'=$n_{R}$] (f1),
 b -- [insertion=0.0] (bb),
 a -- [charged scalar] (g1),
 c-- [charged scalar] (g2),
 };
 \end{feynman}
 \end{tikzpicture} 
\hspace{0.5cm}
 \begin{tikzpicture}
 \begin{feynman}
 \vertex (i1);
 \vertex [right=2cm of i1] (a);
 \vertex [right=2cm of a] (b);
 \vertex [right=2cm of b] (c);
 \vertex [right=2cm of c] (f1);
 \vertex [below=0.001cm of b] (bb);
 \vertex [above=2cm of a] (g1){$\left\langle \phi\right\rangle$};
 \vertex [above=2cm of c] (g2){$\left\langle \phi\right\rangle$};
 \diagram* {
 i1 -- [fermion, edge label'=$N_{aR}$] (a) -- [fermion, edge label'=$F'_L$] (b) -- [fermion, edge label'=$F'_R$] (c) -- [ fermion, edge label'=$n_{L}$] (f1),
 b -- [insertion=0.0] (bb),
 a -- [anti charged scalar] (g1),
 c-- [anti charged scalar] (g2),
 };
 \end{feynman}
 \end{tikzpicture} 
    \caption{Tree-level mass generation mechanism for the new exotic neutral leptons.   \label{fig:diagrams2}}
\end{figure}
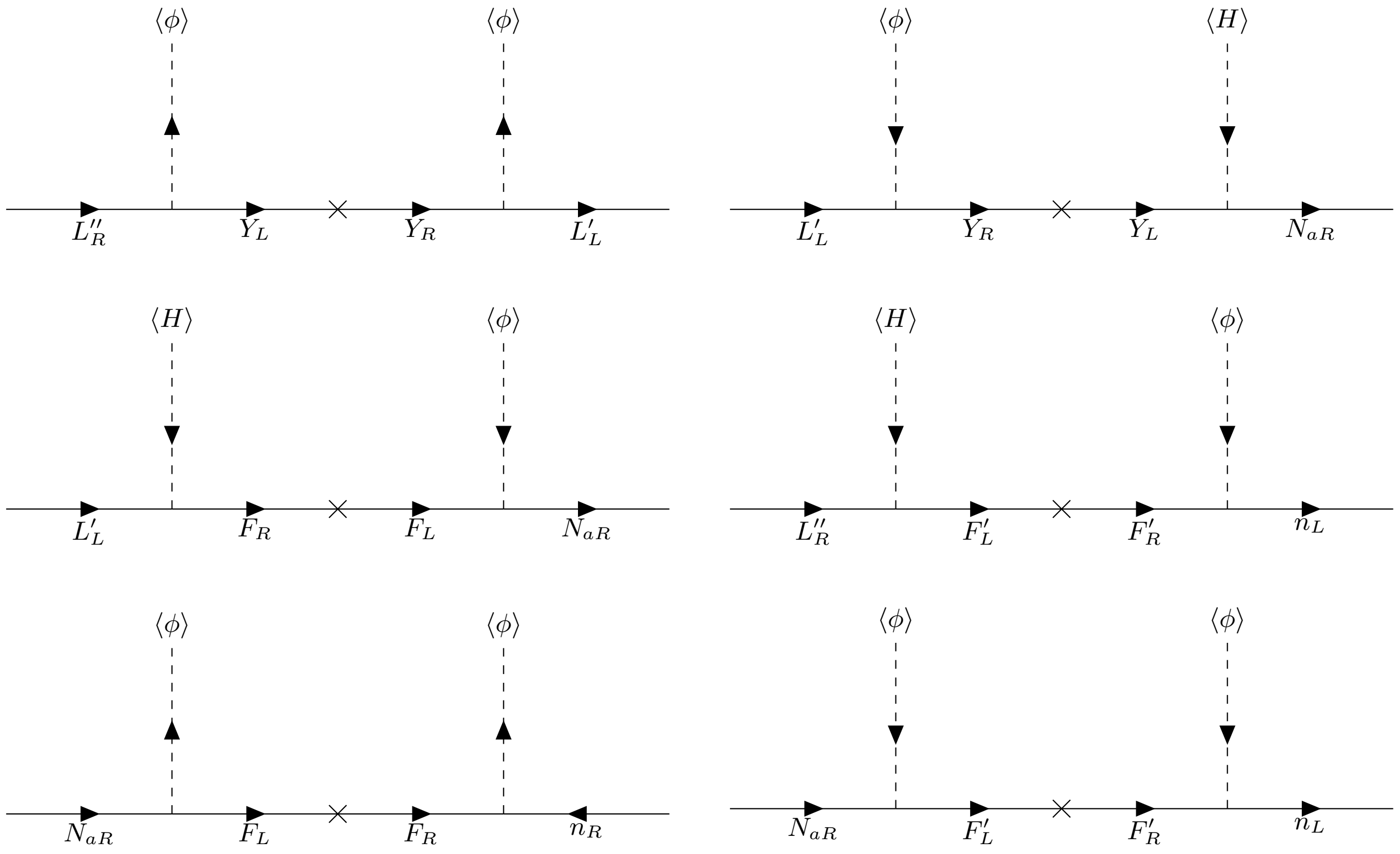

The lightest $D$-carrying state is stable within the perturbative
renormalizable mass-and-decay treatment. There are two scenarios: a light stable charged particle or an electrically neutral WIMP DM candidate. The latter has been studied in the context of gauged lepton number in Refs.~\cite{Schwaller:2013hqa,Debnath:2023akj}. Notice, however, that this stability has a different origin from the
matter parity stabilizing the lightest scotogenic state. The subgroup $P_D\equiv(-1)^D$ makes $N_i,C_j$ odd and all scalars, SM fields
and scotogenic $S$ fields even. In the present model $P_D$ is a
subgroup of the accidental global classical symmetry, not an independently
established residual gauge symmetry.
Gauge invariance preserves the residual symmetry $M_P$ at any operator
dimension, as long as no $M_P$-odd field develops a vacuum expectation
value. It does not, however, protect the accidental parity $P_D$ against
higher-dimensional interactions. For example, it permits the
dimension-five operator
\begin{equation}
 \Delta\mathcal L_{\rm eff}\supset
 \frac{c_a}{\Lambda_{\rm UV}}\,
 \phi\,\sigma^\dagger\,\overline{n_R^c}S_{aR}
 +\mathrm{h.c.},
\end{equation}
with dimensionless coefficients $c_a$ and effective-theory cutoff scale
$\Lambda_{\rm UV}$. This operator violates $P_D$ while preserving $M_P$.
Such interactions may allow exotic-fermion decays, depending on the
spectrum, mixings and effective couplings. In the relic-density analysis we assume a negligible
contribution from the exotic-lepton sector and focus on the scotogenic
scalar component. A quantitative treatment of the exotic
abundance and possible decays is deferred to future work.

%#####################
\section{Phenomenology and constraints \label{sec:pheno}}
%#####################

\subsection{Electroweak precision observables}

The massive $\zp$ boson arising from the spontaneous breaking of the $U(1)_{L_\mu+L_\tau}$ gauge symmetry is a central ingredient in the phenomenology of the model. It is therefore necessary to study the relevant parameter space and the constraints on the gauge sector. 

As discussed in Sec.~\ref{Sec:model}, the SM Higgs doublet is neutral under $U(1)_{L_\mu+L_\tau}$. Therefore, no tree-level $Z$--$\zp$ mass mixing is generated. The only source of mixing considered is the Abelian kinetic mixing as can be seen from Eq.~\eqref{eq:abelian_kinetic_mixing}. The presence of this kinetic mixing modifies the mass and mixing of the $Z$ boson which thereby inducing shifts in EWPOs. It is therefore necessary to identify the regions of the gauge sector parameter space that are compatible with electroweak precision data and can be consistently used in the dark matter analysis.

In this subsection we perform an analysis of a reduced set of EWPOs rather than a complete global electroweak fit. Following Ref.~\cite{Bertuzzo:2025ejw}, we perform a numerical scan over the physical mass of the $Z^\prime$ gauge boson, the $U(1)_{L_\mu+L_\tau}$ gauge coupling, and the kinetic mixing parameter, $\left\{ M_{\zp},\, g^\prime,\, \epsilon \right\}$. The reduced set of EWPOs is defined as

\begin{equation} 
\begin{aligned} 
    \mathcal{O}_{\rm EWPO} = \left\{ M_W,\, \Gamma_Z,\, \Gamma_{\rm inv},\, \sin^2\theta_{\rm eff}^{e} \right\},
\end{aligned} 
\end{equation} 
whose explicit expressions in terms of the $Z^\prime$ parameters are collected in Appendix~\ref{app:EWPO} and the experimental values used in the numerical scan are summarized in Table~\ref{tab:EWPO}.

\begin{table}[t]
\centering
\begin{tabular}{c c}
\hline
Observable & Experimental value \\
\hline
$M_W$ & $80.3692 \pm 0.0133~{\rm GeV}$ \\
$\Gamma_Z$ & $2.4955 \pm 0.0023~{\rm GeV}$ \\
$\Gamma_{\rm inv}$ & $499.2 \pm 1.5~{\rm MeV}$ \\
$\sin^2\theta_{\rm eff}^{e}$ & $0.23153 \pm 0.00016$ \\
\hline
\end{tabular}
\caption{
Experimental inputs at the $Z$ pole used in the electroweak precision analysis \cite{ALEPH:2005ab,ParticleDataGroup:2024cfk}.
}
\label{tab:EWPO}
\end{table}

For each point in the parameter space $\left\{M_{\zp},g^\prime,\epsilon\right\}$, we define
\begin{equation} 
\begin{aligned} 
    \chi^2_{\rm EWPO} \left( M_{\zp},g^\prime,\epsilon \right) = \sum_{i\in\mathcal{O}_{\rm EWPO}} \frac{ \left[ \mathcal{O}_i^{\rm th} \left( M_{\zp},g^\prime,\epsilon \right) - \mathcal{O}_i^{\rm exp} \right]^2 }{ \sigma_{i,\rm exp}^2 }, 
\end{aligned} 
\end{equation} 
where $\sigma_{i,\rm exp}$ denotes the experimental uncertainty of observable $O_i$. We then define
\begin{equation} 
\begin{aligned}
    \Delta\chi^2_{\rm EWPO} = \chi^2_{\rm EWPO} - \chi^2_{\rm EWPO,min}, 
\end{aligned} 
\end{equation} 
where $\chi^2_{\rm EWPO,min}$ is the minimum value found in the scanned region. In the two-dimensional planes shown below, the contours 
\begin{equation} 
\begin{aligned} 
    \Delta\chi^2_{\rm EWPO}=2.30, \qquad \Delta\chi^2_{\rm EWPO}=5.99, 
\end{aligned} 
\end{equation} 
are interpreted as the approximate $68\%$ and $95\%$ confidence-level contours, respectively. 

\begin{figure}[tbp]
    \centering
    \includegraphics[width=0.495\linewidth]{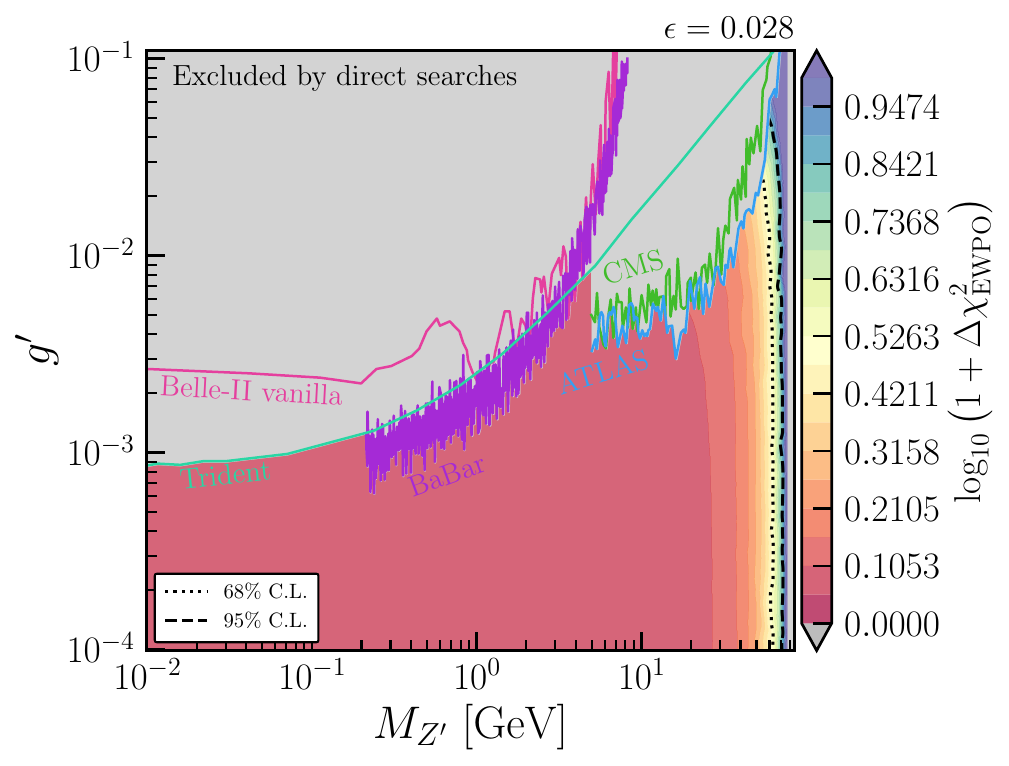}
    \hspace{-0.2cm}\includegraphics[width=0.495\linewidth]{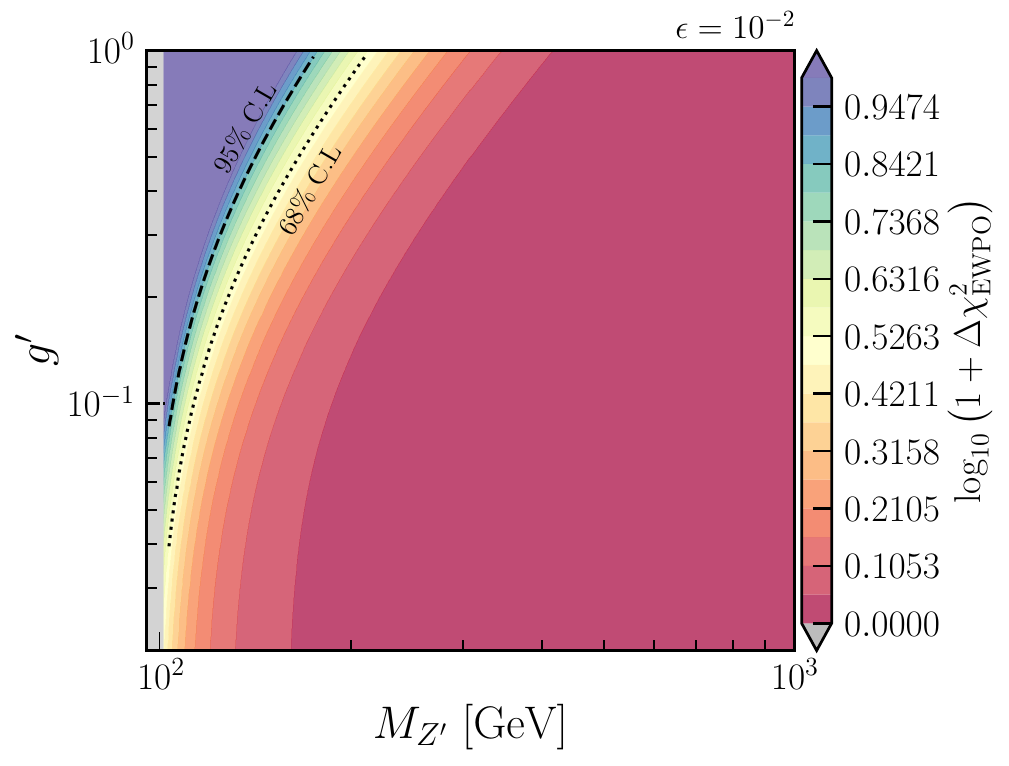}
    \caption{Constraints on the $\left(M_{Z^\prime},g^\prime\right)$ parameter space for fixed values of the kinetic mixing parameter $\epsilon$. The color scale shows $\log_{10}(1+\Delta\chi^2_{\rm EWPO})$, obtained from the electroweak precision analysis, while the solid contours denote $\Delta\chi^2_{\rm EWPO}=2.30$ and $\Delta\chi^2_{\rm EWPO}=5.99$. The vertical gray band indicates the quasi-degenerate region $M_{Z^\prime}\simeq M_Z$, where the small-mixing approximation used in the EWPO analysis is not reliable. External bounds from BaBar, neutrino trident production and direct searches are overlaid as complementary phenomenological constraints.}
    \label{fig:EWPO}
\end{figure}

The results are shown in Fig.~\ref{fig:EWPO}. Both panels display the constraints in the $\left(M_{\zp},g^\prime\right)$ plane for a fixed representative values of the kinetic mixing parameter $\epsilon$. Here we also include, as phenomenological references, direct-search exclusion curves obtained in the context of vanilla $U(1)_{L_\mu-L_\tau}$ models \cite{Belle-II:2022yaw,Belle-II:2023ydz,CMS:2018yxg,ATLAS:2023vxg}. These curves are not part of the EWPO fit and are not used in the definition of $\chi^2_{\rm EWPO}$, nevertheless, they are useful to indicate regions of the parameter space already under tension from direct searches.

For $M_{\zp}<M_Z$, EWPOs mainly constrain the maximum allowed value of the kinetic mixing. Apart from the regions already excluded by $U(1)_{L_\mu-L_\tau}$ direct-search bounds, EWPO effects become appreciable for $\epsilon\sim10^{-2}$. By contrast, for $\epsilon\lesssim 10^{-3}$, the induced deviations in $M_W$, $\Gamma_Z$, $\Gamma_{\rm inv}$, and $\sin^2\theta_{\rm eff}^{e}$ remain small over most of the parameter space considered.

For $M_{\zp}>M_Z$, the EWPO sensitivity is still driven primarily by the kinetic mixing, rather than by the value of the gauge coupling $g^\prime$. Although $g^\prime$ controls the direct couplings of the $\zp$ to second- and third-generation leptons, its impact on $Z$-pole observables is suppressed by the induced mixing angle $\xi$. In the large-mass regime,
\begin{equation} 
\begin{aligned} 
    \xi \simeq -\epsilon s_W \frac{M_Z^2}{M_{\zp}^2}, \qquad M_{\zp}\gg M_Z, 
\end{aligned} 
\end{equation}
so that the electroweak corrections decouple as $M_Z^2/M_{\zp}^2$. In particular, for masses above the TeV scale, no significant tensions with the reduced EWPO set is observed within the ranges of $\epsilon$ and $g^\prime$ considered.

In the dark matter analysis that follows, we therefore restrict ourselves to representative gauge-sector benchmarks lying within the region compatible with EWPOs, excluding both the quasi-degenerate regime $M_{\zp}\simeq M_Z$ and large kinetic mixing.

\subsection{Charged-lepton flavor violation}

Charged-lepton flavor violation (CLFV) has been studied in several gauged $U(1)_{L_\mu-L_\tau}$ models, where its origin depends strongly on the particle content. In particular, models with heavy neutrinos, supersymmetric multiplets, or explicitly non-diagonal gauge interactions can generate different sources of CLFV as can be found in Refs.~\cite{BanerjeeByaktiRoy2018,ExtendedLeftRightLmuLtau2022,HuangSun2024}. The existing literature provide a useful guide for identifying possible sources of CLFV, but their predictions cannot be directly applied to this $U(1)_{L_\mu+L_\tau}$ model.

We now discuss qualitatively the main possible sources of CLFV in this model.

\paragraph{\bf Tree-level gauge contribution}

In gauged $U(1)_{L_\mu-L_\tau}$ models, CLFV can arise at tree level when the rotation from the gauge-charge basis to the charged-lepton mass basis is nontrivial \cite{HuangSun2024}. In that case, the physical $Z'$ couplings are controlled by $Q^{\mathrm{phys}}
=
U_\ell^\dagger Q\,U_\ell$, and the non-universal $\mu$ and $\tau$ charges, $Q_{(L_\mu-L_\tau)}
\propto
\operatorname{diag}(0,1,-1)$ (up to an overall normalization),
allow off-diagonal entries such as $(Q^{\mathrm{phys}})_{\mu\tau}\neq0$. This, in turn, admits tree-level flavor-changing $Z'$ interactions.

The situation is qualitatively different for this model. Here, $Q_{(L_\mu+L_\tau)}
\propto
\operatorname{diag}(0,1,1)$, and consequently, for any unitary rotation acting within the $\mu$--$\tau$ subspace, $U_{\mu\tau}^\dagger
Q_{L_\mu+L_\tau}
U_{\mu\tau}
=
Q_{L_\mu+L_\tau}$. Thus, no off-diagonal $Z'\bar{\mu}\tau$ coupling can be generated by basis rotation. Additionally, the couplings induced by kinetic mixing are also flavor diagonal, so they do not modify this conclusion.

\paragraph{\bf Scotogenic contribution.}
The scotogenic sector provides the standard one-loop contribution to CLFV. The same Yukawa interactions responsible for radiative neutrino-mass generation couple the charged inert scalar $\eta^\pm$ to the SM charged leptons and to the physical Majorana fermions $X_\alpha$. Schematically, using the effective couplings
$\mathcal Y_{i\alpha}$ defined in
Eq.~\eqref{eq:scoto_effective_Y}, the interaction can be written as
\begin{equation}
-\mathcal{L}_{Y}
\supset
-\sum_{i=e,\mu,\tau}
\sum_{\alpha=1}^{4}
\mathcal{Y}_{i\alpha}\,
\overline{X}_\alpha P_L\ell_i\,\eta^+
+\mathrm{h.c.},
\end{equation}
These interactions generate the standard scotogenic dipole contribution \cite{Toma:2013zsa} involving $\eta^\pm$ and $X_\alpha$.
We define the coefficient for
$\ell_i\to\ell_j\gamma$ by
\begin{equation*}
 \mathcal L_{\rm dip}\supset
 e\,m_{\ell_i} C^\gamma_{ji}\,
 \overline{\ell_j}\sigma^{\mu\nu}P_R\ell_i
 F_{\mu\nu}+\mathrm{h.c.},\qquad
 \sigma^{\mu\nu}=\frac{i}{2}[\gamma^\mu,\gamma^\nu].
\end{equation*}
Neglecting the chirality-flipped term proportional to
$m_{\ell_j}$, its flavor structure is
\begin{equation}
C^\gamma_{ji}
\propto
\frac{1}{16\pi^2m_{\eta^+}^2}
\sum_{\alpha=1}^{4}
\mathcal{Y}_{j\alpha}^{*}
\mathcal{Y}_{i\alpha}
F_\gamma\left(
\frac{M_\alpha^2}{m_{\eta^\pm}^2}
\right),
\end{equation}
which induces the radiative decays $\mu\to e\gamma$, $\tau\to e\gamma$ and $\tau\to\mu\gamma$. The same interaction also generates non-dipole photon penguins and box diagrams contributing to three-body decays and $\mu$--$e$ conversion \cite{Toma:2013zsa}.

The Yukawa interactions of the scotogenic sector have a characteristic flavor form. The electron doublet couples to the $S_L$ fermions through $y_{a1}$ couplings, whereas the muon and tau doublets couple to the $S_R$ fermions through $y_{ab}$ couplings. As a consequence, connecting the electron sector to the $\mu$ and $\tau$ sectors require the physical Majorana fermions to be a mixing of $S_L$ and $S_R^c$. This distinguishes $\mu$--$\tau$ transitions from processes involving the electron. The phenomenological consequences of this must be explored with a dedicated numerical analysis.

\paragraph{\bf Radiatively induced $Z'$ contribution.}
Although the tree-level $Z'$ interactions are flavor diagonal, the scotogenic Yukawa couplings can also induce flavor-changing $Z'$ vertices at one-loop level. The effective interaction can be parametrized as \cite{Lindner:2016bgg}
\begin{equation}
\mathcal{L}_{Z'}^{\mathrm{CLFV}}
=
Z'_\mu\,
\overline{\ell_j}\gamma^\mu
\left(
g_{ji}^{L}P_L
+
g_{ji}^{R}P_R
\right)
\ell_i
+\mathrm{h.c.}
\end{equation}
The coefficients $g_{ji}^{L,R}$ depend on the masses and mixing matrix of the Majorana fermions, on the inert scalars, and on the $U(1)_{L_\mu+L_\tau}$ charges of the particles inside the loop. A complete calculation requires the complete gauge-invariant set of diagrams in which the $Z'$ is attached to both the scalar and fermion internal lines, including the non-diagonal $Z'X_\alpha X_\beta$ interactions induced in the mass basis. The phenomenological importance of this contribution requires a dedicated calculation beyond the scope of the present work.

In this model, the additional fermions required for $U(1)_{L_\mu+L_\tau}$ anomaly cancellation do not mix directly with the SM charged leptons. They therefore do not introduce an independent source of CLFV. Additional interactions mixing these fermions with SM charged leptons could generate further CLFV contributions beyond those discussed above.

%\paragraph{\bf Connection with neutrino masses and dark matter.}
The scotogenic contribution shows that CLFV is phenomenologically correlated with the masses and mixing of the heavy Majorana sector responsible for neutrino-mass generation. If one of the $X_\alpha$ particles were a fermionic dark matter candidate, it would therefore be directly connected to the same masses, mixing angles, and Yukawa couplings entering both $\mathbf{m}_\nu$ and the CLFV amplitudes. In such a scenario, neutrino masses, dark matter, and CLFV would be phenomenologically correlated.

The benchmark scenario studied in this work is qualitatively different. The dark matter candidate is taken to be the predominantly singlet CP-even scalar $\phi_2$, whose relic abundance is mainly controlled by gauge and scalar annihilation channels such as
\begin{equation}
\phi_2\phi_2\to Z'Z',
\qquad
\phi_2\phi_2\to h_2h_2.
\end{equation}
 The correlation between the dark matter benchmark considered here and CLFV is therefore expected to be comparatively weak, although it is not strictly absent because the scalar mixing required for neutrino-mass generation connects the singlet and inert-doublet sectors.

Given the four-state Majorana spectrum, the different possible sources of CLFV, the loop-induced $Z'$ contribution, and the potential connection with a fermionic dark matter candidate, a complete analysis would require a simultaneous treatment of neutrino oscillation data, dark matter observables, radiative and three-body charged-lepton decays, and $\mu$--$e$ conversion. Such an analysis is broader than the scalar dark matter benchmark considered here and will be left for a dedicated future work.

\subsection{Dark matter candidate and annihilation channels}
In the general scalar spectrum, the neutral $M_P$-odd mass eigenstates are $\vpo,\vpt$ in the CP-even sector and $\widetilde{\varphi}_{1},\widetilde{\varphi}_{2}$ in the CP-odd sector. They are related to the gauge-basis fields by 

\begin{align} \xi_\eta &= c_{\varphi}\vpo-s_{\varphi}\vpt, & \xi_\sigma &= s_{\varphi}\vpo+c_{\varphi}\vpt, \\ \zeta_\eta &= c_{\wvp}\widetilde{\varphi}_{1} -s_{\wvp}\widetilde{\varphi}_{2}, & \zeta_\sigma &= s_{\wvp}\widetilde{\varphi}_{1} +c_{\wvp}\widetilde{\varphi}_{2}.
\end{align} 
The dark matter candidate is identified with the lightest electrically neutral $M_P$-odd scalar. In the numerical analysis below, we focus on parameter configurations in which the CP-even state $\vpt$ is predominantly singlet-like and is the lightest state of the $M_P$-odd scalar sector. We therefore identify $\vpt$ as the dark matter candidate throughout the relic-density analysis. 

\paragraph{\bf Singlet-like scalar dark matter: $\vpt$.}
The relevant number-depleting reactions are annihilation and
co-annihilation of $M_P$-odd particles into $M_P$-even final states. For
the singlet-like CP-even dark matter candidate $\vpt$, representative
annihilation channels include

\begin{equation} 
\vpt\vpt \to \zp\zp,\; ZZ,\; \zp Z,\; W^+W^-,\;
h_i h_j,\; q\bar q,\; \ell^+\ell^-,
\qquad i,j=1,2,
\end{equation}
whenever allowed by the couplings and phase space. By contrast,
elastic scattering $\vpt\vpt\to\vpt\vpt$ and conversions into other
two-particle $M_P$-odd states, such as $\varphi_i\varphi_j$,
$\widetilde\varphi_i\widetilde\varphi_j$ and $\eta^+\eta^-$, do not
directly change the total odd-particle number and should be distinguished
from these annihilation processes.

\begin{figure}[tbp]
\centering
\includegraphics[alt={Tree-level Feynman diagrams for the scalar dark matter candidate varphi_2}]{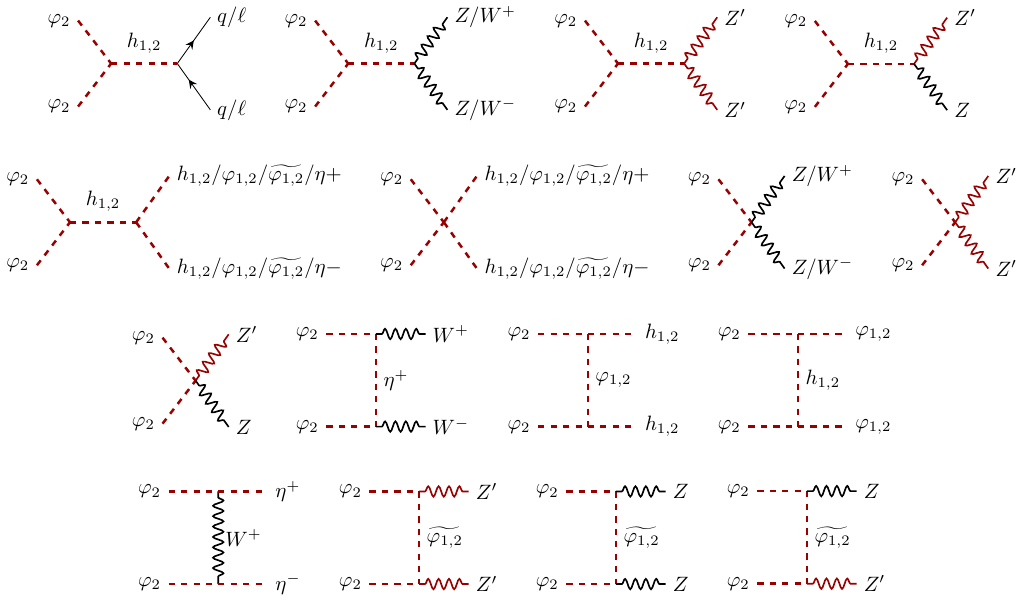}
%\dmdmToffschannel
%\dmdmToVVschannel
%\dmdmToZpZpres
%\dmdmToZpZres
%\dmdmTohhschannel
%\dmdmTohh
%\dmdmToZpZ
%\dmdmToZpZp
%\dmdmToZpZi
%\dmdmTowwtchannel
%\dmdmTohhtchannel
%\dmdmTophiphitchannel
%\dmdmToetaeta
%\dmdmToZpZptchannel
%dmdmToZZtchannel
%\dmdmToZpZtchannel
\caption{Representative tree-level topologies for reactions of the
$M_P$-odd scalar dark matter candidate $\vpt$. Annihilation into fermion
pairs, gauge bosons and $M_P$-even scalar pairs is shown alongside elastic
scattering and conversions into two-particle $M_P$-odd states, including
charged-scalar pairs. The latter processes do not directly deplete the
total odd-particle number. The diagrams include $s$-channel,
$t/u$-channel and quartic contact interactions; their relevance depends
on the benchmark couplings and the available phase space.
\label{fig:DMann}}
\end{figure}

Representative tree-level annihilation, elastic-scattering and conversion
topologies are shown in \Cref{fig:DMann}. They include scalar-mediated
$s$-channel processes, quartic contact interactions, and $t$- and
$u$-channel exchange. In particular, the channel $\vpt\vpt\to\zp\zp$ is present at tree level as a consequence of the $U(1)_{L_\mu+L_\tau}$ charge assignment of the scalar sector. Depending on the parameter region, this final state can receive contributions from quartic contact interactions, $t$- and $u$-channel exchange of $M_P$-odd scalars, and scalar-mediated $s$-channel diagrams. Annihilation into scalar final states is instead controlled by the parameters of the scalar potential. Consequently, the relative importance of the different mechanisms depends on the complete scalar spectrum, the mixing angles, the gauge coupling $\gp$, and the kinematic accessibility of the corresponding final states.
%In particular, the channel $\vpt\vpt\to \zp\zp$ is present already at tree level and is controlled by the $U(1)_{L_\mu+L_\tau}$ charge assignment of the scalar sector. Depending on the parameter region, this final state can receive contributions from scalar-mediated $s$-channel diagrams, quartic contact interactions, and $t$- and $u$-channel exchange of $M_P$-odd scalar states. Likewise, the annihilation channels into scalar final states are controlled by the scalar potential parameters, especially the portal couplings $\lambda_5$, $\lambda_7$, $\lambda_8$, and $\lambda_9$, together with the trilinear coupling $\kappa_2$.

For the parameter configurations considered below, negative values of $\kappa_2$ can generate a spectrum in which the predominantly singlet-like CP-even state $\vpt$ is lighter than the corresponding CP-odd state $\widetilde{\varphi}_{2}$. This mass ordering allows $\vpt$ to constitute the lightest neutral $M_P$-odd state and hence the dark matter candidate in the benchmark regions studied here.

\subsection{Relic density}
\begin{figure}[h!]
\centering
\includegraphics[width=0.49\linewidth]{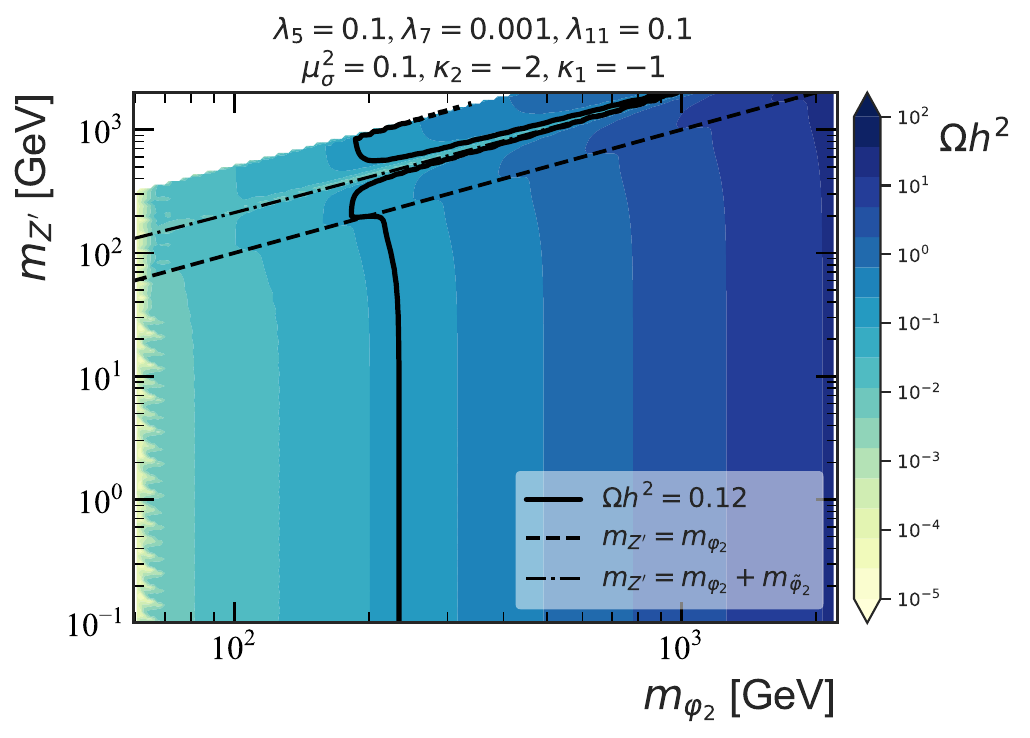}
\includegraphics[width=0.49\linewidth]{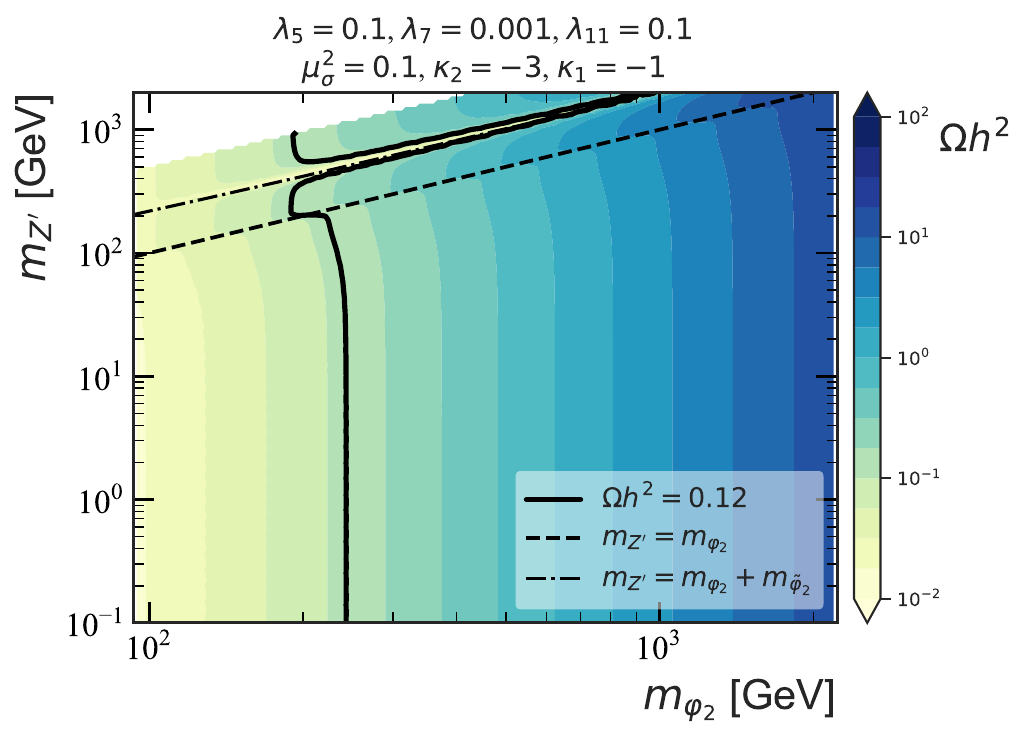}\\
\includegraphics[width=0.49\linewidth]{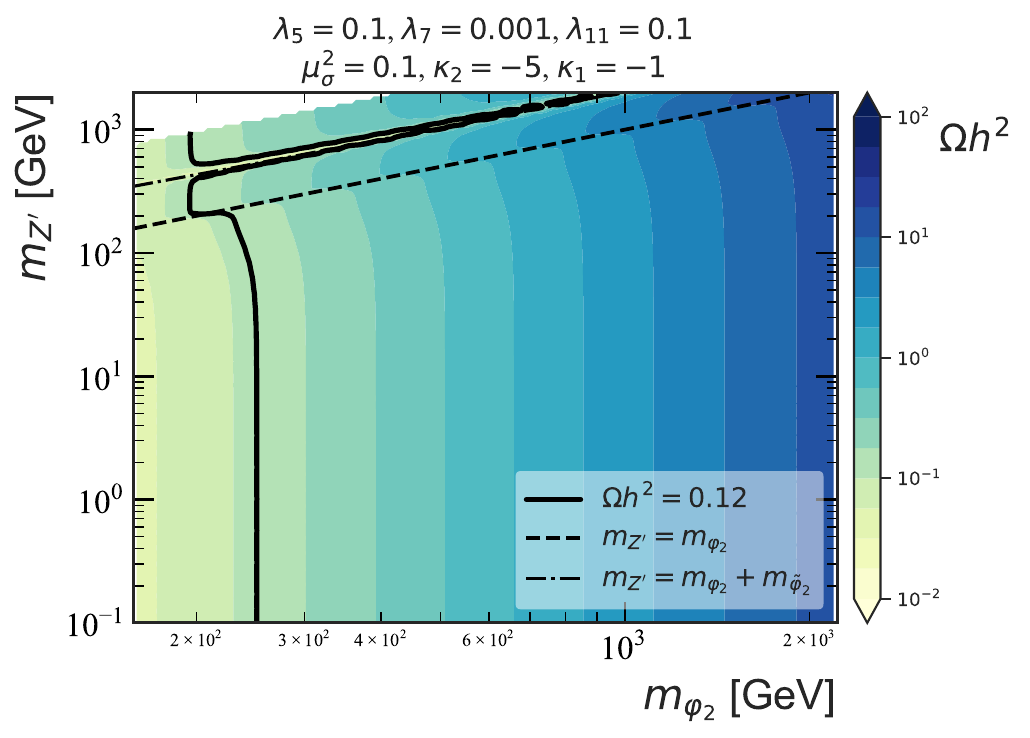}
\caption{Relic abundance $\Ohs$ in the $(\mvpt,\mzp)$ plane for $\kappa_2=-2$ (upper left), $-3$ (upper right) and $-5~\GeV$ (lower), with $\mhp=10~\GeV$ and the parameters of \Cref{eq:bench}. The gauge coupling is $\gp=\mzp/v_\phi$; white regions correspond to $\gp<10^{-5}$ (lower right) or $\gp>1$ (upper left). The solid line denotes $\Ohs=0.12$, the dashed line denotes $\mzp=\mvpt$, above which $\vpt\vpt\to\zp\zp$ is closed, and the dash-dotted line denotes the threshold for the co-annihilation process $\vpt\wvp_2\to{\zp}^{\ast}$, $\mzp=\mvpt+m_{\wvp_2}$.
\label{fig:mzp_msr}}
\end{figure}

\begin{figure}[tbp]
    \centering
    \includegraphics[scale=0.5]{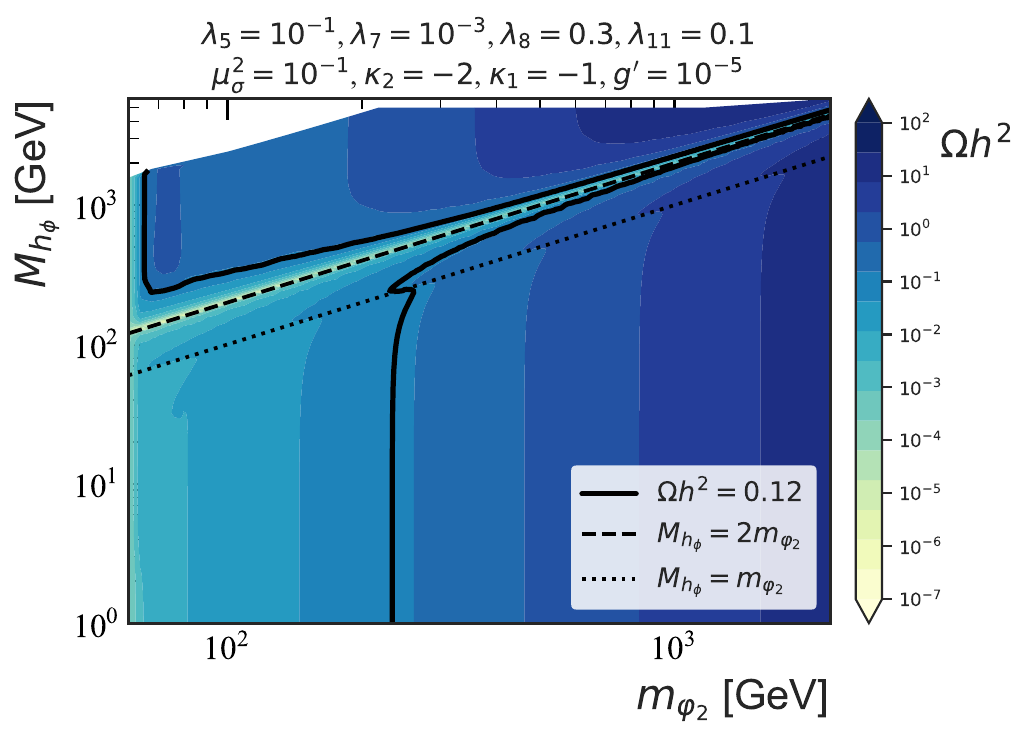}
    \includegraphics[scale=0.5]{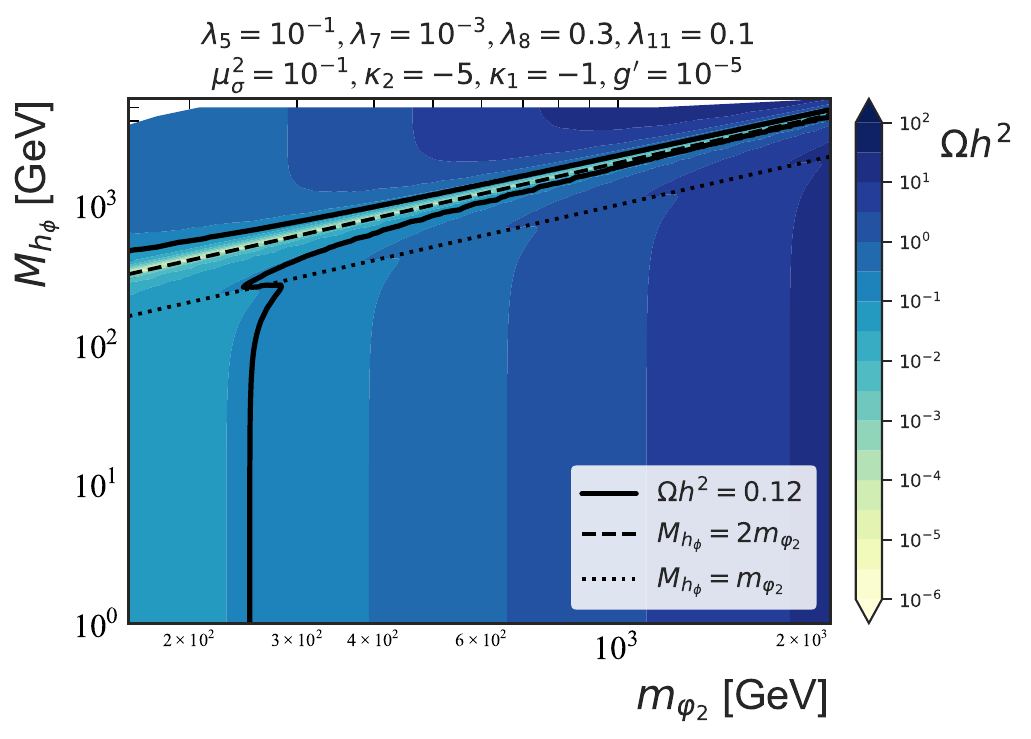}
    \caption{Relic abundance $\Ohs$ in the $(\mvpt,\mhp)$ plane for $\kappa_2=-2$ (left) and $-5~\GeV$ (right), with $\gp=10^{-5}$ and the parameters of \Cref{eq:bench}. The solid line denotes $\Ohs=0.12$, the dashed line the $h_\phi$ resonance $\mhp=2\mvpt$, and the dotted line $\mhp=\mvpt$, above which $\vpt\vpt\to h_\phi h_\phi$ is closed. The white region in the upper left corresponds to $\lambda_3=\mhp^2/(2v_\phi^2)>4\pi$.\label{fig:mhp_msr}}
\end{figure}

We compute the relic abundance of the scalar dark matter candidate $\vpt$ with \texttt{micrOMEGAs~6}~\cite{Alguero:2023zol}, which solves the Boltzmann equation for the total number density of $M_P$-odd particles, including all two-body annihilation and co-annihilation processes. We compare with the Planck value $\Omega_{\rm c}h^2=0.120\pm0.001$~\cite{Planck:2018vyg}: in \Cref{fig:mzp_msr,fig:mhp_msr} the solid black contours denote $\Ohs=0.12$, points with $\Ohs>0.12$ overproduce dark matter, and points with $\Ohs<0.12$ would require $\vpt$ to be only a fraction of it. This work identifies the regions in which thermal freeze-out of $\vpt$ reproduces the observed abundance and explains their structure. We present two projections: the $(\mvpt,\mzp)$ plane, which probes the gauge sector, and the $(\mvpt,\mhp)$ plane, which probes the scalar associated with the breaking of $U(1)_{L_\mu+L_\tau}$. 

In all the scans we take $\theta_h=0$, $\lambda'_{10}=0$ and $\epsilon=0$, so that the SM-like Higgs does not mix with $h_\phi$ and there is no kinetic $Z$--$\zp$ mixing, and we fix
\begin{equation}
\begin{gathered}
\lambda_2=\lambda_4=\lambda_6=\lambda_{10}=0.1\,,\quad \lambda_5=0.1\,,\quad
  \lambda_7=10^{-3}\,,\quad \lambda_8=0.3\,,\quad \lambda_{11}=0.1\,,\\
  \kappa_1=-1~\GeV\,,\quad \mu_\sigma^2=\mu_\eta^2=0.1~\GeV^2\,,
\end{gathered}
\label{eq:bench}
\end{equation}
with the exotic fermions at $3~\TeV$. Non-zero $\kappa_1$ and $\lambda_{11}$ are required by the radiative neutrino-mass mechanism of \cref{scotogenic}. The choice $\lambda_8>\lambda_5$ keeps the inert doublet well separated from the singlet, $m_\eta/\mvpt\simeq\sqrt{\lambda_8/\lambda_5}$, so that the doublet admixture of $\vpt$ induced by $\kappa_1$ and $\lambda_{11}$ remains below about 6\% in the scans. The parameters $\lambda_1$ and $\lambda_3$ are fixed by $m_h=125~\GeV$ and $\mhp$, and the symmetry-breaking scale by $v_\phi=\mzp/\gp$, with $100~\GeV\leq v_\phi\leq10^{4}~\GeV$. All points satisfy positive scalar squared masses, perturbativity, $|\lambda_i|<4\pi$, and the potential's bounded-from-below conditions. For the dark matter selection we require $\vpt$ and $\wvp_2$ to be singlet-like, $c_{\varphi}^2\geq 0.9$ and $c_{\wvp}^2\geq0.9$, the doublet-like states $\vpo$, $\wvp_1$ and $\eta^\pm$ to be at least 20\% heavier than $\vpt$, and

\begin{equation}
  0\leq\frac{m_{\wvp_2}-\mvpt}{\mvpt}<0.20\,,
  \label{eq:split}
\end{equation}

so that co-annihilation with $\wvp_2$ is included whenever it is relevant. For $\kappa_2<0$ the splitting is $m_{\wvp_2}^2-\mvpt^2\simeq2\sqrt2\,|\kappa_2|v_\phi$; its relative size decreases with $\mvpt$ and grows with $|\kappa_2|$, and \Cref{eq:split} sets the lowest dark matter mass allowed in each benchmark. At all the scanned points \texttt{micrOMEGAs} identifies $\vpt$ as the dark matter candidate.

Since $\vpt$ is a real scalar, there is no diagonal
$\zp\vpt\vpt$ vertex. The gauge current instead connects it to the CP-odd
states. At $\epsilon=0$, the equal charges of $\eta^0$ and $\sigma$ give
the exact contact interaction
$\mathcal L\supset({\gp}^2/8)\,\zp_\mu Z'^\mu\vpt^2$,
independently of $\theta_\varphi$. Consequently,
$\vpt\vpt\to\zp\zp$ receives contributions from this contact interaction,
$t/u$-channel exchange of $\widetilde\varphi_{1,2}$, and, in the alignment
benchmark considered here, $s$-channel $h_\phi$ exchange. 
Amplitudes with transversely polarized $\zp$ bosons are of order ${\gp}^2$, while those with longitudinal polarizations are of order $\lambda_5$; their ratio is ${\gp}^2/\lambda_5\simeq\mzp^2/(2\mvpt^2)$, so that for $\mzp\ll\mvpt$ the $\zp\zp$ final state is dominated by longitudinal polarizations. The parameter $\kappa_2$ enters through the splitting $m_{\wvp_2}-\mvpt$, which controls co-annihilation, and through the relation between $\mvpt$ and $v_\phi$. Since $\wvp_2$ annihilates with a rate similar to that of $\vpt$, while $\vpt\wvp_2$ processes are suppressed, co-annihilation slightly reduces the effective annihilation rate. The gauge coupling enters only through kinematics, via the threshold $\mzp=\mvpt$ above which $\zp\zp$ is closed.

\Cref{fig:mzp_msr} shows the relic abundance in the $(\mvpt,\mzp)$ plane for $\kappa_2=-2,-3,-5~\GeV$ and $\mhp=10~\GeV$. Each point has $\gp=\mzp/v_\phi$ with $10^{-5}\leq\gp\leq1$, and since $\mvpt\simeq0.21\,v_\phi$ in this benchmark. For $\mzp\ll\mvpt$ the $\Ohs=0.12$ contour is a vertical line at $\mvpt\simeq  233$,  $243$ and $251~\GeV$ for $\kappa_2=-2$, $-3$ and $-5~\GeV$, respectively. Annihilation into $\zp\zp$ (41--42\%) and $h_\phi h_\phi$ (41\%), both controlled by $\lambda_5$, dominates, and SM final states ($W^+W^-$, $t\bar t$, $hh$, $ZZ$, $hh_\phi$) provide the remaining 17--18\%. Along this line the $(\mzp,\gp)$ plane of \Cref{fig:EWPO} the observed abundance corresponds, for $\mzp\lesssim20~\GeV$, to $\gp\simeq\mzp/v_\phi$. The $\kappa_2$ dependence comes from co-annihilation. On the vertical line $\Delta=5.4\%$, $7.8\%$ and $12.6\%$, and with $x_f=\mvpt/T_f\simeq23.7$ the Boltzmann factor $e^{-x_f\Delta}$ is $0.28$, $0.16$ and $0.05$. Processes with $\wvp_2$ in the initial state contribute less than 1\% each to $(\Ohs)^{-1}$, but since $\wvp_2$ pairs annihilate through the same channels as $\vpt$ pairs while $\vpt\wvp_2$ annihilation is suppressed, a thermal population of $\wvp_2$ lowers the effective cross section~\cite{Griest:1990kh}; the larger splitting for $\kappa_2=-5~\GeV$ reduces this effect and requires a heavier $\vpt$. 

As $\mzp$ approaches $\mvpt$ the phase space of the $\zp\zp$ final state closes and the contour moves to smaller masses; at $\mzp\simeq0.8\,\mvpt$ the $\zp\zp$ channel provides about 30\% of the rate and the contour lies at $\mvpt\simeq216$, $224$ and $230~\GeV$. The threshold $\mzp=\mvpt$ is crossed at $\gp\simeq0.20$--$0.21$. Above it $h_\phi h_\phi$ provides 72--73\% of the rate and SM final states 25--26\%, and for $1.05\leq\mzp/\mvpt\leq1.5$ ($0.22\lesssim\gp\lesssim0.32$) the contour lies at $\mvpt\simeq183$, $190$ and $195~\GeV$. Close to the dash-dotted line, $\mzp\simeq\mvpt+m_{\wvp_2}$, the co-annihilation process $\vpt\wvp_2\to{\zp}^{\ast}\to\ell^+\ell^-,\nu\bar\nu$ becomes resonant, producing charged leptons and neutrinos in the ratio 2:1 expected from the vector coupling of the $\zp$ to $\mu$ and $\tau$ and its coupling to left-handed neutrinos only. At fixed $\mvpt$ the minimum of $\Ohs$ lies slightly above this line, at $\mzp\simeq(1.01$--$1.04)(\mvpt+m_{\wvp_2})$ within the resolution of the scan, i.e.\ at $\gp\simeq0.46$. Inside the resulting narrow funnel $\vpt$ is underabundant, and the $\Ohs=0.12$ contour follows its two edges. In \Cref{fig:mzp_msr} the funnel is visible up to $\mzp=1~\TeV$; in the scan it continues up to $\mvpt\simeq2.1$, $2.0$ and $2.0~\TeV$ ($\mzp\simeq4~\TeV$), close to the upper end of the scanned range. Because the process requires $\wvp_2$ in the initial state, the funnel becomes shallower as $|\kappa_2|$ grows: at $\mvpt\simeq300~\GeV$ its minimum is $\Ohs\simeq0.007$, $0.011$ and $0.029$.

\Cref{fig:mhp_msr} shows the relic abundance in the $(\mvpt,\mhp)$ plane for $\kappa_2=-2$ and $-5~\GeV$; the two panels have the same structure. Since $\theta_h=0$, $h_\phi$ is the singlet-like scalar associated with the breaking of $U(1)_{L_\mu+L_\tau}$, and we fix $\gp=10^{-5}$, so that $\mzp\leq0.1~\GeV$ is negligible in the whole plane. For $\mhp\leq0.2\,\mvpt$ the contour is again a vertical line, at $\mvpt=233.5$ and $251.0~\GeV$ for $\kappa_2=-2$ and $-5~\GeV$, which agrees with \Cref{fig:mzp_msr} although the two scans use different grids. As $\mhp$ grows towards $\mvpt$ the contour moves to larger masses, reaching $\mvpt\simeq245$ ($264$)~GeV at $\mhp\simeq0.6\,\mvpt$ and $261$ ($283$)~GeV at $\mhp\simeq0.92\,\mvpt$ for $\kappa_2=-2$ ($-5$)~GeV, and the $\zp\zp$ share grows from 41\% to about 50\%: for $\mhp<2\mvpt$ the $s$-channel $h_\phi$ exchange adds constructively to the contact interaction in the longitudinal amplitude, while $h_\phi h_\phi$ loses phase space. Once $h_\phi h_\phi$ closes, the contour moves back to $\mvpt\simeq229$ ($244$)~GeV at $\mhp\simeq1.04\,\mvpt$, with $\zp\zp$ providing about 80\% of the rate, so that the threshold appears as a kink of 12--14\% in the dark matter mass. At $\mhp\simeq2\mvpt$, the $s$-channel $h_\phi$ exchange becomes resonant, and for $0.98\lesssim\mhp/(2\mvpt)\lesssim1.08$ the abundance remains below the observed value across the entire scanned range, up to $\mvpt\simeq2.2~\TeV$. The under-abundant band is asymmetric: at $\mvpt\simeq1~\TeV$ it extends from $\mhp\simeq1.89\,\mvpt$ to $2.24\,\mvpt$ for $\kappa_2=-2~\GeV$ ($1.87\,\mvpt$ to $2.26\,\mvpt$ for $\kappa_2=-5~\GeV$), as expected from thermal averaging, since for $\mhp>2\mvpt$ the kinetic energy of the annihilating pair can bring the process on resonance. Along both edges $\zp\zp$ dominates the rate, the remainder being $\wvp_2\wvp_2\to\zp\zp$ at TeV masses and SM final states at low masses. On the far side the contour lies at $\mhp\simeq2.2\,\mvpt$ for $\mvpt\simeq2~\TeV$ and moves up to $\mhp\simeq3.4\,\mvpt$ at $\mvpt\simeq70~\GeV$ for $\kappa_2=-2~\GeV$ ($2.8\,\mvpt$ at $\mvpt\simeq180~\GeV$ for $\kappa_2=-5~\GeV$). For $\mhp\gg2\mvpt$ the contact interaction and the $h_\phi$ exchange cancel in the longitudinal $\zp\zp$ amplitude, which is suppressed as $4\mvpt^2/\mhp^2$; for $\mhp\gtrsim4\,\mvpt$ annihilation then proceeds mainly into SM final states, the abundance varies only slowly with $\mhp$, and $\vpt$ is overproduced, with $\Ohs\simeq0.3$ at $\mvpt\simeq100~\GeV$ and $\Ohs\simeq1.2$--$1.4$ at $\mvpt\simeq300~\GeV$ for $\kappa_2=-2~\GeV$. The exception is the Higgs resonance, where annihilation into $b\bar b$ gives $\Ohs<0.12$ for $\mvpt\lesssim65~\GeV$; this region appears only for $\kappa_2=-2~\GeV$, since for $\kappa_2=-5~\GeV$ \Cref{eq:split} requires $\mvpt\gtrsim158~\GeV$. Finally, the Higgs coupling responsible for the SM final states also mediates spin-independent scattering of $\vpt$ on nuclei; we do not include direct-detection constraints in the present analysis.

\subsection{Collider interpretation}
\label{subsec:collider_constraints}

The relic-density results in the $(\mvpt,\mzp)$ and $(\mvpt,\mhp)$ planes should be interpreted alongside existing searches for leptophilic gauge bosons, which we use to identify the conditions under which the relic-density regions shown above remain phenomenologically viable. The $\zp$ associated with $U(1)_{L_\mu+L_\tau}$ couples directly to second- and third-generation leptons and to their corresponding neutrinos. In the small-kinetic-mixing limit, its couplings to electrons and quarks are suppressed, and the dominant signatures are controlled by muons, taus, neutrinos, and possible invisible decays into the dark sector. The relevant experimental information, therefore, comes from searches for leptophilic $\zp$ bosons in visible and invisible final states.
Selected partial decay widths of the $Z'$ boson and
the singlet-like scalar $h_2$ are collected in
Appendix~\ref{sec:decays}, together with their normalization
conventions and domains of validity.

At low mediator masses, BaBar conducted a dedicated search for a muon-philic dark gauge boson in $e^+e^-\to\mu^+\mu^-\zp$, with $\zp\to\mu^+\mu^-$ \cite{BaBar:2016sci}. This study used an integrated luminosity of $514~\mathrm{fb}^{-1}$ and investigated $\zp$ masses from the dimuon production threshold, $\mzp\simeq0.212~\GeV$, up to $10~\GeV$. BaBar derived 90\% C.L. upper bounds on $\sigma(e^+e^-\to\mu^+\mu^-\zp)\times{\rm BR}(\zp\to\mu^+\mu^-)$ and translated them into limits on the $\zp$ coupling to muons, for the benchmark in which the $\zp$ has equal-magnitude couplings to muons, taus and their neutrinos. The highest sensitivity is near the dimuon threshold, where couplings as small as $\gp\simeq7\times10^{-4}$ are excluded, while in the least sensitive part of the mass range the limits weaken to $\gp\sim\mathcal{O}(10^{-1})$.

At intermediate masses, CMS searched for a narrow $L_\mu-L_\tau$ gauge boson in $Z\to\zp\mu^+\mu^-\to4\mu$ using $77.3~\mathrm{fb}^{-1}$ of proton-proton collision data at $\sqrt{s}=13~\TeV$, for $\zp$ masses between 5 and $70~\GeV$~\cite{CMS:2018yxg}. No significant excess was found, and the 95\% C.L. upper limits on $\mathcal{B}(Z\to\zp\mu^+\mu^-)\mathcal{B}(\zp\to\mu^+\mu^-)$, at the level of $10^{-8}$--$10^{-7}$, exclude in the minimal $L_\mu-L_\tau$ model couplings to muons above $0.004$--$0.3$, depending on $\mzp$. ATLAS performed the same search with $139~\mathrm{fb}^{-1}$ for $5\leq\mzp\leq81~\GeV$ and excludes couplings above $0.003$--$0.2$ at 95\% C.L.~\cite{ATLAS:2023vxg}.

Belle II provides complementary sensitivity to invisible $\zp$ decays through $e^+e^-\to\mu^+\mu^-\zp$, reconstructing the $\zp$ as missing energy~\cite{Belle-II:2022yaw}. The search used $79.7~\mathrm{fb}^{-1}$ collected at $\sqrt{s}=10.58~\GeV$ and set 90\% C.L. upper limits on the production cross section and the coupling in the vanilla $L_\mu-L_\tau$ model, where the $\zp$ decays invisibly only into neutrinos. These limits range from $\gp\simeq3\times10^{-3}$ at low $\zp$ masses to $\gp\simeq1$ at $\mzp=8~\GeV$. Since the $\zp$ partial widths are the same in the present model, this interpretation applies directly, as discussed below.

Belle II also searched for a $\tau^+\tau^-$ resonance in $e^+e^-\to\mu^+\mu^-\tau^+\tau^-$ events, using $62.8~\mathrm{fb}^{-1}$ collected at $\sqrt{s}=10.58~\GeV$ and probing masses between 3.6 and $10~\GeV$~\cite{Belle-II:2023ydz}. No evidence for a signal was found, and the 90\% C.L. upper limits on $\sigma(e^+e^-\to\mu^+\mu^-X)\times{\rm BR}(X\to\tau^+\tau^-)$ range from 0.7 to $24~\mathrm{fb}$, depending on the resonance mass and the signal hypothesis. For the $L_\mu-L_\tau$ benchmark, Belle II translates these cross-section limits into upper bounds on $\gp$ using the quadratic scaling of the production rate; the result, shown in Fig.~S10 of Ref.~\cite{Belle-II:2023ydz}, reaches values as low as $\gp\simeq2.5\times10^{-2}$. In the present model, this search constrains the $\zp\to\tau^+\tau^-$ mode above the $\tau^+\tau^-$ threshold.

We can compare these limits with the usual $L_\mu-L_\tau$ scenario. Although $L_\mu+L_\tau$ and $L_\mu-L_\tau$ differ only by the sign of the tau charge, the production rates and partial widths depend only on the squared magnitudes of the lepton couplings, which are the same in both models. For $\epsilon=0$, and as long as decays into $M_P$-odd states are closed, the $\zp$ branching fractions are therefore those of the minimal $L_\mu-L_\tau$ model: ${\rm BR}(\zp\to\nu\bar\nu)=1$ below the dimuon threshold and, up to phase-space effects, ${\rm BR}(\zp\to\mu^+\mu^-)={\rm BR}(\zp\to\nu\bar\nu)=1/2$ between the $\mu^+\mu^-$ and $\tau^+\tau^-$ thresholds and $1/3$ for each of $\mu^+\mu^-$, $\tau^+\tau^-$ and $\nu\bar\nu$ above the $\tau^+\tau^-$ threshold. The decay $\zp\to\vpt\wvp_2$, which would reduce the visible branching fractions, requires $\mzp>\mvpt+m_{\wvp_2}$; on the relic-density contours of \Cref{fig:mzp_msr} this happens only for $\mzp\gtrsim500~\GeV$. For the light-$\zp$ solutions, the published $L_\mu-L_\tau$ limits therefore apply directly, and a model-dependent recast with the explicit $\zp$ branching ratios is needed only for a heavy $\zp$ close to and beyond the co-annihilation pole.

Fixed-target missing-energy experiments are also sensitive to the very light $\zp$ regime. The NA64-e Collaboration performed a search for a light $\zp$ in the gauged $L_\mu-L_\tau$ framework using a 100 GeV electron beam incident on an active thick target, analysing a dataset of $2.84\times10^{11}$ electrons on target and probing $\zp$ masses from 1 to 600 MeV \cite{NA64:2022rme}. In the minimal $L_\mu-L_\tau$ model, the $\zp$ has no tree-level coupling to electrons, and muon and tau loops induce an effective electron coupling. The ultraviolet-divergent contributions cancel because the muon and tau have opposite $U(1)$ charges, leaving a finite, momentum-dependent contribution. By contrast, in the gauged $L_\mu+L_\tau$ model, the muon and tau contributions have the same sign, and their divergent parts do not cancel. The effective $\zp$ coupling to electrons therefore depends on the renormalized kinetic-mixing parameter, together with the radiative contributions of the charged particle spectrum. Consequently, the NA64-e exclusion limits obtained for the minimal $L_\mu-L_\tau$ scenario cannot be directly applied to the present model without specifying the effective kinetic mixing and the relevant invisible branching fraction. We therefore treat these results as complementary experimental information and as motivation to reinterpret fixed-target missing-energy searches within the gauged $L_\mu+L_\tau$ framework.  

For the present $L_\mu+L_\tau$ model, a muon-beam search provides a more direct probe. The $\zp$ couples at tree level to muons and their corresponding neutrinos, so it can be radiated from an incoming muon via $\mu N\to\mu N\zp$ and then decay invisibly into neutrinos or into kinematically accessible $M_P$-odd states, leaving a scattered muon accompanied by missing energy. NA64-$\mu$ has already reported first results for an invisibly decaying $L_\mu-L_\tau$ $\zp$, using $1.98\times10^{10}$ muons of $160~\GeV$ on target~\cite{NA64:2024nwj}. Since its production rate is controlled directly by the gauge coupling to muons, rather than by a loop-induced or kinetic-mixing-induced electron coupling, this is the most direct probe of the light-$\zp$ region of the $L_\mu+L_\tau$ model. A dedicated analysis for the present model, including the $\mu N\to\mu N\zp$ production rate, the $\zp$ invisible branching fraction and the NA64-$\mu$ acceptance for the specific $L_\mu+L_\tau$ spectrum, is left for future work.

These limits can be compared directly with the relic-density solutions. In the light-$\zp$ regime of \Cref{fig:mzp_msr}, the observed abundance requires $v_\phi\simeq1.07$--$1.21~\TeV$, i.e.\ $\gp\simeq\mzp/v_\phi$. In the $(\mzp,\gp)$ plane of \Cref{fig:EWPO}, which also shows the limit from neutrino trident production measured by \cite{CCFR:1991trident,Altmannshofer:2014pba}, this line lies below the trident, BaBar and Belle II limits for $\mzp\lesssim5~\GeV$, but above the ATLAS and CMS $4\mu$ limits from $\mzp\simeq5~\GeV$, the lower end of these searches, up to $\mzp\simeq60~\GeV$ and $55~\GeV$, respectively. Since points with $\Ohs<0.12$ have smaller $v_\phi$ and thus larger $\gp$ at fixed $\mzp$, this window allows no points with $ \Ohs\leq0.12$ in the benchmark of \Cref{eq:bench}. The solutions of \Cref{fig:mhp_msr}, with $\gp=10^{-5}$ and $\mzp\leq0.1~\GeV$, lie far below all these limits, and for $\theta_h=0$ the scalar $h_\phi$ does not mix with the Higgs boson, so they are not constrained by the searches discussed here.

%#####################
\section{Summary and conclusions \label{sec:conclusions}}
%#####################

In this work, we have introduced the first scotogenic model based on a gauged mu plus tau lepton number $L_\mu+L_\tau$ symmetry, which is spontaneously broken by two units, yielding
a residual matter parity $M_P$. In this scenario, the active neutrinos acquire Majorana masses radiatively at one-loop level and both, 
the lightest electrically neutral $M_P$-odd state and the lightest neutral exotic-lepton  provide a WIMP dark matter candidate. 

The gauge sector analysis shows that the $Z'$ phenomenology is compatible with a broad range of parameters relevant for DM. In particular, electroweak precision constraints remain compatible with the reduced EWPO set considered here for kinetic mixing values up to $\epsilon\sim 10^{-2}$, both in the light-mass regime and for masses above the $Z-$pole, provided that the quasi-degenerate region $M_{Z'}\simeq M_Z$ is avoided. In the $\left(M_{Z'},g^\prime\right)$ plane, the most relevant external constraints are provided by published searches for leptophilic $U(1)_{L_\mu-L_\tau}$ gauge bosons, which we use as phenomenological reference bounds, such as BaBar, neutrino trident production and collider searches, interpreted in the context of searches for vanilla $U(1)_{L_\mu-L_\tau}$ models. For $M_{Z'}\gg M_Z$, the EWPO effects decouple as $M_Z^2/M_{Z'}^2$, while for $M_{Z'}\gtrsim 1~{\rm TeV}$ the model remains only weakly constrained within the scanned range.

Taking the singlet-like CP-even scalar $\vpt$ as the dark matter candidate, we identified the regions where thermal freeze-out reproduces $\Ohs=0.12$. Since $\vpt$ is a real field, there is no $\zp\vpt\vpt$ vertex, and the portal coupling $\lambda_5$ controls annihilation, fixing both $\mvpt^2\simeq\lambda_5v_\phi^2/2$ and, via the Goldstone-boson equivalence theorem, the rate into longitudinal $\zp$ bosons. For $\mzp\ll\mvpt$ the observed abundance is obtained, independently of $\gp$, at $\mvpt\simeq234$--$251~\GeV$, i.e.\ $v_\phi\simeq1.1$--$1.2~\TeV$, through the $\zp\zp$ and $h_\phi h_\phi$ final states; $\kappa_2$ enters through the splitting $m_{\wvp_2}-\mvpt$, which controls the co-annihilation with $\wvp_2$. Above the $\zp\zp$ threshold, annihilation into $h_\phi h_\phi$ and SM final states requires $\mvpt\simeq183$--$195~\GeV$, and the resonant co-annihilation $\vpt\wvp_2\to{\zp}^{\ast}\to\ell^+\ell^-,\nu\bar\nu$ opens a narrow funnel at $\mzp\simeq\mvpt+m_{\wvp_2}$, $\gp\simeq0.46$, which extends up to $\mvpt\simeq2~\TeV$ and becomes shallower as $|\kappa_2|$ grows. In the $(\mvpt,\mhp)$ plane, the $h_\phi h_\phi$ threshold changes the required mass by 12--14\%, and the $s$-channel resonance $\mhp\simeq2\mvpt$ gives an underabundant band over the whole scanned range, up to $\mvpt\simeq2.2~\TeV$. For $\mhp\gtrsim4\mvpt$ the $\zp\zp$ rate is suppressed and $\vpt$ is overproduced, except near the Higgs resonance, $\mvpt\lesssim65~\GeV$. The observed abundance therefore arises from a combination of gauge, scalar, and co-annihilation processes, all tied to the $U(1)_{L_\mu+L_\tau}$ breaking scale $ v_\phi$.

This connection makes the light-$\zp$ solutions directly testable, since they satisfy $\gp\simeq\mzp/v_\phi$ with $v_\phi\simeq1.1$--$1.2~\TeV$. They lie below the trident, BaBar and Belle II limits for $\mzp\lesssim5~\GeV$, but are excluded by the $Z\to4\mu$ searches of ATLAS and CMS for $5~\GeV\lesssim\mzp\lesssim60~\GeV$. The solutions with $\gp=10^{-5}$ in the $(\mvpt,\mhp)$ plane lie far below all the $\zp$ limits considered. A model-specific collider recast is needed only where the decay $\zp\to\vpt\wvp_2$ is open, which on the relic-density contours happens for $\mzp\gtrsim500~\GeV$.

%The collider interpretation of these regions depends on the decay pattern of the $\zp$. When decays into the dark sector are closed, visible searches involving muons and taus provide useful constraints. Once invisible or exotic decay channels become accessible, the visible branching fractions can be reduced, and missing-energy searches become increasingly relevant. A dedicated collider recast including the model-dependent $\zp$ branching fractions would be required for a fully quantitative exclusion of the relic-density-compatible parameter space. 

The present study establishes the existence and main structure of thermal dark matter solutions in this model, but does not constitute a global analysis of dark matter. In particular, direct- and indirect-detection constraints are not included here, and a simultaneous numerical fit of the neutrino sector and the relic abundance remains to be performed. Such a combined analysis, together with lepton-flavor-violation constraints and a model-specific collider recast, would provide the next step toward a global phenomenological analysis of the gauged $L_\mu+L_\tau$ scotogenic scenario.

\acknowledgments 

\noindent
C.A.V.-A. is supported by the Secretar\'ia de Ciencia, Humanidades, Tecnolog\'ia e Innovaci\'on (Secihti) Investigadoras e Investigadores por M\'exico project 749 and SNII 58928, and acknolwledges the Universidad Andrés Bello (UNAB) for hospitality and support during the final stage of this work. Y.V. received partial support by Grant ANR-21-CE31-0013, Project DMwithLLPatLHC, from the French \textit{Agence Nationale de la Recherche} (ANR), Simons Foundation (Award Number:1023171-RC) and Fundação de Amparo e Promoção da Ciência, Tecnologia e Inovação do Rio Grande do Norte (FAPERN) via the call number 06/2026 - FAPERN/IIP, with process number 10910022.000493/2024-04. J. E. P. was supported by ANID (Chile) under FONDECYT Grant No. 3260491. S.K was supported by ANID-Chile FONDECYT 1230160.

\appendix

\section{Anomaly Cancellation}
\label{App:anomaly}
In this Appendix we explicitly show how the potential anomalies from gauged $X\equiv L_\mu+L_\tau$ vanish with the inclusion of the extra fields (vector-like fermions are ommited):
\begin{itemize}
\item $[SU(2)_W]^2U(1)_X$
\begin{equation}
\begin{split}
\sum_{\text{ferm doub}}X_L-\sum_{\text{ferm doub}}X_R&=2\times 1\times 2+1\times1\times 6-1\times1\times 10=0.
\end{split}
\end{equation}
\item $[U(1)_X]^3$
\begin{equation}
\begin{split}
\sum_{\text{fermions}}X^3_L-\sum_{\text{fermions}}X^3_R=&2\times 1\times 2\times 2^3+1\times 1\times 2\times 6^3+1\times 1\times 1\times 10^3\\&+1\times 1\times 1\times 12^3-2\times 1\times 1\times 2^3-2\times 1\times 1\times 8^3\\&-1\times1\times1\times 6^3-1\times1\times1\times\left(-4\right)^3-1\times1\times2\times 10^3=0.
\end{split}
\end{equation}
\item $[\text{Grav}]^2U(1)_X$
\begin{equation}
\begin{split}
\sum_{\text{fermions}}X_L-\sum_{\text{fermions}}X_R=&2\times 1\times 2\times 2+1\times 1\times 2\times 6+1\times 1\times 1\times10\\&+1\times 1\times 1\times 12-2\times 1\times 1\times 2-2\times 1\times 1\times 8\\&-1\times1\times1\times 6-1\times1\times1\times\left(-4\right)-1\times1\times2\times10=0.
\end{split}
\end{equation}
\item $[U(1)_Y]^2U(1)_X$
\begin{equation}
\begin{split}
\sum_{\text{fermions}}Y^2_LX_L-\sum_{\text{fermions}}Y^2_RX_R=&2\times 1\times2\times(-1/2)^2 (2) + 
  1\times1\times2\times (-1/2) ^2 (6) \\&+ 1\times1\times1\times(-1)^2 (10)- 1\times1\times2\times (-1/2)^2 (10)\\& - 2\times1\times1\times (-1)^2 (2)  - 1\times1\times1\times(-1)^2 (6)  =0.
\end{split}
\end{equation}
\item $[U(1)_Y]U(1)_X^2$
\begin{equation}
\begin{split}
\sum_{\text{fermions}}Y_LX^2_L-\sum_{\text{fermions}}Y_RX^2_R=&2\times 1\times2\times(-1/2) (2)^2 + 
  1\times1\times2\times (-1/2) (6)^2 \\&+ 1\times1\times1\times(-1) (10)^2- 1\times1\times2\times (-1/2) (10)^2\\& - 2\times1\times1\times (-1) (2)^2  - 1\times1\times1\times(-1) (6)^2  =0.
\end{split}
\end{equation}
\end{itemize}

%%%%%%%%%%%%%%%%%%%%%%%%%%%%%%%%%%%%%
\section{Diphoton width}\label{sec:diphoton}
%%%%%%%%%%%%%%%%%%%%%%%%%%%%%%%%%%%%
In this appendix we give the one-loop $h\to\gamma\gamma$ amplitude
and discuss the possibility of suppressing its new charged-particle
contributions in the tree-level alignment limit $\theta_h=0$.
We use the CP-conserving charged-fermion parameters specified in
Sec.~\ref{new_leptons}. 
For on-shell photons of momenta $k_1,k_2$, we define the dimensionless
CP-even diphoton amplitude as
\begin{equation}
 \mathcal M=\frac{\alpha_{\rm em}}{2\pi v}\,
 \mathcal A_{\gamma\gamma}
 \big[(k_1\!\cdot k_2)g_{\mu\nu}-k_{2\mu}k_{1\nu}\big]
 \epsilon_1^{*\mu}\epsilon_2^{*\nu}.
\label{eq:app-hgg-tensor}
\end{equation}
The tensor is transverse to both photon momenta.  The corresponding unpolarized decay width is given by
\begin{equation}
 \Gamma(h\to\gamma\gamma)
 =\frac{1}{2!}\frac{1}{16\pi m_h}\sum_{\rm pol}|\mathcal M|^2
 =\frac{\alpha_{\rm em}^2m_h^3}{256\pi^3v^2}
   |\mathcal A_{\gamma\gamma}|^2.
\label{eq:app-hgg-width}
\end{equation}
In the alignment limit, realized for $\lambda_9=0$ with the
doublet state identified as the $125~\GeV$ Higgs boson, the
rotation in Eq.~\eqref{eq:CP_even_scalars} reduces to
\begin{equation}
 h\equiv h_1=\xi,\qquad h_\phi\equiv h_2=\xi_\phi.
 \label{eq:rev_h_rotation}
\end{equation}
The dimensionless Higgs couplings of the physical charged
fermions defined in Eq.~\eqref{eq:new_charged} are
\begin{equation}
    \kappa_i=\frac{v(G_h^C)_{ii}}{m_{C_i}},
\end{equation}
with
\begin{equation}
 G_h^C=U_{CL}^\dagger
 \frac{\partial\mathcal M_C}{\partial v}U_{CR}.
 \label{eq:rev_higgs_derivatives}
\end{equation}
Adopting the notation and conventions  of \cite{Djouadi:2005gi}, the one-loop diphoton amplitude can be written as 
\begin{equation}
 \mathcal A_{\gamma\gamma}=
 \mathcal A^{\mathrm{SM}}_{\gamma\gamma}
 +\sum_{i=1}^4\kappa_i A_{1/2}(\tau_{C_i})
 +\frac{v g_{h\eta^+\eta^-}}{2m_{\eta^\pm}^2}A_0(\tau_\eta),
 \qquad \tau_j=\frac{m_h^2}{4m_j^2}.
 \label{eq:rev_diphoton_amplitude}
\end{equation}
Here $\mathcal A^{\mathrm{SM}}_{\gamma\gamma}$ is the SM amplitude evaluated
with the same input convention, and
\begin{equation}
 A_{1/2}(\tau)=\frac{2[\tau+(\tau-1)f(\tau)]}{\tau^2},\qquad
 A_0(\tau)=\frac{f(\tau)-\tau}{\tau^2},
\label{eq:rev_loop_functions}
\end{equation}
with
\begin{equation}
    f(\tau)= \begin{cases} \arcsin^2\!\sqrt{\tau}, & \tau \leq 1, \\[3mm] -\dfrac{1}{4} \left[ \ln\!\left( \dfrac{1+\sqrt{1-\tau^{-1}}} {1-\sqrt{1-\tau^{-1}}} \right) -i\pi \right]^2, & \tau > 1. \end{cases}
\end{equation}
The heavy-particle limit is
$\tau_j\to0$, in which $A_{1/2}(\tau_j)\to4/3$ and
$A_0(\tau_j)\to1/3$.

The scalar-coupling convention is
$-\mathcal L\supset g_{h\eta^+\eta^-}h\eta^+\eta^-$, with
\begin{equation}
g_{h\eta^+\eta^-}=\lambda'_{10}v .
 \label{eq:rev_eta_higgs}
\end{equation}
Thus alignment alone does not remove the charged-scalar loop.
With the additional choice $\lambda'_{10}=0$, as adopted in the
relic-density scans, this coupling vanishes and $\eta^\pm$ does
not contribute to the one-loop $h\to\gamma\gamma$ amplitude.

For heavy charged fermions, $m_{C_i}\gg m_h$, the contribution
in the same alignment limit becomes
\begin{equation}
 \mathcal A_C\equiv\sum_{i=1}^4\kappa_i A_{1/2}(\tau_{C_i})
 \simeq\frac{4v}{3}\operatorname{Tr}\!\left[
 \mathcal M_C^{-1}\partial_v\mathcal M_C\right]
 =\frac{4v}{3}\partial_v\ln|\det\mathcal M_C|.
 \label{eq:rev_trace_identity}
\end{equation}
Here the derivative is taken at fixed $v_\phi$, Yukawa couplings
and bare mass parameters. Thus, suppression of the heavy-fermion
contribution requires a small logarithmic dependence of the
determinant on the electroweak VEV.

The determinant of the charged mass matrix in
Eq.~\eqref{eq:disc-charged-matrix} can be organized as
\begin{equation}
 \det\mathcal M_C=D_0+D_2+D_4,
 \qquad D_n\propto v^n,\quad n=0,2,4.
\label{eq:rev_determinant}
\end{equation}
Here $D_n$ collects the terms homogeneous of degree $n$ in $v$,
with $v_\phi$, Yukawa couplings and bare masses held fixed.
Consequently $v\partial_v\det\mathcal M_C=2D_2+4D_4$, and
the leading heavy-fermion contribution is
\begin{equation}
 \mathcal A_C^{\rm heavy}
 =\frac43\frac{2D_2+4D_4}{D_0+D_2+D_4}.
 \label{eq:rev_diphoton_Dn}
\end{equation}
Only the electroweak-independent term is needed explicitly:
\begin{equation}
 \begin{aligned}
 D_0&=\left.\det\mathcal M_C\right|_{v=0}
 =-m^Y_1m^Y_2m^\rho_Lm^\rho_R\\
 &=-\frac{y^Y_1y^Y_2\rho_L\rho_R}{4}\,v_\phi^4.
 \end{aligned}
 \label{eq:rev_determinant_vzero}
\end{equation}
When $D_0\neq0$, all four charged fermions remain massive in
the electroweak-symmetric limit. In particular,
\begin{equation*}
 |\det\mathcal M_C|=\prod_{i=1}^4 m_{C_i},
\end{equation*}
where the physical masses are the singular values of
$\mathcal M_C$.
A sufficient condition for suppressing the
leading heavy-fermion amplitude is
\begin{equation}
 |D_0|\gg|D_2|,\ |D_4|.
 \label{eq:rev_diphoton_hierarchy}
\end{equation}
The extended texture therefore permits a nearly SM diphoton
partial width in alignment, with $\lambda'_{10}=0$ and
sufficiently heavy charged fermions. 
The hierarchy above suppresses the logarithmic electroweak-VEV
derivative of the determinant but does not generally suppress the dependence of the
charged-fermion masses on $v_\phi$. Consequently, their
contributions to radiative decays of $h_\phi$ need not be small.
The same Yukawa couplings also affect the exotic-lepton
spectrum and mixings, and hence collider signatures,
$Z\gamma$ amplitudes and electroweak precision observables.
These quantities are not determined by $D_0$ alone; their
simultaneous phenomenological analysis lies beyond the scope
of this work.

\section{Contributions of gauged $U(1)_{L_\mu+L_\tau}$ to EWPOs}
\label{app:EWPO}

In this appendix we summarize the expressions used to compute the electroweak precision observables considered in Sec.~\ref{sec:pheno}. The gauge sector conventions, the diagonalization of the kinetic mixing, the definition of the mixing angle $\xi$, and the physical $\zp$ couplings are presented in
Sec.~\ref{Sec:model}.

We work at leading order in the kinetic mixing parameter $\epsilon$ and away from the quasi-degenerate region $M_{\zp}\simeq M_Z$.

\subsection{Shift of the neutral electroweak mass}

The diagonalization of the neutral gauge sector modifies the relation between the physical $Z$-boson mass and the mass $\widehat M_Z$ that the neutral electroweak gauge boson would have before mixing with the new gauge boson. We
parametrize this effect as
\begin{equation}
\begin{aligned}
\delta_Z
\equiv
\frac{M_Z^2-\widehat M_Z^2}{\widehat M_Z^2}.
\end{aligned}
\end{equation}
At leading order in $\epsilon$, and away from $M_{\zp}\simeq M_Z$, one obtains
\begin{equation}
\begin{aligned}
\delta_Z
\simeq
\epsilon^2 s_W^2
\frac{M_Z^2}{M_Z^2-M_{\zp}^2}.
\end{aligned}
\label{eq:EWPO_deltaZ}
\end{equation}

\subsection{$W$-boson mass}

Although the charged gauge sector is not directly modified in the model, the relation between $M_W$ and the electroweak parameters receives a correction induced by the mixing in the neutral sector. In the analysis we use
\begin{equation}
\begin{aligned}
M_W
=
M_W^{\rm SM}
\left[
1
-
\frac{c_W^2}{2(c_W^2-s_W^2)}
\delta_Z
\right],
\end{aligned}
\label{eq:EWPO_MW}
\end{equation}
where $M_W^{\rm SM}$ denotes the reference SM prediction.

\subsection{Effective $Z$-boson couplings}

The chiral couplings of the SM $Z$ boson are written as
\begin{equation}
\begin{aligned}
G_{Zf,L}^{\rm SM}
=
g_Z z_f^L,
\qquad
G_{Zf,R}^{\rm SM}
=
g_Z z_f^R,
\qquad
g_Z=\sqrt{g^2+g_Y^2},
\end{aligned}
\end{equation}
with
\begin{equation}
\begin{aligned}
z_f^{L,R}
=
T_{f_{L,R}}^3
-
s_W^2 q_f.
\end{aligned}
\end{equation}
Here $q_f$ is the electric charge of the fermion $f$, and $T_{f_{L,R}}^3$ is the corresponding weak isospin.

In the presence of $Z$--$\zp$ mixing, the physical $Z$ couplings are shifted as
\begin{equation}
\begin{aligned}
G_{Zf,L}
&=
G_{Zf,L}^{\rm SM}
+
\Delta G_{Zf,L},
\\
G_{Zf,R}
&=
G_{Zf,R}^{\rm SM}
+
\Delta G_{Zf,R}.
\end{aligned}
\end{equation}
At leading order in the mixing angle, these shifts are given by
\begin{equation}
\begin{aligned}
\Delta G_{Zf,L}
\simeq
\xi\, h_f^L,
\qquad
\Delta G_{Zf,R}
\simeq
\xi\, h_f^R,
\end{aligned}
\label{eq:EWPO_deltaG}
\end{equation}
where $h_f^{L,R}$ denotes the coupling of the canonical Abelian field $V_\mu$ to the corresponding chiral fermion. In the conventions of
Sec.~\ref{sec:pheno},
\begin{equation}
\begin{aligned}
h_f^{L,R}
=
\frac{g^\prime}{2}
Q_{f_{L,R}}^{(\mu+\tau)}
-
\epsilon g_YY_{f_{L,R}}.
\end{aligned}
\label{eq:EWPO_hf}
\end{equation}
The first term corresponds to the direct coupling associated with $U(1)_{L_\mu+L_\tau}$, whereas the second term arises from kinetic mixing with
hypercharge.

\subsection{$Z$-boson partial widths}

For a SM fermion $f$, the partial width of the $Z$ boson can be written as
\begin{equation}
\begin{aligned}
\Gamma_f
=
N_c^f
\frac{M_Z}{24\pi}
\left[
\left|G_{Zf,L}\right|^2
+
\left|G_{Zf,R}\right|^2
\right],
\end{aligned}
\label{eq:EWPO_partial_width}
\end{equation}
where $N_c^f=1$ for leptons and $N_c^f=3$ for quarks.

Expanding to first order in the coupling shifts,
\begin{equation}
\begin{aligned}
\Gamma_f
=
\Gamma_f^{\rm SM}
\left(
1+\delta\Gamma_f
\right),
\end{aligned}
\end{equation}
with
\begin{equation}
\begin{aligned}
\delta\Gamma_f
\simeq
2
\frac{
G_{Zf,L}^{\rm SM}\Delta G_{Zf,L}
+
G_{Zf,R}^{\rm SM}\Delta G_{Zf,R}
}{
\left(G_{Zf,L}^{\rm SM}\right)^2
+
\left(G_{Zf,R}^{\rm SM}\right)^2
}.
\end{aligned}
\label{eq:EWPO_deltaGammaf}
\end{equation}

The total $Z$-boson width is obtained by summing over all kinematically accessible SM fermions,
\begin{equation}
\begin{aligned}
\Gamma_Z
=
\Gamma_Z^{\rm SM}
+
\sum_f
\Gamma_f^{\rm SM}
\delta\Gamma_f.
\end{aligned}
\label{eq:EWPO_GammaZ}
\end{equation}

\subsection{Invisible $Z$-boson width}

The invisible width is computed from the neutrino partial widths,
\begin{equation}
\begin{aligned}
\Gamma_{\rm inv}
=
\Gamma_{\rm inv}^{\rm SM}
+
\sum_{\alpha=e,\mu,\tau}
\Gamma_{\nu_\alpha}^{\rm SM}
\delta\Gamma_{\nu_\alpha}.
\end{aligned}
\label{eq:EWPO_Gammainv}
\end{equation}
Since SM neutrinos do not have a right-handed chiral coupling to the $Z$ boson,
\begin{equation}
\begin{aligned}
G_{Z\nu_\alpha,R}^{\rm SM}=0,
\end{aligned}
\end{equation}
the relative correction to the partial width reduces to
\begin{equation}
\begin{aligned}
\delta\Gamma_{\nu_\alpha}
\simeq
2
\frac{
\Delta G_{Z\nu_\alpha,L}
}{
G_{Z\nu_\alpha,L}^{\rm SM}
}.
\end{aligned}
\label{eq:EWPO_deltaGammav}
\end{equation}
In the present model, $\nu_\mu$ and $\nu_\tau$ receive direct contributions proportional to $\xi g^\prime$, due to their nonzero charges under $U(1)_{L_\mu+L_\tau}$. By contrast, $\nu_e$ is affected only through the effects
induced by kinetic mixing.

\subsection{Effective weak mixing angle}

The effective weak mixing angle is extracted from the ratio between the vector and axial couplings of the $Z$ boson to electrons. We define
\begin{equation}
\begin{aligned}
G_V^e
=
G_{Ze,L}
+
G_{Ze,R},
\qquad
G_A^e
=
G_{Ze,L}
-
G_{Ze,R}.
\end{aligned}
\end{equation}
The electron effective weak mixing angle is then given by
\begin{equation}
\begin{aligned}
\sin^2\theta_{\rm eff}^{e}
=
\frac{1}{4|q_e|}
\left(
1
-
\frac{G_V^e}{G_A^e}
\right).
\end{aligned}
\label{eq:EWPO_sinthetaeff}
\end{equation}
In the reduced EWPO analysis we use only $\sin^2\theta_{\rm eff}^{e}$, avoiding the combination of different leptonic extractions of the effective weak angle.

%#####################
\section{Decays \label{sec:decays}}
%#####################

In this appendix we collect selected partial widths of the new gauge
boson $\zp$ and the singlet-like scalar $h_2\simeq h_\phi$. We state
the physical interaction vertices, their normalization and the
kinematic assumptions before giving the expanded expressions in
Tables~\ref{tab:Zp_decays} and~\ref{tab:Hphi_decays}. 
These decays are central to characterizing the theory's collider signatures, shaping the dark sector's thermal evolution, and assessing the consistency of the benchmark scenarios in our study. 

\paragraph*{Conventions and accuracy.}
We use the integer charges of Table~\ref{tab:3211}, with direct
coupling $g^\prime Q/2$, and the sign conventions of
Eq.~\eqref{eq:Zp_chiral_couplings}. The displayed expressions are
tree-level, on-shell two-body widths. The expanded tables use
perturbatively truncated mixing-dependent couplings. For channels
induced first at $\mathcal O(\epsilon)$, squaring the leading
amplitude gives the leading $\mathcal O(\epsilon^2)$ width.
For channels with a direct coupling, retaining the displayed
$\epsilon^2$ terms is not a complete expansion of the width through
$\mathcal O(\epsilon^2)$. The small-mixing expressions require
$|\epsilon s_W m_Z^2/\Delta_Z|\ll1$, where
$\Delta_Z=m_Z^2-\mzp^2$; near $Z$--$\zp$ degeneracy the neutral
gauge sector must instead be diagonalized without this expansion.
Every two-body width below is understood to vanish below its
physical mass threshold. QCD and other radiative corrections,
as well as decays through virtual intermediate states, are not
included in the tabulated tree-level expressions.
%\endgroup

\subsection*{Decays of the gauge boson $\zp$}
The $\zp$ couplings follow the charge normalization and sign
conventions of Eq.~\eqref{eq:Zp_chiral_couplings}. For $\epsilon=0$
and below all new-particle thresholds, its two-body decays are into
$\nu_\mu\bar\nu_\mu$, $\nu_\tau\bar\nu_\tau$, and the charged-lepton
pairs $\mu^+\mu^-$ and $\tau^+\tau^-$ when kinematically allowed.
Kinetic mixing induces additional SM channels.

The exotic scalar-pair channel $\zp\to\vpt\widetilde{\varphi}_2$
opens for $\mzp>m_{\vpt}+m_{\widetilde{\varphi}_2}$.
Remnant matter parity stabilizes only the lightest odd particle,
taken here to be $\vpt$. The heavier partner may decay as
$\widetilde{\varphi}_2\to\vpt f\bar f$ through virtual $Z$ or $\zp$
exchange, when allowed by the couplings and phase space. Thus this
channel is not automatically invisible: its signature depends on
the secondary branching fractions, the mass splitting and the
partner's decay length. An invisible event requires no detectable
visible secondary decay products, or a partner that escapes the
detector before decaying.

At tree level, for $\epsilon=0$, $\mzp\gg2m_\tau$, and with all other
exotic channels closed, the sum of SM lepton widths is
$\Gamma_{\rm SM}={g^\prime}^2\mzp/(4\pi)$, whereas
$\Gamma(\zp\to\vpt\widetilde{\varphi}_2)
\leq {g^\prime}^2\mzp/(192\pi)$. Consequently,
\[
 \BR(\zp\to\vpt\widetilde{\varphi}_2)
 \leq\frac{1}{49}\simeq2.04\%.
\]
Here SM lepton masses are neglected; scalar mixing and phase-space
suppression can only reduce the bound. This mode therefore cannot
dominate in this regime. 

Here, we use
\begin{equation}
r_i\equiv \frac{m_i}{\mzp}, \qquad \Delta_Z\equiv m_Z^2-\mzp^2,
\end{equation}
\begin{equation}
\lambda(x,y,z) = x^2+y^2+z^2-2xy-2xz-2yz,
\end{equation}
and
\begin{equation}
A_\ell = 8c_W^4m_Z^4(1+2r_\ell^2) -12c_W^2m_Z^2\mzp^2(1+2r_\ell^2) +\mzp^4(5+7r_\ell^2).
\end{equation}
The partial widths in Table~\ref{tab:Zp_decays} follow from the physical vertices and master formulas below.

\paragraph*{Dirac fermions.}
For a physical Dirac mass eigenstate $f$ we define
\begin{equation}
 \mathcal L\supset\zp_\mu\bar f\gamma^\mu
 \left(g_f^L P_L+g_f^R P_R\right)f .
 \label{eq:decay_dirac_vertex}
\end{equation}
The couplings $g_f^{L,R}\equiv g_{\zp f}^{L,R}$ already include
the gauge couplings and charges. With
$P_{L,R}=(1\mp\gamma_5)/2$, the partial width per flavor is
\begin{equation}
 \begin{aligned}
 \Gamma(\zp\to f\bar f)={}&
 \frac{N_c\mzp}{24\pi}\sqrt{1-4r_f^2}
 \\ &\times\left[
 (|g_f^L|^2+|g_f^R|^2)(1-r_f^2)
 +6r_f^2\operatorname{Re}(g_f^Lg_f^{R*})
 \right].
 \end{aligned}
 \label{eq:decay_dirac_width}
\end{equation}
Here $N_c=1$ for leptons and $3$ for quarks, and the initial
spin average is included. Substitution of
Eq.~\eqref{eq:Zp_chiral_couplings} yields the SM-fermion rows of
Table~\ref{tab:Zp_decays}; its quark entries already include color.

\paragraph*{Scalar pairs.}
The derivative-current convention is
\begin{equation}
 \begin{split}
 \mathcal L\supset{}&
 i g_\eta\zp_\mu\eta^-
 \overleftrightarrow{\partial^\mu}\eta^+\\
 &+\sum_{i,j}G_{ij}\zp_\mu
 \left(\varphi_i\partial^\mu\widetilde\varphi_j
 -\widetilde\varphi_j\partial^\mu\varphi_i\right),\\
 g_\eta={}&-\frac12\left[
 g^\prime+\frac{e\epsilon(2c_W^2m_Z^2-\mzp^2)}
 {c_W\Delta_Z}\right].
 \end{split}
 \label{eq:decay_scalar_vertices}
\end{equation}
Using the scalar rotations defined in the model section gives
\begin{equation}
 G=O^\varphi
 \begin{pmatrix}
 \dfrac{g^\prime}{2}-\dfrac{e\epsilon\mzp^2}{2c_W\Delta_Z}&0\\
 0&\dfrac{g^\prime}{2}
 \end{pmatrix}
 (O^{\widetilde\varphi})^T .
 \label{eq:decay_scalar_current}
\end{equation}
At the stated coupling accuracy,
\begin{align}
 \Gamma(\zp\to\eta^+\eta^-)
 &=\frac{|g_\eta|^2\mzp}{48\pi}
 (1-4r_{\eta^+}^2)^{3/2},
 \label{eq:decay_charged_scalar_width}\\
 \Gamma(\zp\to\varphi_i\widetilde\varphi_j)
 &=\frac{|G_{ij}|^2}{48\pi\mzp^5}
 \lambda^{3/2}(\mzp^2,m_{\varphi_i}^2,
 m_{\widetilde\varphi_j}^2).
 \label{eq:decay_neutral_scalar_width}
\end{align}

\paragraph*{Majorana singlet fermions.}
The general scotogenic sector contains four Majorana mass
eigenstates.
% , not two Dirac particles. 
Keeping the convention of
the neutrino-mass section, define
\begin{equation}
 \begin{gathered}
 n_L=(S_{2L},S_{3L},S_{2R}^c,S_{3R}^c)^T,\qquad
 X_L=U^X n_L,\qquad X_\alpha=X_{\alpha L}+X_{\alpha L}^c,\\
 Q_S=\operatorname{diag}(1,1,-1,-1),\qquad
 \mathcal G_X=\frac{g^\prime}{2}U^XQ_S(U^X)^\dagger,\qquad
 G^X_{\alpha\beta}=(\mathcal G_X)_{\alpha\beta}.
 \end{gathered}
 \label{eq:decay_majorana_current}
\end{equation}
In four-component notation the interaction is
\begin{equation}
 \mathcal L_X=\frac12\zp_\mu\sum_{\alpha,\beta}
 \bar X_\alpha\gamma^\mu
 \left(G^X_{\alpha\beta}P_L
       -G^{X*}_{\alpha\beta}P_R\right)X_\beta .
 \label{eq:decay_majorana_vertex}
\end{equation}
For each unordered pair $\alpha\leq\beta$, put
\begin{equation}
 \begin{gathered}
 r_\alpha=\frac{M_\alpha}{\mzp},\qquad
 \Lambda_{\alpha\beta}=\lambda(1,r_\alpha^2,r_\beta^2),\\
 F_{\alpha\beta}=1-\frac{r_\alpha^2+r_\beta^2}{2}
 -\frac{(r_\alpha^2-r_\beta^2)^2}{2}.
 \end{gathered}
 \label{eq:decay_majorana_kinematics}
\end{equation}
For $\mzp>M_\alpha+M_\beta$ the width is
\begin{equation}
 \Gamma^X_{\alpha\beta}=
 \frac{\mzp\sqrt{\Lambda_{\alpha\beta}}}
 {24\pi(1+\delta_{\alpha\beta})}
 \left[
 2|G^X_{\alpha\beta}|^2F_{\alpha\beta}
 -6r_\alpha r_\beta
 \operatorname{Re}\!\left((G^X_{\alpha\beta})^2\right)
 \right],
 \label{eq:decay_majorana_width}
\end{equation}
and it is zero otherwise. The factor $1/(1+\delta_{\alpha\beta})$
accounts for identical final particles; the total is summed over
$\alpha\leq\beta$ without doubling off-diagonal channels.
The $1/2$ in Eq.~\eqref{eq:decay_majorana_vertex} compensates the
Majorana contractions and is not an extra suppression of the
vertex. Since $\mathcal G_X$ is Hermitian, its diagonal entries
are real, and
\begin{equation}
 \Gamma^X_{\alpha\alpha}
 =\frac{(G^X_{\alpha\alpha})^2\mzp}{24\pi}
 (1-4r_\alpha^2)^{3/2}.
 \label{eq:decay_majorana_diagonal}
\end{equation}
The diagonal vector current vanishes; diagonal couplings, if
nonzero, are axial, whereas off-diagonal channels are generally
present.

\begin{table}[h!]
\centering
\renewcommand{\arraystretch}{2.5}
\resizebox{\textwidth}{!}{%
\begin{tabular}{c|c}
\hline
Channel & $\Gamma$ \\
\hline
$\zp\to e^+e^-$&
$\cfrac{
e^2\epsilon^2\,\mzp\sqrt{1-4r_e^2}
}{
96\pi\,c_W^2\left(\Delta_Z\right)^2
}
\left[
8c_W^4m_Z^4(1+2r_e^2)
-12c_W^2m_Z^2\mzp^2(1+2r_e^2)
+\mzp^4(5+7r_e^2)
\right]$\\
$\zp \to \mu^+\mu^{-}, \tau^+\tau^-$&
$\cfrac{
\mzp\sqrt{1-4r_\ell^2}
}{
96\pi\,c_W^2\Delta_Z^2
}
\left[
e^2\epsilon^2 A_\ell
+
4c_We\epsilon g^{\prime}\Delta_Z
(4c_W^2m_Z^2-3\mzp^2)(1+2r_\ell^2)
+
8c_W^2{g^{\prime}}^2\Delta_Z^2(1+2r_\ell^2)
\right]$\\
$\zp\to \nu_e\bar\nu_e$&
$\cfrac{
e^2\epsilon^2\,\mzp^5
}{
96\pi\,c_W^2\left(m_Z^2-\mzp^2\right)^2
}$\\
$\zp\to \nu_{\mu,\tau}\bar\nu_{\mu,\tau}$&
$\cfrac{
\mzp
}{
96\pi c_W^2
\left(m_Z^2-\mzp^2\right)^2
}
\left[
2g^{\prime}c_W
\left(m_Z^2-\mzp^2\right)
-
e\epsilon \mzp^2
\right]^2$\\
$\zp\to q\bar q\;(q=u,c,t)$&
$\cfrac{
e^2\epsilon^2\,\mzp\sqrt{1-4r_q^2}
}{
288\pi\,c_W^2\left(\Delta_Z\right)^2
}
\left[
32c_W^4m_Z^4(1+2r_q^2)
-40c_W^2m_Z^2\mzp^2(1+2r_q^2)
+\mzp^4(17+7r_q^2)
\right]$\\
$\zp\to q\bar q\;(q=d,s,b)$&
$\cfrac{
e^2\epsilon^2\,\mzp\sqrt{1-4r_q^2}
}{
288\pi\,c_W^2\left(\Delta_Z\right)^2
}
\left[
8c_W^4m_Z^4(1+2r_q^2)
-
4c_W^2m_Z^2\mzp^2(1+2r_q^2)
+
\mzp^4(5-17r_q^2)
\right]$\\
$\zp\to X_\alpha X_\beta\;(\alpha\leq\beta)$&
$\Gamma^X_{\alpha\beta}\quad
\text{[Eq.~\eqref{eq:decay_majorana_width}]}$\\
$\zp\to W^+W^-$&
$\cfrac{
\epsilon^2c_W^2s_W^4\,m_Z^4\,\mzp^5
}{
12\pi e^2v^4\left(\Delta_Z\right)^2
}
\sqrt{1-4r_W^2}
\left(
1+16r_W^2-68r_W^4-48r_W^6
\right)$\\
$\zp\to \eta^+\eta^-$&
$\cfrac{
\mzp
}{
192\pi
}
\left(1-4r_{\eta^+}^2\right)^{3/2}
\left[
g^{\prime}
+
\cfrac{
e\epsilon\left(2c_W^2m_Z^2-\mzp^2\right)
}{
c_W\left(\Delta_Z\right)
}
\right]^2$\\
$\zp\to h_1 Z$&
$\cfrac{
\epsilon^2\,
\lambda^{1/2}
\left(\mzp^2,m_{h_1}^2,m_Z^2\right)
}{
768\pi\,c_W^4s_W^2\,m_Z^2\mzp^5
\left(\Delta_Z\right)^2
}
\left[
m_{h_1}^4+m_Z^4+\mzp^4
+10m_Z^2\mzp^2
-2m_{h_1}^2(m_Z^2+\mzp^2)
\right]
\left[
c_H e^2 v\,\mzp^2-4s_H{g^{\prime}}^2s_W^2c_W^2v_\phi\,m_Z^2
\right]^2$\\
$\zp\to h_2 Z$&
$\cfrac{
\epsilon^2\,
\lambda^{1/2}
\left(\mzp^2,m_{h_2}^2,m_Z^2\right)
}{
768\pi\,c_W^4s_W^2\,m_Z^2\mzp^5
\left(\Delta_Z\right)^2
}
\left[
m_{h_2}^4+m_Z^4+\mzp^4
+10m_Z^2\mzp^2
-2m_{h_2}^2(m_Z^2+\mzp^2)
\right]
\left[
e^2s_Hv\,\mzp^2
+
4c_H{g^{\prime}}^2s_W^2c_W^2v_\phi\,m_Z^2
\right]^2$\\
$\zp\to \vpo\widetilde{\varphi}_1$&
$\cfrac{
\lambda^{3/2}
\left(\mzp^2,m_{\vpo}^2,m_{\widetilde{\varphi}_1}^2\right)
}{
192\pi\,\mzp^5
}
\left[
g^{\prime}
\left(
c_\varphi c_{\widetilde{\varphi}}
+
s_\varphi s_{\widetilde{\varphi}}
\right)
-
\cfrac{
e\epsilon \mzp^2
}{
c_W\left(\Delta_Z\right)
}
c_\varphi c_{\widetilde{\varphi}}
\right]^2$\\
$\zp\to \vpt\widetilde{\varphi}_2$&
$\cfrac{
\lambda^{3/2}
\left(
\mzp^2,
m_{\vpt}^2,
m_{\widetilde{\varphi}_2}^2
\right)
}{
192\pi\,\mzp^5
}
\left[
g^{\prime}
\left(
c_\varphi c_{\widetilde{\varphi}}
+
s_\varphi s_{\widetilde{\varphi}}
\right)
-
\cfrac{
e\epsilon \mzp^2
}{
c_W\left(m_Z^2-\mzp^2\right)
}
s_\varphi s_{\widetilde{\varphi}}
\right]^2$\\
$\zp\to \vpo\widetilde{\varphi}_2$&
$\cfrac{
\lambda^{3/2}
\left(
\mzp^2,
m_{\vpo}^2,
m_{\widetilde{\varphi}_2}^2
\right)
}{
192\pi\,\mzp^5
}
\left[
g^{\prime}
\left(
c_\varphi s_{\widetilde{\varphi}}
-
s_\varphi c_{\widetilde{\varphi}}
\right)
-
\cfrac{
e\epsilon \mzp^2
}{
c_W\left(m_Z^2-\mzp^2\right)
}
c_\varphi s_{\widetilde{\varphi}}
\right]^2$\\
$\zp\to \vpt\widetilde{\varphi}_1$&
$\cfrac{
\lambda^{3/2}
\left(
\mzp^2,
m_{\vpt}^2,
m_{\widetilde{\varphi}_1}^2
\right)
}{
192\pi\,\mzp^5
}
\left[
g^{\prime}
\left(
c_\varphi s_{\widetilde{\varphi}}
-
s_\varphi c_{\widetilde{\varphi}}
\right)
+
\cfrac{
e\epsilon \mzp^2
}{
c_W\left(m_Z^2-\mzp^2\right)
}
s_\varphi c_{\widetilde{\varphi}}
\right]^2$\\
\hline
\end{tabular}%
}
\caption{Partial decay widths of $\zp$. Each SM-fermion entry is per flavor; color is included for quarks. The $X_\alpha X_\beta$ row refers to physical Majorana states and is counted once for each unordered pair.}
\label{tab:Zp_decays}
\end{table}

\subsection*{Decays of the singlet-like scalar $h_2$}

The tree-level couplings of the singlet-like scalar
$h_2\simeq h_\phi$ to SM states arise through scalar and gauge
mixing, while its interactions with the symmetry-breaking sector
allow decays into new particles. The channel
$h_2\to\vpt\vpt$ is invisible when open, with a width controlled by
the corresponding trilinear coupling. Closing this channel does
not, however, imply that SM final states dominate. SM-like decays
induced by scalar mixing are suppressed by $\sin^2\theta_h$ and
compete with all open new-particle channels.

In the scan benchmark $\theta_h=\epsilon=0$, the tree-level widths
into SM fermions, $W^+W^-$ and $ZZ$ vanish. The decay
$h_2\to\zp\zp$ nevertheless remains allowed for
$m_{h_2}>2\mzp$. In this limit $\mzp=g^\prime v_\phi$, and its
tree-level width reduces to
\[
 \Gamma(h_2\to\zp\zp)
 =\frac{m_{h_2}^3}{32\pi v_\phi^2}
 \sqrt{1-4x}\,(1-4x+12x^2),
 \qquad x=\frac{\mzp^2}{m_{h_2}^2}.
\]
For $m_{h_2}\gg\mzp$, this approaches
$m_{h_2}^3/(32\pi v_\phi^2)$: longitudinal $\zp$ polarizations
prevent a suppression by small $g^\prime$ at fixed $v_\phi$.
The subsequent $\zp$ decays determine whether the cascade yields
visible particles, missing energy, or a combination of both.

Other odd-scalar and new-fermion pairs must also be included when
open. If the leading two-body channels are closed, loop-induced
decays and decays through virtual intermediate states can become
important. Examples include $h_2\to\gamma\gamma$ and
$h_2\to\zp Z^{\prime *}\to\zp f\bar f$, provided the relevant couplings
and final-state phase space allow them; no universal branching
hierarchy follows from alignment alone.

Near $m_{h_2}\simeq2m_{\vpt}$, the resonant annihilation discussed
in \cref{sec:pheno} is sensitive to the total scalar width.
The propagator must therefore use a width computed from the same
parameters and all open channels, including those not displayed
in the selected two-body widths of Table~\ref{tab:Hphi_decays}.
The resonance does not modify the individual decay formulas,
but its quantitative effect cannot be inferred from the pole
mass alone.

\paragraph*{Scalar trilinears and symmetry factors.}
For real CP-even mass eigenstates we use
\begin{equation}
 -\mathcal L\supset\frac12h_2\sum_{i,j}
 g^{(2)}_{ij}\varphi_i\varphi_j,\qquad
 g^{(2)}_{ij}=g^{(2)}_{ji}.
 \label{eq:decay_h2_trilinears}
\end{equation}
The CP-odd sector has the analogous interaction with
$\widetilde g^{(2)}_{ij}$ and $\widetilde\varphi_i$.
The two-body width is
\begin{equation}
 \Gamma(h_2\to\varphi_i\varphi_j)=
 \frac{|g^{(2)}_{ij}|^2}
 {16\pi m_{h_2}^3(1+\delta_{ij})}
 \lambda^{1/2}(m_{h_2}^2,m_{\varphi_i}^2,m_{\varphi_j}^2),
 \label{eq:decay_h2_scalar_master}
\end{equation}
with the same expression after replacing the fields and couplings
by their CP-odd counterparts. Thus identical real particles give
a factor $1/2$, while distinct states do not.

For an independent vertex definition, let
$\mathcal M_{\rm even}^2$ and $\mathcal M_{\rm odd}^2$ denote
the inert $2\times2$ subblocks of
Eqs.~\eqref{MCPeven2} and~\eqref{MCPodd2}. With
$c_H=\cos\theta_h$ and $s_H=\sin\theta_h$,
\begin{align}
 g^{(2)}
 &=O^\varphi
 \left(-s_H\partial_v+c_H\partial_{v_\phi}\right)
 \mathcal M_{\rm even}^2\,(O^\varphi)^T,
 \label{eq:decay_h2_even_vertex}\\
 \widetilde g^{(2)}
 &=O^{\widetilde\varphi}
 \left(-s_H\partial_v+c_H\partial_{v_\phi}\right)
 \mathcal M_{\rm odd}^2\,(O^{\widetilde\varphi})^T .
 \label{eq:decay_h2_odd_vertex}
\end{align}
The derivatives act on the weak-basis mass matrices at fixed
Lagrangian parameters, before the rotations. In terms of the
auxiliary quantities defined below Table~\ref{tab:Hphi_decays},
\begin{align}
 g^{(2)}_{22}&=c_H A_2-s_H B_2,\qquad
 g^{(2)}_{12}=\frac{\mathcal C_{12}}2,
 \label{eq:decay_h2_even_mapping}\\
 \widetilde g^{(2)}_{22}
 &=-\left(c_{\widetilde\varphi}^2A_{\widetilde\varphi_2}
 -c_{\widetilde\varphi}s_{\widetilde\varphi}B_{\widetilde\varphi_2}
 +s_{\widetilde\varphi}^2D_{\widetilde\varphi_2}\right),
 \label{eq:decay_h2_odd_diagonal_mapping}\\
 \widetilde g^{(2)}_{12}
 &=-\frac12\left[
 \cos(2\theta_{\widetilde\varphi})\widetilde A_{12}
 +\sin(2\theta_{\widetilde\varphi})\widetilde B_{12}
 \right].
 \label{eq:decay_h2_odd_mixed_mapping}
\end{align}
Consequently the mixed-pair entries correctly contain
$\mathcal C_{12}^2/(64\pi m_{h_2}^3)$ and its CP-odd analogue:
their displayed combinations are twice the physical vertices
(up to an overall sign), not the vertices themselves.

For the distinct charged particles we instead define
\begin{equation}
 \begin{gathered}
 -\mathcal L\supset g_{h_2\eta^+\eta^-}h_2\eta^+\eta^-,
 \qquad
 g_{h_2\eta^+\eta^-}=-s_H\lambda'_{10}v+c_H\lambda_8v_\phi,\\
 \Gamma(h_2\to\eta^+\eta^-)
 =\frac{|g_{h_2\eta^+\eta^-}|^2}{16\pi m_{h_2}}
 \sqrt{1-\frac{4m_{\eta^+}^2}{m_{h_2}^2}} .
 \end{gathered}
 \label{eq:decay_h2_charged_vertex}
\end{equation}
No identical-particle factor is present here. The Yukawa
normalization in the SM-fermion rows is $y_f=\sqrt2m_f/v$.
All the trilinear couplings above have mass dimension one.
%\endgroup

\begin{table}[h!]
\centering
\renewcommand{\arraystretch}{2.5}
\resizebox{\textwidth}{!}{%
\begin{tabular}{c|c}
\hline
Channel & $\Gamma$ \\
\hline
$h_2\to W^+W^-$ & $\cfrac{s_H^2\,m_{h_2}^3}{16\pi v^2}
\sqrt{1-4\beta_W^2}
\left(1-4\beta_W^2+12\beta_W^4\right)$\\
$h_2\to q\bar q$&$\cfrac{N_{c}\,s_H^2\,y_{q}^2\,m_{h_2}}{16\pi}
\left(1-4\beta_{q}^2\right)^{3/2}$\\
$h_2\to \ell\bar \ell$&$\cfrac{s_H^2\,y_{\ell}^2\,m_{h_2}}{16\pi}
\left(1-4\beta_{\ell}^2\right)^{3/2}$\\
$h_2\to \eta^+\eta^-$&
$\cfrac{
\sqrt{1-4\beta_{\eta^+}^2}
\left(\bl{\lambda'_{10}}s_H v-\lambda_8 c_H v_\phi\right)^2
}{
16\pi\,m_{h_2}
}$\\
$h_2\to \zp\zp$&
$\cfrac{
\sqrt{1-4\beta_{\zp}^2}
\left(1-4\beta_{\zp}^2+12\beta_{\zp}^4\right)
}{
512\pi\,c_W^4\,m_{h_2}\,\beta_{\zp}^4
} 
\left[\cfrac{e^2\epsilon^2 s_{H} v\,\mzp^4}{\left(m_Z^2-\mzp^2\right)^2} - 4 c_{H} c_W^2 {g^{\prime}}^2 v_{\phi} \right]^2$\\
$h_2\to ZZ$&
$\cfrac{
\sqrt{1-4\beta_Z^2}
\left(1-4\beta_Z^2+12\beta_Z^4\right)
}{
512\pi\,c_W^4s_W^4\,m_{h_2}\,\beta_Z^4
}
\left[
e^2s_Hv
-
\cfrac{
4c_Hc_W^2s_W^4{g^{\prime}}^2\epsilon^2v_\phi\,m_Z^4
}{
\left(m_Z^2-\mzp^2\right)^2
}
\right]^2$\\
$h_2\to \vpt\vpt$&
$\cfrac{
\sqrt{1-4\beta_{\vpt}^2}
}{
32\pi\,m_{h_2}
}
\left(
c_H A_2-s_H B_2
\right)^2$\\
$h_2\to \widetilde{\varphi}_2\widetilde{\varphi}_2$&
$\cfrac{
\sqrt{1-4\beta_{\widetilde{\varphi}_2}^2}
}{
32\pi\,m_{h_2}
}
\left(
c_{\widetilde{\varphi}}^2 A_{\widetilde{\varphi}_2}
-
c_{\widetilde{\varphi}}s_{\widetilde{\varphi}} B_{\widetilde{\varphi}_2}
+
s_{\widetilde{\varphi}}^2 D_{\widetilde{\varphi}_2}
\right)^2$\\
$h_2\to h_1h_1$&
$\cfrac{
\sqrt{1-4\beta_{h_1}^2}
}{
32\pi\,m_{h_2}
}
\left(
6c_Hs_H
\left(
\lambda_1 c_Hv-\lambda_3s_Hv_\phi
\right)
+
\lambda_9
\left[
s_Hv(1-3c_H^2)
+
c_Hv_\phi(3s_H^2-1)
\right]
\right)^2$\\
$h_2\to \zp Z$&
$\cfrac{
\lambda^{1/2}
\left(m_{h_2}^2,m_Z^2,\mzp^2\right)
}{
256\pi\,c_W^4s_W^2\,m_{h_2}^3\,m_Z^2\mzp^2
}
\left[
m_{h_2}^4+m_Z^4+\mzp^4
+10m_Z^2\mzp^2
-2m_{h_2}^2(m_Z^2+\mzp^2)
\right]
\left[
\cfrac{
\epsilon
\left(
e^2s_Hv\,\mzp^2
+
4c_H{g^{\prime}}^2s_W^2c_W^2v_\phi\,m_Z^2
\right)
}{
m_Z^2-\mzp^2
}
\right]^2$\\
$h_2\to \vpo\vpt$&
$\cfrac{
\lambda^{1/2}
\left(m_{h_2}^2,m_{\vpo}^2,m_{\vpt}^2\right)
}{
64\pi\,m_{h_2}^3
}
\left(\mathcal{C}_{12}\right)^2$\\
$h_2\to \widetilde{\varphi}_1\widetilde{\varphi}_2$&
$\cfrac{
\lambda^{1/2}
\left(m_{h_2}^2,m_{\widetilde{\varphi}_1}^2,m_{\widetilde{\varphi}_2}^2\right)
}{
64\pi\,m_{h_2}^3
}
\left(
\cos(2\theta_{\widetilde{\varphi}})\,\widetilde A_{12}
+
\sin(2\theta_{\widetilde{\varphi}})\,\widetilde B_{12}
\right)^2$\\
&\\
\hline
\end{tabular}%
}
\caption{Selected two-body partial widths of $h_2$, with the vertex and symmetry-factor conventions stated in the text.} \label{tab:Hphi_decays}
\end{table}

The auxiliary combinations used in Table~\ref{tab:Hphi_decays} are defined by

$$
\beta_{i}\equiv \frac{m_{i}}{m_{h_2}},
\qquad
c_H\equiv \cos\theta_h,
\qquad
s_H\equiv \sin\theta_h,
\qquad
c_\varphi\equiv \cos\theta_\varphi,
\qquad
s_\varphi\equiv \sin\theta_\varphi ,
\qquad
c_{\widetilde{\varphi}}\equiv \cos\theta_{\widetilde{\varphi}},
\qquad
s_{\widetilde{\varphi}}\equiv \sin\theta_{\widetilde{\varphi}}.
$$

$$
A_2
=
\sqrt{2}\,c_\varphi^2\kappa_2
-c_\varphi s_\varphi \lambda_{11}v
+c_\varphi^2\lambda_5v_\phi
+s_\varphi^2\lambda_8v_\phi .
$$

$$
B_2
=
v\left[
c_\varphi^2\lambda_7
+s_\varphi^2\left(\lambda_{10}+\lambda'_{10}\right)
\right]
-c_\varphi s_\varphi
\left(
\sqrt{2}\kappa_1+\lambda_{11}v_\phi
\right).
$$

$$
\mathcal{C}_{12}
=
\cos(2\theta_\varphi)
\left[
\lambda_{11}
\left(
c_Hv-s_Hv_\phi
\right)
-
\sqrt{2}\,\kappa_1s_H
\right]
+
\sin(2\theta_\varphi)
\left[
c_H
\left(
\sqrt{2}\,\kappa_2
+
(\lambda_5-\lambda_8)v_\phi
\right)
+
s_Hv
\left(
\lambda_{10}+\lambda'_{10}-\lambda_7
\right)
\right].
$$

$$
A_{\widetilde{\varphi}_2}
=
\sqrt{2}\,c_H\kappa_2
+\lambda_7s_Hv
-c_H\lambda_5v_\phi.
$$

$$
B_{\widetilde{\varphi}_2}
=
\sqrt{2}\,\kappa_1s_H
+c_H\lambda_{11}v
-\lambda_{11}s_Hv_\phi.
$$

$$
D_{\widetilde{\varphi}_2}
=
(\lambda_{10}+\lambda'_{10})s_Hv
-c_H\lambda_8v_\phi.
$$

$$
\widetilde A_{12}
=
\lambda_{11}
\left(
c_Hv-s_Hv_\phi
\right)
+
\sqrt{2}\,\kappa_1s_H ,
$$

$$
\widetilde B_{12}
=
c_H
\left[
\sqrt{2}\,\kappa_2
-
(\lambda_5-\lambda_8)v_\phi
\right]
-
s_Hv
\left(
\lambda_{10}+\lambda'_{10}-\lambda_7
\right).
$$

%\endgroup % End Appendix D revision-link color scope.
\bibliographystyle{utphys}
\bibliography{bibliography}
\end{document}